\documentclass[iop]{emulateapj}

\usepackage{xspace}
\usepackage{color}
\usepackage{rotating}
\usepackage{subfigure}
\usepackage{afterpage}
\usepackage{morefloats}
\def\imagebox#1#2{\vtop to #1{\null\hbox{#2}\vfill}} 

\def\mearth{{\rm\,M_\oplus}}
\def\rearth{{\rm\,R_\oplus}}
\def\msun{{\rm\,M_\odot}}
\def\rsun{{\rm\,R_\odot}}
\def\lsun{{\rm\,L_\odot}}
\def\gsim{~\rlap{$>$}{\lower 1.0ex\hbox{$\sim$}}}
\def\lsim{~\rlap{$<$}{\lower 1.0ex\hbox{$\sim$}}}

\def\eg{{\it e.g.\ }}

\def\ie{{\it i.e.\ }}

\def\acen{{$\alpha$~Cen}}

\def\kepler{{\it Kepler}}

\def\vplanet{\texttt{\footnotesize{VPLANET}}\xspace}
\def\atmesc{\texttt{\footnotesize{ATMESC}}\xspace}
\def\distorb{\texttt{\footnotesize{DISTORB}}\xspace}
\def\distrot{\texttt{\footnotesize{DISTROT}}\xspace}
\def\eqtide{\texttt{\footnotesize{EQTIDE}}\xspace}

\def\radheat{\texttt{\footnotesize{RADHEAT}}\xspace}
\def\thermint{\texttt{\footnotesize{THERMINT}}\xspace}
\def\stellar{\texttt{\footnotesize{STELLAR}}\xspace}
\def\galhabit{\texttt{\footnotesize{GALHABIT}}\xspace}

\usepackage{amsmath}
\begin{document}

\title{The Habitability of Proxima Centauri b I: Evolutionary Scenarios}
\author{Rory Barnes\altaffilmark{1,2,3}, Russell Deitrick\altaffilmark{1,2}, Rodrigo Luger\altaffilmark{1,2}, Peter E. Driscoll\altaffilmark{4,2}, Thomas R. Quinn\altaffilmark{1,2}, David P. Fleming\altaffilmark{1,2}, Benjamin Guyer\altaffilmark{1,2}, Diego V. McDonald\altaffilmark{1,2}, Victoria S. Meadows\altaffilmark{1,2}, Giada Arney\altaffilmark{1,2}, David Crisp\altaffilmark{5,2}, Shawn D. Domagal-Goldman\altaffilmark{6,2}, Daniel Foreman-Mackey\altaffilmark{1,7}, Nathan A. Kaib\altaffilmark{8}, Andrew Lincowski\altaffilmark{1,2}, Jacob Lustig-Yaeger\altaffilmark{1,2}, Eddie Schwieterman\altaffilmark{1,2}}
\altaffiltext{1}{Astronomy Department, University of Washington, Box 951580, Seattle, WA 98195}
\altaffiltext{2}{NASA Astrobiology Institute -- Virtual Planetary Laboratory Lead Team, USA}
\altaffiltext{3}{E-mail: rory@astro.washington.edu}
\altaffiltext{4}{Department of Terrestrial Magnetism, Carnegie Institution for Science, Washington, DC}
\altaffiltext{5}{Jet Propulsion Laboratory, California Institute of Technology, M/S 183-501, 4800 Oak Grove Drive, Pasadena, CA 91109}
\altaffiltext{6}{Planetary Environments Laboratory, NASA Goddard Space Flight Center, 8800 Greenbelt Road, Greenbelt, MD 20771}
\altaffiltext{7}{NASA Sagan Fellow}
\altaffiltext{8}{Department of Physics and Astronomy, University of Oklahoma, 440 W. Brooks St, Norman, OK 73019}

\begin{abstract}
We analyze the evolution of the potentially habitable planet Proxima
Centauri b to identify environmental factors that affect its long-term
habitability. We consider physical processes acting on size scales
ranging from the galactic to the stellar system to the planet's core.
We find that there is a significant probability that
Proxima Centauri has had encounters with its companion stars, Alpha
Centauri A and B, that are close enough to destabilize an extended planetary system. If the system has an additional
planet, as suggested by the discovery data, then it may perturb
planet b's eccentricity and inclination, possibly driving those
parameters to non-zero values, even in the presence of strong tidal
damping. We also model the internal evolution of the planet, evaluating
the roles of different radiogenic abundances and tidal heating and
find that magnetic field generation is likely for billions of years. We
find that if planet b formed {\it in situ}, then it
experienced $169\pm 13$ million years in a runaway greenhouse as the star
contracted during its formation. This early phase could remove up to 5 times as much water as in the modern Earth's oceans, possibly producing a large abiotic
oxygen atmosphere. On the other hand, if Proxima Centauri b formed
with a substantial hydrogen atmosphere (0.01 -- 1\% of the planet's mass), then this envelope could have shielded
the water long enough for it to be retained before being
blown off itself. After modeling this wide range of processes we conclude that water retention during the host star's pre-main sequence phase is the biggest obstacle for Proxima b's habitability. These results are all obtained with a new
software package called \vplanet.

\end{abstract}

\section{Introduction\label{sec:intro}}

The discovery of Proxima Centauri b, hereafter Proxima b, revealed the closest
possible exoplanet and one of the most observationally accessible planets
orbiting a late-type M dwarf host.    Very little is currently known about
Proxima b and its environment, but the planet is likely terrestrial and
receives an incident flux that places it in the ``habitable zone'' (HZ)
\citep{Kasting93,Selsis07,Kopparapu13}.   Although it does not transit
\citep{Kipping17}, its proximity and favorable star-planet contrast ratio
make Proxima b an exciting target in the near term for phase curve observations
with JWST \citep{Turbet16,KreidbergLoeb16,Meadows18}, and for early direct
imaging characterization efforts, especially with ground-based Extremely Large
Telescopes \citep{Meadows18}.  Modifications to instrumentation on the Very
Large Telescope (VLT) may combine high-contrast imaging with high-resolution
spectroscopy to enable the search for O$_2$ in Proxima b's atmosphere in the next
few years \citep{Lovis17}.

The interpretation of these spectra require a firm understanding of
the history of Proxima b and its host system. Proxima b exists in an
environment that is significantly different from Earth and has likely
experienced different phenomena that could preclude or promote the
development of life. When viewed across interstellar distances,
biology is best understood as a planetary process: life is a global
phenomenon that alters geochemical and photochemical
processes \citep{Lovelock65}. Unambiguous spectroscopic indicators of life, \ie biosignatures, can
only be identified if the abiotic processes on a planet are understood
-- no single feature in a spectrum is a ``smoking gun'' for life. A
robust detection of extraterrestrial life requires that all plausible
non-biological sources for an observed spectral feature can be ruled
out. This requirement is a tall order in light of the expected
diversity of terrestrial exoplanets in the galaxy and the plethora of
mechanisms capable of mimicking biosignatures
\citep{Schwieterman16,Meadows17}.  With these challenges in mind,
Proxima b may still offer the best opportunity to search for unequivocal signs of life beyond the Solar System.

In this study, we leverage the known (but sparse) data on Proxima b and
its host system to predict the range of evolutionary pathways that the
planet may have experienced. As we show below, many evolutionary histories are
possible and depend on factors ranging from the cooling rate of b's
core to the orbital evolution of the stellar system through the Milky
Way galaxy, and everything in between. The evolution of Proxima b, and by
extension its potential habitability, depends on physical processes
that tend to be studied by scientists from different
fields, such as geophysics and astrophysics. However, for the purpose
of interpreting Proxima b, these divisions must be overcome. A critical examination of the potential habitability of Proxima b necessitates
a cohesive model that can fold in the impact of the many factors that shape its evolutionary
history. Our examination of Proxima b will draw on simple, but realistic, models
that have been developed in the fields of geophysics, planetary
science, atmospheric science and astrophysics. From this synthesis, we
identify numerous opportunities and obstacles for life to develop on
Proxima b, as well as numerous possible uninhabitable states. These calculations lay a foundation for future interpretation
of spectroscopic observations, which are explored in the companion paper \citep{Meadows18}. Additionally, many of the principles described are relevant to any potentially habitable planet orbiting low mass stars, such as TRAPPIST-1 d--f \citep{Gillon17} and LHS~1140~b~\citep{Dittman17}.

This paper is organized as follows. In $\S$~\ref{sec:obs} we review
the observational data on the system and the immediate implications
for habitability. In $\S$~\ref{sec:models} we describe models to
simulate the evolution of the system, with a focus on habitability. In
this section we introduce a new software package, \vplanet, which
couples physical models of planetary interiors, atmospheres, spins and
orbits, stellar evolution, and galactic effects. In
$\S$~\ref{sec:results} we present results of these models. An
exhaustive analysis of all histories is too large to present here, so we only present suites for phenomena that are well-constrained and/or have a large impact on habitability. In $\S$~\ref{sec:disc} we discuss the
results and identify additional observations that could improve
modeling efforts and connect our results to the companion paper
\citep{Meadows18}. Finally, in $\S$~\ref{sec:concl} we draw our
conclusions.

\section{Observational Constraints} \label{sec:obs}

In this section we review known features of the triple star system
Alpha Centauri (hereafter \acen) of which Proxima Centauri is likely a
third member. This star system has been studied carefully for centuries as
it is the closest to the Sun. We will first review the direct
observational data, then we will make inferences from those data, and finally we qualitatively consider how these data
constrain the possibility for life to exist on Proxima b, which guide the quantitative modeling described in the subsequent sections. Transit searches have failed to turn up definitive evidence of one \citep{Kipping17,Li17,Liu18}.

\subsection{Properties of the Proxima Planetary System}
\label{sec:obs:planetsys}
Very little data exist for Proxima b. The radial velocity data
reveal a planet with a minimum mass $m$ of 1.27~$\mearth$, an orbital period $P$
of 11.186 days, and an orbital eccentricity $e$ less than 0.35 \citep{AngladaEscude16}. Aside from the orbital phase, these
data are the extent of the direct observational data on the planet,
but even the minimum mass relies on uncertain estimates of the mass of
the host star, described below.

Proxima b may not be the only planet orbiting Proxima
Centauri. The Doppler data suggest the presence of another planetary
mass companion with an orbital period near 215 days, but it is not
definitive yet \citep{AngladaEscude16,DamassoDelSordo17}. If present, the second planet has a
projected mass of $\lesssim~10~\mearth$, consistent with previous
non-detections \citep{EndlKurster08,Barnes14,Lurie14}. Recent ALMA observations have revealed a dust disk located from $\sim$1--4 AU \citep{Anglada17}, which is significantly farther out then the putative second planet. The orbital
eccentricity and its relative inclination to Proxima b's orbit are also
unknown, but as described below, could take any value that permits
dynamical stability of planet b and the dust belt. Additionally, lower mass and/or more distant
planetary companions could also be present in the system.

\subsection{Properties of the Host Star}
\label{sec:obs:star}
As Proxima Centauri is the closest star to the Sun, it has been studied
extensively since its discovery 100 years ago \citep{Innes1915}.  It
has a radius $R_*$ of $0.14~\rsun$, a temperature $T_{eff}$ of 3050 K, a
luminosity $L_*$ of $0.00165~\lsun$~\citep{Demory09,Boyajian12}, and a rotation
period $P_*$ of 83 days \citep{Benedict98}. \cite{AngladaEscude16} find a
spectral type of M5.5V. \cite{Wood01} searched for
evidence of stellar winds, but found none, indicating mass loss rates
$\dot{M}_*$ less than 20\% of our Sun's, \ie
$<4~\times~10^{-15}~\msun/\textrm{yr}$. Proxima Centauri possesses a
much larger magnetic field ($B~\sim~600$~G) than our Sun ($B~=~1$~G)
\citep{ReinersBasri08}, but somewhat low compared to the majority of
low mass stars.

Like our Sun, Proxima Centauri's luminosity varies slowly with time
due to starspots \citep{Benedict93}. HST observations of Proxima
Centauri found variations up to 70 milli-magnitudes (mmag) in $V$
\citep{Benedict98}, which, if indicative of the bolometric luminosity (which is unlikely),
corresponds to about a 17.5\% variation, with a
period of 83.5 days (\ie the rotation period). Moreover,
\cite{Benedict98} found evidence for two discrete modes of
variability, one lower amplitude mode ($\Delta~V~\sim~30$~mmag) with a
period of $\sim 42$ days, and a larger amplitude mode
($\Delta~V~\sim~60$~mmag) with a period of 83 days. These periods are
a factor of 2 apart, leading \cite{Benedict98} to suggest that
sometimes a large spot (or cluster of spots) is present on one
hemisphere only, while at other times smaller spots exist on opposite
hemispheres. Regardless, incident stellar radiation (``instellation'')
variations of 17\% could impact atmospheric evolution and surface
conditions of a planet (the sun's variation is of order 0.1\%
\citep{Willson81}).

Additionally, the magnetic field strength may vary with
time. \cite{Cincunegui07} monitored the Ca II H and K lines, which are
indicators of chromospheric activity, as well as H$\alpha$ for 7 years
and found modest evidence for a 442 day cycle in stellar
activity. Their result has recently been corroborated by \cite{Wargelin17}, and modeled by \cite{Yadav16}. Although the strength of Proxima's magnetic field at the
orbit of planet b is uncertain, it could affect the stability of
b's atmosphere and potentially affect any putative life on b \citep{Garraffo16,Airapetian17,Atri17}.

Proxima Centauri is a known flare star
\citep{Shapley51}\footnote{Although Shapley is the sole author of his
  1951 manuscript, the bulk of the work was performed by two
  assistants, acknowledged only as Mrs. C.D. Boyd, and Mrs. V.M. Nail.}  and indeed several flares
were reported during the Pale Red Dot campaign
\citep{AngladaEscude16}. \cite{Walker81} performed the first study of
the frequency of flares as a function of energy, finding that high
energy events ($\sim~10^{30}$ erg) occurred about once per day, while
lower energy events ($\sim~10^{28}$ erg) occurred about once per
hour. Numerous observational campaigns since then have continued to
find flaring at about this frequency
\citep{Benedict98,AngladaEscude16,Davenport16}.

\subsection{Properties of the Stellar System}
\label{sec:obs:stellarsys}
Many of the properties of Proxima Centauri are inferred from its
relationship to \acen~A and B, thus a discussion of the current
knowledge of \acen~is warranted here.
Proxima Centauri is $\sim 13,000$ AU from \acen~A and B, but all three
have the same motion through the galaxy. The proper motion and radial
velocity of the center of mass of \acen~A and B permit the calculation
of the system's velocity relative to the sun. \cite{Poveda96} find the
three velocities are ($U, V, W$) = (-25, -2, 13) km/s for the center
of mass. This velocity implies the system is currently moving in the
general direction of the Sun, and is on a roughly circular orbit around the galaxy with an
eccentricity of 0.07 \citep{AllenHerrera98}.

A recent, careful analysis of astrometric and HARPS RV data by
\cite{PourbaixBoffin16} found the masses of the two stars to be 1.133
and $0.972~\msun$, respectively, with an orbital eccentricity of 0.52
and a period of 79.91 years. The similarities between both A and B and
the Sun, as well as their low apparent magnitudes, has allowed
detailed studies of their spectral and photometric properties. These
two stars (as well as Proxima) form a foundation in stellar astrophysics,
and hence a great deal is known about A and B. However, as we describe
below, many uncertainties still remain regarding these two stars.

The spectra of \acen~A and B provide information about the stellar
temperature, gravitational acceleration in the photosphere, rotation
rate, and chemical composition. That these features can be
measured turns out to be critical for our analysis of the evolution of
Proxima b. Proxima Centauri is a low mass star with strong molecular
absorption lines and non-local thermal equilibrium effects, which make it extraordinarily
difficult to identify elemental abundances; its composition
is far more difficult to measure than for G and K dwarfs like \acen~ A
and B \citep{Johnson2009}.  Recently, \cite{HinkelKane13} completed a
reanalysis of published compositional studies, rejecting the studies
of \cite{Laird85} and \cite{NeuforgeMagain97} because they were too
different from the other 5 they considered.  \cite{HinkelKane13}
  found the mean metallicity [Fe/H] of each of the two stars to be
+0.28 and +0.31 and with a large spread of 0.16 and 0.11,
respectively. While it is frustrating that different groups have
arrived at significantly different iron abundances, it is certain the
stars are more metal-rich than the Sun.

\cite{HinkelKane13} go on to examine 21 other elements, including C,
O, Mg, Al, Si, Ca, and Eu. These elements can be important for the
bulk composition and/or are tracers of other species that are relevant
to planetary processes. In nearly all cases, the relative abundance of
these elements to Fe is statistically indistinguishable from the solar
ratios. Exceptions are Na, Zn and Eu in \acen~A, and V, Zn, Ba
and Nd in \acen~B. The discrepancies between the two stars is
somewhat surprising given their likely birth from the same molecular
cloud. On the other hand, the high eccentricity of their orbit could
point toward a capture during the open cluster phase \citep[\eg][]{Malmberg07}. For all
elements beside Eu, the elemental abundances relative to Fe are larger
than in the Sun. In particular, it seems likely that the stars are
significantly enriched in Zn.

\acen~A and B are large and bright enough for asteroseismic
studies that can reveal physical properties and ages of stars to a few
percent, for high enough quality data \citep{Chaplin2014}. Indeed, these two stars
are central to the field of asteroseismology, and have been studied in
exquisite detail \citep[e.g.][]{Bouchy01,Bouchy02}. However, significant uncertainties persist in our
understanding of these stars, despite all the observational
advantages.

A recent study undertook a comprehensive Bayesian analysis of \acen~A
with priors on radius, composition, and mass derived from
interferometric, spectroscopic and astrometric measurements,
respectively \citep{Bazot16}. Their adopted metallicity constraint
comes from \cite{NeuforgeMagain97} via \cite{Thoul03}, which was
rejected by the \cite{HinkelKane13} analysis. They also used an older
mass measurement from \cite{Pourbaix02}, which is slightly smaller
than the updated mass from \cite{PourbaixBoffin16}. They then used an
asteroseismic code to determine the physical characteristics of
A. Although the mass of A is similar to the Sun at $1.1~\msun$, the
simulations of \cite{Bazot16} found that \acen~A's core lies at the
radiative/convective boundary and the transition between pp- and
CNO-dominated energy production chains in the core. Previous results
found the age of \acen~A to be 4.85 Gyr with a convective core
\citep{Thevenin02}, or 6.41 Gyr without a convective core
\citep{Thoul03}. The ambiguity is further increased by uncertainty in
the efficiency of the $^{14}$N(p,$\gamma$)$^{15}$O reaction rate in
the CNO cycle, and by the possibility of
convective overshooting of hydrogen into the core. They also consider
the role of ``microscopic diffusion,'' the settling of heavy
elements over long time intervals. All these uncertainties
prevent a precise and accurate measurement of \acen~A's
age. Combining the different model predictions and including 1$\sigma$
uncertainties, the age of \acen~A is likely to be between 3.4 and 5.9 Gyr,
with a mean of 4.78 Gyr.

\acen~B has also been studied via asteroseismology, but as with A, the
results have not been consistent. \cite{Lundkvist14} find a nominal age of 1.5
Gyr with ``Asteroseismology Made Easy,'' but with uncertainties in excess of 4 Gyr. The
asteroseismic oscillations on B are much smaller than on A, which make
analyses more difficult \citep[see, \it{e.g.},][]{CarrierBourban03},
leading to the large uncertainty. Combining studies of A and B, we
must conclude that the ages of the two stars are uncertain by at least
25\%. Given the difficulty in measuring B's asteroseismic pulsations,
we will rely on A's asteroseismic data and assume the age of A and B (and Proxima)
to be $4.8^{+1.1}_{-1.4}$ Gyr.

\clearpage
\subsection{Inferences from the Observational Data}
\label{sec:obs:inf}
Because Proxima b was discovered indirectly, its properties and
evolution depend critically on our knowledge of the host star's
properties. Although many properties of Proxima Centauri are known, the mass $M_{Prox}$,
age, and composition are not. The spectra and luminosity suggest
the mass of Proxima is $\sim~0.12~\msun$ \citep{Delfosse00}. If we
adopt this value, then the semi-major axis of b's orbit is 0.0485~AU
and the planet receives 65\% of the instellation Earth receives
from the Sun \citep{AngladaEscude16}.  Note that \cite{Sahu14}
suggested that Proxima's proper motion sent it close enough to two
background stars for the general relativistic deflection of their
light by Proxima to be detectable with HST and should allow the determination of
$M_{Prox}$ to better than 10\%, but those results are not yet available.

Additional inferences rely on the assumption that Proxima formed with
the \acen~binary.  The similarities in the proper motion and parallax
between Proxima and \acen~immediately led to speculation as to whether
the stars are ``physically connected or members of the same drift''
\citep{Voute1917}, \ie are they bound or members of a moving group?
If Proxima is just a random star in the solar neighborhood,
\cite{MatthewsGilmore93} concluded that the probability that Proxima would
appear so close to \acen~is about 1 in a million, suggesting it is
very likely the stars are somehow associated with each other. Using
updated kinematic information, \cite{Anosova94} concluded that Proxima
is not bound, but instead part of a moving group consisting of about a
dozen stellar systems. \cite{WertheimerLaughlin06}'s reanalysis found
that the observational data favor a configuration that is at the
boundary between bound and unbound orbits. However, their best fit
bound orbit is implausibly large as the semi-major axis is 1.31 pc,
\ie larger than the distance from Earth to
Proxima. \cite{MatvienkoOrlov14} also failed to unequivocally resolve
the issue, and concluded that RV precision of better than 20 m/s is
required to determine if Proxima is bound, which should be available
in the data from \cite{AngladaEscude16}. Most recently,
\cite{Kervella17} improved upon previous RV measurements and
found a probability of $4 \times 10^{-8}$
that Proxima is \emph{not} gravitationally bound to \acen,
and obtained a reasonable best fit orbit
($a=8700^{+0.7}_{-0.4}$ AU,
$e=0.5^{+0.08}_{-0.09}$), and that Proxima is currently near apoastron.

Regardless of whether or not Proxima is bound to \acen, the very low
probability that the stars would be so close to each other strongly
supports the hypothesis that the stars formed in the same star
cluster. We will assume that they are associated and have
approximately equal ages and similar compositions. An age near 5 Gyr
for Proxima is also consistent with its slow rotation period and relatively
modest activity levels and magnetic field \citep{ReinersBasri08}.

Planet formation around M dwarfs is still relatively understudied. Few
observations of disks of M dwarfs exist
\citep[\eg][]{Hernandez07,WilliamsCieza11,Luhman12,Downes15}, but
these data point to a wide range of lifetimes for the gaseous
disks of 1--15~Myr. This timescale is likely longer than the time to form
terrestrial planets in the HZs of late M dwarfs
\citep{Raymond07,Lissauer07}, and hence Proxima b may have been fully
formed before the disk dispersed. For Proxima, the lifetime of the
protoplanetary disk is unknown, and could have been altered by the
presence of \acen~A and B, so any formation pathway or evolutionary
process permitted within this constraint is plausible \citep{Coleman17,AlibertBenz17}.

The radial velocity data combined with $M_{Prox}$ only provide a
minimum mass for the planet, but significantly larger planet masses are geometrically
unlikely, and very large masses can be excluded because they would
incite detectable astrometric signals (note that the minimum mass
solution predicts an astrometric orbit of the star of $\sim$1
microsecond of arc), or would be detectable via  direct imaging \citep{Mesa17}. Assuming the probability distribution of Proxima b's orbital plane is isotropic, there is a 50\% chance that its mass is  $<2.84~\mearth$ \citep{Luger17}. \cite{BixelApai17} convolve planet occurrence rates for M dwarfs and mass-radius relationships from \kepler~\citep{DressingCharbonneau13,WeissMarcy14,Rogers15} to arrive at mass and radius estimates of $1.63^{+1.66}_{-0.72}~\mearth$ and $1.07^{+0.38}_{-0.31}~\rearth$, respectively, and that there is at least an 80\% probability the planet is rocky.  However, even at the minimum mass, we
cannot exclude the possibility that Proxima b possesses a significant
hydrogen envelope, and is better described as a ``mini-Neptune,''
which is unlikely to be habitable \citep[but
  see][]{PierrehumbertGaidos11}.

If non-gaseous, the composition is still highly uncertain and depends
on the unknown formation process. Several possibilities exist
according to recent theoretical studies: 1) the planet formed {\it in
  situ} from local material; 2) the planet formed at a larger
semi-major axis and migrated in while Proxima still possessed a
protoplanetary disk; or 3) an instability in the system occurred that
impulsively changed b's orbit. For case 1, the planet may be depleted
in volatile material \citep{Raymond07,Lissauer07,Coleman17}, but could still
initially possess a significant water reservoir \citep{Ciesla15,Mulders15,AlibertBenz17}. For case 2,
the planet would have likely formed beyond the snow line and hence
could initially be very water-rich \citep{CarterBond12}. Such a formation-then-migration scenario may be likely as previous studies of in situ planet formation about M dwarfs have found it is difficult to form Earth-mass and greater planets (Raymond et al. 2007). For case 3, the planet
could be formed either volatile-rich or poor depending on its initial
formation location as well as the details of the instability, such as
the frequency of impacts that occurred in its aftermath. We conclude
that all options are possible given the data and for simplicity will
assume the planet is Earth-like in composition. If we adopt the minimum mass from \cite{AngladaEscude16} as a fiducial, the
silicate planet scaling law of \cite{Sotin07}, and  then the radius of a
$1.27~\mearth$ planet is $1.07~\rearth$, assumptions consistent with \cite{BixelApai17}.

\subsection{Implications for Proxima b's Evolution and Habitability}
\label{sec:obs:imp}

Given the above observations and their immediate implications, this
planet may be able to support life. All life on Earth requires three
basic ingredients: Water, energy, and the bioessential elements C, H,
O, N, S and P. Additionally, these ingredients must be present in an
environment that is stable in terms of temperature, pressure, and pH
for long periods of time. Proxima b clearly receives significant energy from its host star, and the the bioessential elements are some of the most common in the galaxy. Thus, we assume that liquid water is the limiting factor for Proxima b to be habitable, and we adopt a working definition of ``habitable'' to be that the planet has liquid surface water.

Proxima's luminosity and effective temperature combined with b's
orbital semi-major axis place the planet in the HZ of
Proxima and nearly in the same relative position of Earth in the Sun's
HZ in terms of instellation. Specifically, the planet receives about 65\% of Earth's
instellation, which, due to the redder spectrum of Proxima, places b comfortably in the ``conservative'' HZ of
\cite{Kopparapu13}. Even allowing for observational uncertainties,
\cite{AngladaEscude16} find that the planet is in this
conservative HZ.

However, its habitability depends on many more factors than just the
instellation. The planet must form with sufficient water and maintain
it over the course of the system's age. Additionally, even if water is present,
the evolution and potential habitability of Proxima b depends on many
other factors involving stellar effects, the planet's internal properties, and the
gravitational influence of the other members of the stellar system.

The host star is about 10 times smaller and less massive than the Sun,
the temperature is about half that of the Sun, and the luminosity is
just 0.1\% that of the Sun. These differences are significant and can
have a profound effect on the evolution of Proxima b. Low mass stars
can take billions of years to begin fusing hydrogen into helium in
their cores, and the star's luminosity can change dramatically during that
time. Specifically, the star contracts at roughly constant temperature and so
the star's luminosity drops with time. For the case of Proxima, this
stage lasted $\sim~1$~Gyr \citep{Baraffe15} and hence Proxima b
could have spent significant time interior to the HZ. This ``pre-main sequence'' (pre-MS) phase could
either strip away a primordial hydrogen atmosphere to reveal a
``habitable evaporated core'' \citep{Luger15}, or, if b formed as a
terrestrial planet with abundant water, it could desiccate that planet during an early runaway greenhouse phase and build up an
oxygen-dominated atmosphere \citep{LugerBarnes15}. Thus, the large
early luminosity of the star could either help or hinder
b's habitability.

Low mass stars also show significant activity, \ie flares, coronal
mass ejections, and bursts of high energy radiation
\citep[\eg][]{West08}. This activity can change the composition of the
atmosphere through photochemistry, or even completely strip the
atmosphere away \citep{Raymond06}. The tight orbit of Proxima b places
it at risk of atmospheric stripping by these phenomena. A planetary
magnetic field could increase the probability of atmospheric retention
by deflecting charged particles, or it could decrease it by funneling
high energy particles into the magnetic poles and provide enough
energy to drive atmospheric escape \citep{Strangeway10}. Either
way, knowledge of the frequency of flaring and other high energy
events, as well as of the likelihood that Proxima b possesses a
magnetic field, would be invaluable information in assessing the
longevity of Proxima b's atmosphere.

The close-in orbit also introduces the possibility that tidal effects
are significant on the planet. Tides can affect the planet in five
ways. First, they could cause the rotation rate to evolve to a
frequency that is equal to or similar to the orbital frequency, a
process typically called tidal locking
\citep{Dole64,Kasting93,Barnes17}, or into a spin-orbit resonance or other super-synchronous state \citep{Ribas16,ZanazziLai17}. Second, they can drive the planet's
obliquity $\psi$ to zero or $180^{\circ}$, such that the planet has no seasons
\citep{Heller11}. Third, they can cause the orbital eccentricity to
change, usually (but not always) driving the orbit toward a circular
shape \citep{Darwin1880,FerrazMello08,Barnes17}. Fourth, they can cause
frictional heating in the interior, known as tidal heating
\citep{Peale79,Jackson08c,Barnes13}. Finally, they can cause the
semi-major axis to decay as orbital energy is transformed into
frictional heat, possibly pulling a planet out of the HZ
\citep{Barnes08}. Except in extreme cases, these processes
are unlikely to sterilize a planet, but they can profoundly affect the
planet's evolution \citep{DriscollBarnes15}.

Many researchers have concluded that tidally locked planets of M
dwarfs are unlikely to support life because their atmospheres would
freeze out on the dark side \citep{Kasting93}. However, numerous
follow-up calculations have shown that tidal locking is not likely to
result in uninhabitable planets
\citep{Joshi97,Pierrehumbert11,Wordsworth11,Yang13,Shields16,Kopparapu16}. These
models all find that winds are able to transport heat to the back side
of the planet for atmospheres larger than about 0.3 bars. Even below 0.3 bar, it may be possible for ocean currents to transport sufficient energy to ameliorate the temperature difference \citep{Yang14}. In fact,
synchronous rotation may actually allow habitable planets to exist
closer to the host star because cloud coverage develops at the
sub-stellar point and increases the planetary albedo
\citep{Yang13}. Thus, tidal locking may increase a planet's potential
to support life. \cite{Turbet16} find with 3-D models that Proxima b could be habitable regardless of its current spin state.

Although the abundances of elements relative to iron in \acen~A and B,
and, by assumsion Proxima, are similar to the Sun's, there is no
guarantee that the abundance pattern is matched in Proxima
b. Planet formation is often a stochastic process and composition
depends on the impact history of a given world. The planet could have
formed near its current location, which would have been relatively hot
early on and the planet could be relatively depleted in volatiles
\citep{Raymond07,Mulders15}. These studies may even overestimate
volatile abundances as they ignored the high luminosities that late M
dwarfs have during planet formation. Alternatively, the planet could
have formed beyond the snow line and migrated in either while the gas
disk was still present, or later during a large scale dynamical
instability. In those cases, the planet could be
volatile-rich.

If the abundances of Proxima are indeed similar to \acen~A and B, then
the depletion of Eu in \acen~A is of note as it is often a tracer
of radioactive material like $^{232}$Th and $^{238}$U
\citep{Young14}. These isotopes are primary drivers of the internal
energy of Earth, and hence if they are depleted in Proxima b, its
internal evolution could be markedly different than Earth's. However,
since no depletion is observed in \acen~B, it is far from clear that
such a depletion exists. One interesting radiogenic possibility is
that the planet formed within 1~Myr \citep{Raymond07}, then
$^{26}$Al could still provide energy to the planet's interior, possibly altering the planets thermal evolution. Hence any
prediction of b's evolution should also consider its role.

The presence of additional planets can change the orbit and obliquity
of planet b through gravitational perturbations. These interactions
can change the orbital angular momentum of b and drive oscillations in the eccentricity
$e$, the inclination $i$, longitude of periastron $\varpi$, and
longitude of ascending node $\Omega$. Changes in inclination can lead
to changes in $\psi$ as the planet's rotational axis is likely fixed
in inertial space, except for precession caused by the stellar torque,
while the orbital plane precesses. These variations can significantly
affect climate evolution and possibly even the planet's potential to
support life \citep{Armstrong14}.

If Proxima is bound to \acen~A and B, then perturbations by passing
stars and torques by the galactic tide can cause drifts in Proxima's
orbit about A and B \citep{Kaib13}. During epochs of high
eccentricity, Proxima may swoop so close to A and B that their gravity
is able disrupt Proxima's planetary system. This could have occurred
at any time in Proxima's past and can lead to a total rearrangement of
the system. Thus, should additional planets exist in the Proxima
planetary system, these could be present on almost any orbit consistent with the dust ring, possibly with
large eccentricities and large mutual inclinations relative to b's
orbital plane \citep[\eg][]{Barnes11}. The dust ring could even be a result of a recent collision between bodies orbiting Proxima that were destabilized by a recent encounter between Proxima and \acen~A and B.

The inferred metallicity of Proxima Centauri is quite high for the
solar neighborhood, which has a mean of -0.11 and standard deviation
of 0.18 \citep{AllendePrieto04}. Indeed, recent simulations of stellar
metallicity distributions in the galaxy find that at the sun's
galactic radius $R_{gal}$ of $\sim$8 kpc, stars cannot form with
[Fe/H] $> +0.15$ \citep{Loebman16}. The discrepancy can be resolved by
invoking radial migration \citep{SellwoodBinney02}, in which stars on
nearly circular orbits are able to ride corotation resonances with
spiral arms either inward and outward. \cite{Loebman16} find that with
migration, the metallicity distribution of stars in the Sloan Digital
Sky Survey III's Apache Point Observatory Galactic Evolution
Experiment \citep{Hayden15} is reproduced. Furthermore, Loebman et
al.\ find that stars in the solar neighborhood with [Fe/H] $> +0.25$
must have formed at $R_{gal} < 4.5$~kpc. Similar conclusions were
reached in an analysis of the RAVE survey by \cite{Kordopatis15}.  We
conclude that this system has migrated outward at least 3.5~kpc, but
probably more. As the surface density scale length of the galaxy is
$\sim$2.5 kpc, this implies that the density of stars at their
formation radius was at least 5 times higher than at the Sun's current
Galactic radius.

The observed and inferred constraints for the evolution of Proxima b
are numerous, and the plausible range of evolutionary pathways is
diverse. The proximity of two solar-type stars complicates the
dynamics, but allows the extension of their properties to Proxima
Centauri. In the next sections we apply quantitative models of the
processes described in this section to the full stellar system in order to
explore the possible histories of Proxima b in detail. The outcomes of these histories serve as a foundation for the modeling efforts in Paper II \citep{Meadows18}.

\section{Models}\label{sec:models}

In this section we describe the models we use to consider the
evolution and potential habitability of Proxima b. We generally use
published models that are common in different disciplines of
science. Although the models come from disparate sources, we have
compiled them all into a new software program called \vplanet. This
code is designed to simulate planetary system evolution, with a focus on
habitability. The problem of habitability is interdisciplinary, but we
find it convenient to break the problem down into more manageable
chunks, which we call ``modules,'' which are incorporated when
applicable. At this time, \vplanet~consists of simple models that are
all representable as sets of ordinary differential equations. Below we
describe qualitatively the modules and direct the reader to the
references for the quantitative description. We then briefly describe
how \vplanet~unifies these modules and integrates the system forward.

\subsection{Stellar Evolution: \stellar}
\label{sec:models:stellar}

Of the many stellar evolutionary tracks available in the literature
\citep[\eg][]{Baraffe98,Dartmouth08,Baraffe15}, we find that the
Yonsei-Yale tracks for low-mass stars \citep{YonseiYale13} provide the
best match to the stellar parameters of Proxima Centauri. We selected
the [Fe/H] = +0.3 track with mixing length parameter
$\alpha_\mathrm{MLT} = 1.0$ and linearly interpolated between the $0.1
\msun$ and $0.15 \msun$ tracks to obtain a track at $M_{Prox} = 0.12
\msun$.  While these choices yield a present-day radius within
$1\sigma$ of $0.1410 \pm 0.0070 \rsun$ \citep{Demory09,Boyajian12}, the model
predicts a luminosity $L_\star$ at $t = 4.78\ \mathrm{Gyr}$ that is $\sim 20\%$
higher than the value reported in \cite{Demory09} (a $\sim 2.2\sigma$
discrepancy). Such a discrepancy is not unexpected, given both the
inaccuracies in the evolutionary models for low mass stars and the
large intrinsic scatter of the luminosity and radius of M dwarfs at
fixed mass and metallicity, likely due to unmodeled magnetic effects
\citep{YonseiYale13}. Moreover, the large uncertainties in the age,
mass, and metallicity of Proxima Centauri (\S\ref{sec:obs}) further
contribute to the inconsistency.

Nevertheless, since we are concerned with the present-day habitability
of Proxima b, it is imperative that our model match the present-day
luminosity of its star. We therefore scale the Yonsei-Yale luminosity
track down to match the observed value, adjusting the evolution of the
effective temperature to be consistent with the radius evolution
(which we do not change). We note that this choice results in a
\emph{lower} luminosity for Proxima Centauri at all ages, which yields
conservative results (``optimistic'' in terms of habitability) for the total amount of water lost from the planet (\S\ref{sec:results:atmesc}). Moreover, this adjustment likely has a smaller effect on our results than the large uncertainties on the properties of the star and the planet.

We also model the fiducial evolution of the XUV luminosity of the star as in
\cite{LugerBarnes15}. We use the power-law of \cite{Ribas05} with
power law exponent $\beta = -1.23$, a saturation fraction
$L_\mathrm{xuv}/L_\mathrm{bol} = 10^{-3}$ and a saturation time of 1
Gyr. These choices yield a good match to the present-day value,
$L_\mathrm{xuv}/L_\mathrm{bol} = 2.83\times 10^{-4}$
\citep{Boyajian12}.

We use the fiducial $L_\star$ and $L_\mathrm{xuv}$ evolution tracks discussed above to obtain the evolution of Proxima b's water content (\S\ref{sec:models:atmesc}). Although this procedure yields the maximum likelihood estimate of the planet's present-day water content conditioned on our assumptions, the large uncertainties associated with each of the parameters in our model result in even larger uncertainties on our result. To account for this, we run a suite of Markov Chain Monte Carlo (MCMC) simulations conditioned on the observational uncertainties associated with the model parameters to derive posterior probability distributions for Proxima b's water content and other quantities of interest. We discuss this procedure in detail in \S\ref{sec:results:atmesc}.

\begin{figure*}[ht]
\centering
\includegraphics[width=\textwidth]{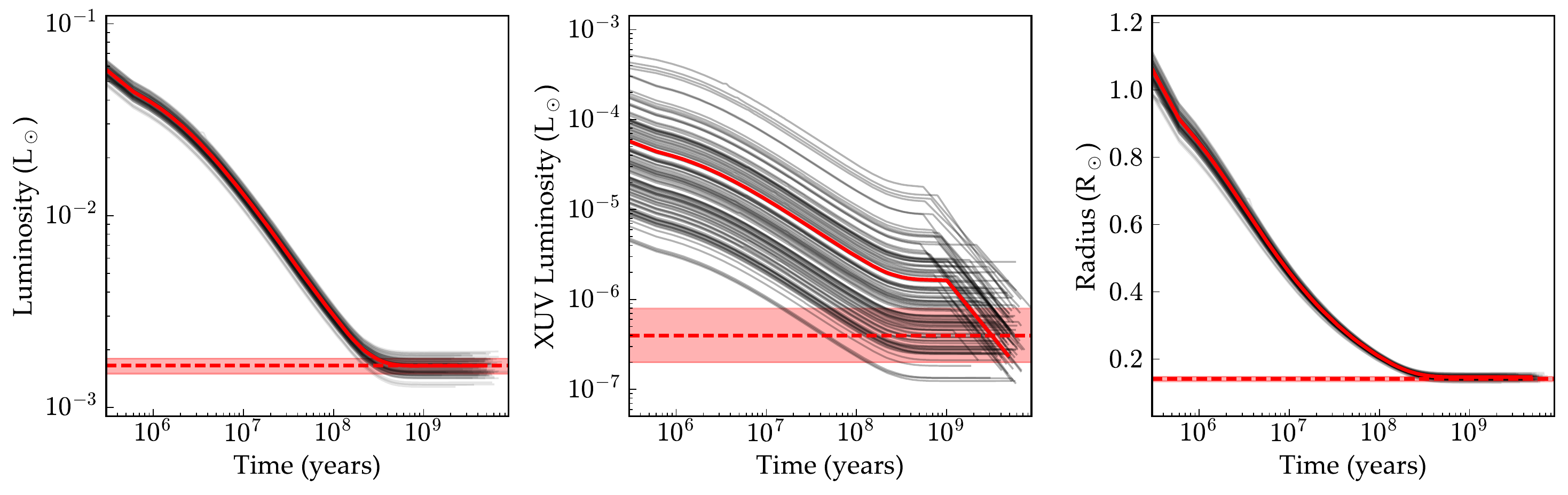}
\caption{Luminosity, XUV luminosity, and radius evolution of Proxima
  Centauri. The dashed red lines
  indicate the measured values of each parameter (see text). 1$\sigma$
  uncertainties are shaded in light red. The mean (fiducial) evolution is indicated by the solid red tracks. The black lines correspond to 100 randomly drawn posterior samples from our MCMC chains (see \S\ref{sec:results:atmesc}.) By construction, all tracks
  match the observed values at the present day within 1$\sigma$. }
\label{fig:stellar:evol}
\end{figure*}

In Fig.~\ref{fig:stellar:evol} we plot the evolution of the stellar luminosity, XUV luminosity, and radius as a function of time from $t_0$ = 1
Myr to the mean system age of 4.78 Gyr. We show the mean tracks (solid red lines) along with 100 samples randomly drawn from the posterior distributions of our MCMC runs (black lines). The present-day values are indicated by the dashed red lines, with 1$\sigma$ uncertainties shaded. Note the long ($\sim$ 1 Gyr) pre-MS phase studied by \cite{LugerBarnes15}, and the high XUV luminosity during the first several 100 Myr.

\subsection{Atmospheric Escape: \atmesc}
\label{sec:models:atmesc}

We model atmospheric escape under the energy-limited
\citep{Watson81,Erkaev07} and diffusion-limited \citep{Hunten73}
parameterizations, closely following \cite{Luger15} and
\cite{LugerBarnes15}. We refer the reader to those papers for a
detailed description of the equations and methodology. In this section we outline the main adaptations and improvements to the models therein.

We model both the escape of hydrogen from a putative thin primordial
envelope and the escape of hydrogen and oxygen from photolysis of
water during an early runaway greenhouse. As in \cite{LugerBarnes15},
we set water escape rates to zero once the planet enters the HZ, since
the establishment of a cold trap should limit the availability of
water in the upper atmosphere. We further assume that if Proxima b forms with a hydrogen envelope, it must be completely lost before water can escape,
given the expected large diffusive separation between light H atoms
and heavy H$_2$O molecules. We shut off hydrodynamic escape at 1 Gyr,
the approximate time at which the star reaches the main sequence, to
account for the transition to ballistic escape predicted by
\cite{OwenMohanty16}. We assume fiducial XUV escape efficiencies
$\epsilon_\mathrm{xuv}$ for hydrogen and water of 0.15 but also consider values
ranging down to 0.01. Finally, for hydrogen-rich cases, we use the radius evolution tracks for super-Earths of \cite{Lopez12} and \cite{LopezFortney14}; when no envelope is present, we use the terrestrial mass-radius relation of \cite{Sotin07} to compute the planet's radius.

The rate of escape of a steam atmosphere closely depends on the fate
of photolytically-produced oxygen. We compute the hydrodynamic drag on
oxygen atoms using the formalism of \cite{Hunten87} to obtain oxygen
escape rates, tracking the buildup of O$_2$ in the atmosphere. As in
\cite{Tian15} and \cite{Schaefer16}, we account for the increasing
mixing ratio of O$_2$ at the base of the hydrodynamic flow, which
slows the escape of hydrogen. \cite{Tian15} finds that as oxygen
becomes the dominant species in the upper atmosphere, the
\cite{Hunten87} formalism predicts that an oxygen-dominated flow can
rapidly lead to the loss of all O$_2$ from planets around M
dwarfs. However, because an oxygen atom's mass, $m_O$, is larger than a hydrogen's, $m_H$,
hydrodynamic oxygen-dominated escape requires exospheric temperatures
$\sim m_\mathrm{O}/m_\mathrm{H} = 16$ times higher than that for a
hydrogen-dominated flow, which is probably unrealistic for Proxima
b. Following the prescription of \cite{Schaefer16}, we therefore
shut off oxygen escape once its mixing ratio exceeds $X_\mathrm{O} =
0.6$, switching to the diffusion-limited escape rate of
hydrogen. Finally, as in \cite{LugerBarnes15}, we also consider the
scenario in which sinks at the surface are efficient enough to remove
O$_2$ from the atmosphere at the rate at which it is produced,
resulting in an atmosphere that never builds up substantial amounts of
oxygen. Recently, \cite{Schaefer16} used a magma ocean model to
calculate the rate of O$_2$ absorption by the surface, showing that
final atmospheric O$_2$ pressures may range from zero to hundreds or
even thousands of bars for the hot Earth GJ 1132b. Our two scenarios
(no O$_2$ sinks and efficient O$_2$ sinks) should therefore bracket
the atmospheric evolution of Proxima b.

We caution, finally, that the energy-limited formalism we adopt here is a very approximate description of the escape of an atmosphere to space. The heating of the upper atmosphere that drives hydrodynamic escape is strongly wavelength dependent and varies with both the composition and the temperature structure of the atmosphere. Moreover, line cooling mechanisms such as recombination radiation scale nonlinearly with the incident flux. Nonthermal escape processes, such as those controlled by magnetic fields, flares, and/or coronal mass ejections, lead to further departures from the simple one-dimensional energy-limited escape rate. Nevertheless, several studies show that for small planets the escape rate does indeed scale with the stellar XUV flux and inversely with the gravitational potential energy of the gas \citep[e.g.,][]{Lopez12,Lammer2013,OwenWu13,OwenWu17} and that $\epsilon_\mathrm{xuv} \approx 0.1$ is a reasonable median value that predicts the correct escape fluxes within a factor of a few. Since presently we have no information about Proxima b's atmospheric composition, we choose to employ the energy-limited approximation and fold all of our uncertainty regarding the physics of the escape process into the XUV escape efficiency $\epsilon_\mathrm{xuv}$, which we vary between 0.01 and 0.15. This is roughly the range of escape efficiencies predicted by \cite{Bolmont16} for the XUV fluxes received by Proxima b between its formation and the present day and should thus bracket reasonable escape rates for Proxima b.

\subsection{Tidal Evolution: \eqtide}
\label{sec:models:eqtide}
To model the tidal evolution of the Proxima system, we will use a
simple, but commonly-used model called the ``constant-phase-lag''
model \citep{Goldreich66,Greenberg09,Heller11}. This model reduces the
evolution to two parameters, the ``tidal quality factor'' $Q$ and the
Love number of degree 2, $k_2$. While this model has known
shortcomings \citep{ToumaWisdom94,EfroimskyMakarov13}, it provides a
qualitatively accurate picture of tidal evolution, and produces
similar results as the competing constant-time lag model
\citep{Heller10,Barnes13,Barnes17}. Moreover, \cite{Kasting93} used
CPL to calculate the ``tidal lock radius.'' For this study, we use the
model described in \cite{Heller11}, and validated it by reproducing the
tidal evolution of the Earth-Moon orbit \citep{MacDonald64,Barnes17} and the
tidal heating of Io \citep{Peale79}. We call this moduke \eqtide~as it is nearly identical to the code by the same name \citep{Barnes17}\footnote{\eqtide is available at https://github.com/RoryBarnes}, and which served as a template for \vplanet~development.

The values of $Q$ and $k_2$ for Earth are well-constrained by lunar
laser ranging \citep{Dickey94} to be $12~\pm~2$ and 0.299, respectively
\citep{Williams78,Yoder95}. However, their values for celestial bodies
are poorly constrained because the timescales for the evolution are
very long. Values of $Q$ for
stars are typically estimated to be of order $10^6$
\citep[\eg][]{Jackson09}; dry terrestrial planets have $Q~\sim~100$
\citep{Yoder95,Henning09}, and gas giants have $Q=10^4-10^6$
\citep{AksnesFranklin01,Jackson08a}. In $\S$~\ref{sec:results} we will
consider the possibility that Proxima b began with a hydrogen envelope
and was perhaps more like Neptune than Earth. There is some debate
regarding the location of tidal dissipation in gaseous exoplanets,
whether it is in the envelope (high $Q$) or in the core (low $Q$)
\citep[\eg][]{StorchLai14}. We will consider planets with very thin
hydrogen envelopes, so we will make this latter assumption and use the $Q$
value computed by \thermint (see $\S$~\ref{sec:models:thermint}) in simulations which track orbital, internal, atmospheric and stellar evolution for habitable evaporated core scenarios in $\S$~\ref{sec:results:tidal_hec}.

\subsection{Orbital Evolution: \distorb}
\label{sec:models:distorb}
The model for orbital evolution, \distorb (for ``disturbing function orbit evolution''),
uses the 4th order secular disturbing function from \cite{MurrayDermott99}
(see their Appendix B), with equations of motion given by Lagrange's planetary
equations \citep[again, see][]{MurrayDermott99} and presented in their entirety in \cite{Deitrick2018}. This secular (\ie long-term
averaged) model does not account for the effects of mean-motion
resonances; however, since we apply it to well-separated planets
here, it is adequate for much of our parameter space. Since the model
is 4th order in $e$ and $i$, it can account for coupling of
eccentricity and inclination, although it does begin to break down at
higher eccentricity ($\gsim 0.3$) or mutual inclination ($\gsim 30^\circ$). We have compared our model to the
{\footnotesize \texttt{HNBody}}\footnote{Publicly available at
 https://janus.astro.umd.edu/HNBody/}
integrator \citep{RauchHamilton02} and find that for modest values of
$e$ and $i$ the two methods are nearly indistinguishable. We apply this model to Proxima b and a possible longer period
companion, hinted at in the discovery data.

\subsection{Rotational Evolution from Orbits and the Stellar Torque: \distrot}
\label{sec:models:distrot}
The planetary obliquity is a primary driver of climate, and hence we
also track planet b's spin-axis evolution carefully. Not only is it responsible
for seasons, but a non-zero obliquity can result in tidal heating
\citep{Heller11}, which can change outgassing rates and atmospheric
properties. Proxima b's obliquity is affected by two key processes:
tidal damping and perturbations from other planets. The \eqtide~module
handles the former, \distrot~the latter.

Our obliquity evolution model, \distrot (for ``disturbing function
rotation evolution''), uses the equations of motion originally developed in
\cite{Kinoshita1975, Kinoshita1977} and utilized in numerous studies
including \cite{Laskar1986}, \cite{Laskar1993a,Laskar1993b}, and
\cite{Armstrong14}.  It treats the planet as an oblate spheroid
(having an axisymmetric equatorial bulge), with a shape controlled by
the rotation rate (see below). The planet is subject to a torque
from the host star, which causes axial precession, and changes in its
orbital plane due to perturbations from a companion planet, which
directly change the obliquity angle.  This model is thus dependent on
\distorb through the eccentricity, the inclination, the longitude of
ascending node $(\Omega)$, and the derivatives $dp/dt$ and $dq/dt$. This model is described in its entirety in \cite{Deitrick2018}.

Since we couple obliquity evolution in \distrot to tidal evolution in
\eqtide, it is necessary to account for changes in the planet's shape
(its dynamical ellipticity) as its rotation rate changes due to
tides. Following the examples of \cite{Atobe2007} and \cite{Brasser2014}, we
scale the planet's oblateness coefficient, $J_2$ (from which the
dynamical ellipticity, $E_d$, can be derived), with the radius $R_p$,
rotation rate $\omega_{rot}$, and mass $M$, as
\begin{equation}
J_2 \propto \frac{\omega_{rot}^2 R_p^3}{M}.
\end{equation}
This is equivalent to assuming hydrostatic equilibrium, i.e.,
there is no frozen-in ``fossil figure''.
We use the Earth's $J_2$ as a proportionality factor. As pointed out
by \cite{Brasser2014}, around a rotation period of 13 days, $J_2$
calculated in this way reaches the $J_2$ of Venus, which maintains
this shape at a much slower rotation period, however, the
slowest rotation rate for planet b is $\sim$ 11 days, thus this
inconsistency is not encountered in our simulations.

In the presence of strong tidal effects, as we would expect at Proxima
b's orbital distance, the obliquity damps extremely quickly (in a few
hundred kyr). However, if another planetary mass companion is present,
then gravitational perturbations can prevent the obliquity from
damping completely. Furthermore, this equilibrium configuration,
called a Cassini state, is confined to a configuration in
which the total angular momentum vector of the planetary system,
$\hat{k}$, the rotational angular momentum vector of the planet,
$\hat{s}$, and the planet's own orbital angular momentum vector,
$\hat{n}$, all lie in the same plane \citep{Colombo1966}.

To identify Cassini states, we use the formula
\begin{equation}
\sin{\Psi} = \frac{(\hat{k}\times \hat{n}) \times (\hat{s} \times \hat{n})}{ | \hat{k}
\times \hat{n}  | \left | \hat{s} \times \hat{n} \right |},
\label{eqn:cassini}
\end{equation}
suggested by \cite{Hamilton2004}.  In a Cassini state, the angle $\Psi$
will oscillate (with small amplitude) about $0^{\circ}$ or
$180^{\circ}$, so $\sin({\Psi})$ will approach zero.  We will refer to
$\sin({\Psi})$ as the ``Cassini parameter''. If a planet is in a Cassini
state, its obliquity will not be damped to 0.

\subsection{Radiogenic Heating: \radheat}
\label{sec:models:radheat}

The first of two geophysical modules tracks the abundances of
radioactive elements in the planet's core, mantle and crust. We
consider 5 radioactive species: $^{26}$Al, $^{40}$K, $^{232}$Th,
$^{235}$U, and $^{238}$U. These elements have measured half-lifes of
$7.17 \times 10^5$, $1.251 \times 10^9$, $1.405 \times 10^{10}$,
$7.038 \times 10^8$, and $4.468 \times 10^9$ years, respectively. The
energy associated with each decay is $6.415 \times 10^{-13}$,
$2.134 \times 10^{-13}$, $6.834 \times 10^{-12}$,
$6.555 \times 10^{-12}$ and $8.283 \times 10^{-12}$ J, respectively.

We will consider four different abundance ratios. First, we consider an
Earth-like case with standard abundance concentrations
\citep[\eg][]{Korenaga03,Arevalo09,Huang13}. Note that geoneutrino
experiments are only able to measure the decay products of $^{232}$Th and
$^{238}$U inside Earth \citep{Raghavan98,Araki05,Dye10}.

The second case uses chondritic abundances, in which we augment the
mantle's $^{40}$K budget by a factor of 30 in number to match the
potassium abundance seen in chondritic meteorites
\citep{AndersGrevesse89,Arevalo09}. Such high potassium abundances could be
present if the planet formed beyond the snowline where potassium, a
volatile, is more likely to become embedded in solids, see $\S$~\ref{sec:obs:planetsys}.

The third case is a planet containing an initial abundance of 1 part
per trillion (ppt) of $^{26}$Al. If the planet formed within 1 Myr and
the planetary disk was enriched by a nearby supernova (or kilonova),
 then not all the $^{26}$Al
would have decayed. A planet that formed quickly, either by planetesimal accumulation or a direct collapse in the outer
regions of Proxima's protoplanetary disk, would likely have
more than 1 ppt of $^{26}$Al, but as we will see in
$\S$~\ref{sec:results}, this case provides so much heating that our model breaks down. The decay of $^{26}$Al at $t=0$ produces enough heat to melt 1 g
of a CI meteorite, preventing their solidification for several
half-lives \citep{HeveySanders06}. Note that Earth required tens to
hundreds of millions of years to form, so all the primordial $^{26}$Al
in the Solar System had already decayed.

The final case is an inert planet with no radioactive particles. This
case is very unlikely in reality, but serves as a useful end-member
case to bound the evolution of Proxima b.

\subsection{Geophysical Evolution: \thermint}
\label{sec:models:thermint}

We model the coupled core-mantle evolution of Proxima b with a
1-dimensional model that has been calibrated to modern-day Earth
\citep{DriscollBercovici14,DriscollBarnes15}; the
reader is referred to those studies for a comprehensive description.
Briefly, the model solves for the average core and mantle
temperatures as determined by energy balance in the two layers and
temperature-dependent parameterizations for heat loss. The code
includes heat transport across the mantle-surface and core-mantle boundaries (CMB),
mantle melt production and eruption
rates, latent heat production by mantle and core solidification, and radiogenic and tidal heating,
see $\S$~\ref{sec:models:eqtide}. Given the thermodynamic state of the core and the pressure of
the stellar wind at the orbit of Proxima b, a magnetic moment scaling law is used to predict the core
generated magnetic field and the resulting magnetopause radius \citep{DriscollBercovici14}.
However, we note that the host star's strong magnetic field may compress the planet's magnetosphere close to
its surface \citep{Vidotto13,Cohen14}.

Our model has been validated by simulating the history of Earth since the moon-forming impact and reproduces 8 features of the modern Earth: upper mantle temperature, core/mantle boundary temperature, core/mantle boundary heat flux, average eruptive mass, upper mantle viscosity, inner core radius, surface heat flow, and magnetic moment. Fiducial values and uncertainties for these parameters were taken from \cite{DriscollBercovici14}, \cite{DriscollBarnes15}, and \cite{Jaupart15}.
This was accomplished by calibrating with observed values for each
of these parameters and an error scale, such that a chi-squared deviation could
be defined. The model was then sampled over the \thermint and \radheat parameter
spaces, with a given parameter's range bounded by empirical and theoretical
constraints. Sampling was performed using a
Markov-Chain Monte Carlo (MCMC) sampler \citep{ForemanMackey13}, which is well-suited to deal with the complexity of the model and the large number of parameters.

This model produces the expected divergent evolution
of Venus and Earth under the assumption that they formed with similar compositions
and temperatures, but that Venus has had a stagnant lid and Earth a mobile
lid \citep{DriscollBercovici14}. While
this model is generic in many ways, it does assume an Earth-like composition, structure, mass and
radius. The minimum mass for Proxima b is close enough to
Earth's for this model to produce preliminary predictions for its thermal evolution.  We note that
 \thermint is limited to initial mantle temperatures above $\sim$1500~K, below which point
differentiation may not occur, and below 8000~K, where additional phase changes would require additional physics.

The \thermint modules can be directly coupled to \eqtide as demonstrated in
\cite{DriscollBarnes15}. All the tidal power
is deposited in the mantle using a visco-elastic Maxwell dissipation model \citep{Henning09}.  Heating of the mantle changes viscosity, shear modulus, and
in turn the dissipation efficiency (or tidal $Q$). The dissipation model predicts a maximum dissipation rate for mantle
temperature near 1800~K, thus cooling planets that pass
through this temperature can experience a spike in tidal power
generation.

\subsection{Galactic Effects: \galhabit}
\label{sec:models:galhabit}
Proxima Centauri appears to be tenuously bound to the binary $\alpha$ Cen A
and B, with a semi-major axis of $\sim$
8,700 AU \citep{Kervella17}, and is therefore susceptible to perturbations from the Milky Way. We model the changes to Proxima's orbit produced by galactic tides
and stellar encounters using the equations and prescriptions
developed to study the Oort cloud \citep{Heisler1986, Heisler1987, Rickman2008},
as Proxima has a similar orbit about $\alpha$ Cen A and B. These formulations
have also been shown to be accurate in modeling groups of stars \citep{Aguilar1985}.
We utilize an updated galactic
density of $\rho_0 = 0.102~\msun \rm{pc}^{-3}$ \citep{Holmberg2000} and treat
$\alpha$ Cen A and B as a single point mass with $M = 2.1$
M$_{\odot}$ (\ie the recently updated masses given by
\cite{PourbaixBoffin16}). This approach is somewhat crude ---
the two stars produce a significant quadrupole moment associated
with their orbit --- as a back-of-the-envelope calculation indicates
that the torque associated with this quadrupole potential would
be equal to the galactic tidal torque at distance less than $\sim 2000$ AU. Hence,
the effect of the binary host should be minor
at Proxima's current semi-major axis. However, the quadrupole's importance is increased if Proxima
has a significant eccentricity. Since modeling the
triple system in a comprehensive way is significantly more complicated
\citep[see, \eg][]{Harrington1968, Ford2000, Breiter2015},  we restrict ourselves to the
secular effects of galactic tides and passing stars
and will revisit the triple star dynamics in future work.

At Proxima's semi-major axis, galactic tides and stellar
encounters can pump its eccentricity to values large enough to
cause disruption from the system, and/or a periastron distance
so close to the binary $\alpha$ Cen that we would expect
consequences for any planetary system, such as eccentricity
excitation or destabilization \citep{Kaib13}.
In such situations, Proxima b may currently have significant tidal
heating due to recent ecentricity excitation.

Following \cite{Heisler1987} and \cite{Rickman2008}, we model stellar
encounters with a stochastic Monte Carlo approach, estimating times
of encounters from the stellar density and velocity dispersion,
and then randomly drawing stellar magnitudes and velocities
from the distributions published in \cite{Garciasanchez2001}.
The impact parameter and velocity are calculated from the relative
velocities (stellar velocity relative to the apex velocity, \ie the velocity of the star with respect to
the Local Standard of Rest,
see \cite{Rickman2008}), and then a $\Delta v$ is applied to
Proxima's orbit according to the impulse approximation
\citep{Remy1985}. The masses of passing stars are calculated
using the empirical relations from \cite{Reid2002}. We have tested
this impulse approximation against an N-Body model (\texttt{HNBody})
with Bulirsch-Stoer integration
in cases where errors are expected to be largest (smaller semi-major
axes for Proxima), for a total of 337,235 comparison simulations.
In all cases the errors (\ie the difference between the resulting
orbital elements in the N-Body and impulse approximation scenarios) were
less than 1\%, and the systematic offset in semi-major axis was
$1.1 \times 10^{-4}$, which agrees well with the errors found
in using the impulse approximation by \cite{Rickman2005}.

As previously noted, the metallicities of $\alpha$ Cen A and B
suggest that the system formed at a galactocentric distance of
$\lesssim 4.5$ kpc \citep{Loebman16}. To model the potential
effects of radial migration on the triple star system, we
scale the stellar density and gas density of the galaxy
according to the radial scale lengths ($R_{\star}, R_{gas}$) found
by \cite{Kordopatis15}. The dark matter density at each distance
is estimated from their spheroidal model---unlike the disk models
used for stars and gas, this model is not axisymmetric. However,
as the dark matter near the midplane of the disk makes up
$\lesssim 1\%$ of the total density, it is a decent approximation
to assume axisymmetry of the total mass density, as the
\cite{Heisler1986} tidal model assumes. We scale the velocity
dispersions of the nearby stars as a decaying exponential
with twice the stellar scale length, $2R_{\star}$, multiplied by
$\sqrt{t}$, where $t$ is the time since galactic formation, as found
to be broadly true in galactic simulations
\citep{Minchev2012, Roskar2012}. In this fashion, the velocity
dispersion grows slowly in time at all galactic radii, and it grows
larger closer to the galactic center.
The apex velocity will vary according to the detailed
orbital motion
of Proxima through the galaxy, including the radial migration.  For the
purposes of this study, we
simply keep the apex velocity constant in time and space
(though it is dependent on the spectral class of the perturbing star),
assuming the current Solar
value is typical.

With these scaling laws in place, we model radial migration
as a single, abrupt jump in the galactocentric distance of the
system. The reasoning behind this approximation is that N-Body
simulations show migration typically occurs over the
course of a single galactic orbit \citep{Roskar2010}; hence, the
migration time is short compared to the age of the stellar system.
We then randomly choose formation distances over the range
$(1.5,4.5)$ kpc and migration times over the range $(1,5)$ Gyr
since formation.

\subsection{The Coupled Model: \vplanet}
\label{sec:models:vplanet}
The previously described modules are combined into a single software program
called \vplanet. This code, written in C, is designed to be
modular so that for any given body, only specific modules are applied
and specific parameters integrated in the forward time direction.
Parameters are integrated
using a 4th order Runge-Kutta scheme with a timestep equal to $\eta$
times the shortest timescale for all active parameters, \ie
$x/(dx/dt)$, where $x$ is a parameter. In general, we obtain convergence if
$\eta~\le~0.01$. A more complete and quantitative description of
\vplanet~will be presented soon (Barnes et al., in prep.).

Each individual model is validated against observations in our Solar
System or in stellar systems, as described above. When possible, conserved quantities are
also tracked and required to remain within acceptable limits. With
these requirements met, we model the evolution of Proxima Centauri b
for predicted formation models to identify plausible evolutionary
scenarios, focusing on cases that allow the planet to be habitable. As
Proxima b is near the inner edge of the HZ, we are primarily concerned
with transitions into or out of a runaway greenhouse. For water-rich
planets, this occurs when the outgoing flux from a planet is $\sim
300$~W/m$^2$ \citep{Kasting93,Abe93} and for dry planets it is at
415~W/m$^2$ \citep{Abe11}. For water-rich planets, we use the
relationship between HZ limits, luminosity and effective temperature
as defined in \cite{Kopparapu13}.

\section{Results}\label{sec:results}

\subsection{Galactic Evolution}
\label{sec:results:galactic}
\subsubsection{Proxima's orbit about \acen}
\begin{figure*}
\centering
\includegraphics[width=\textwidth]{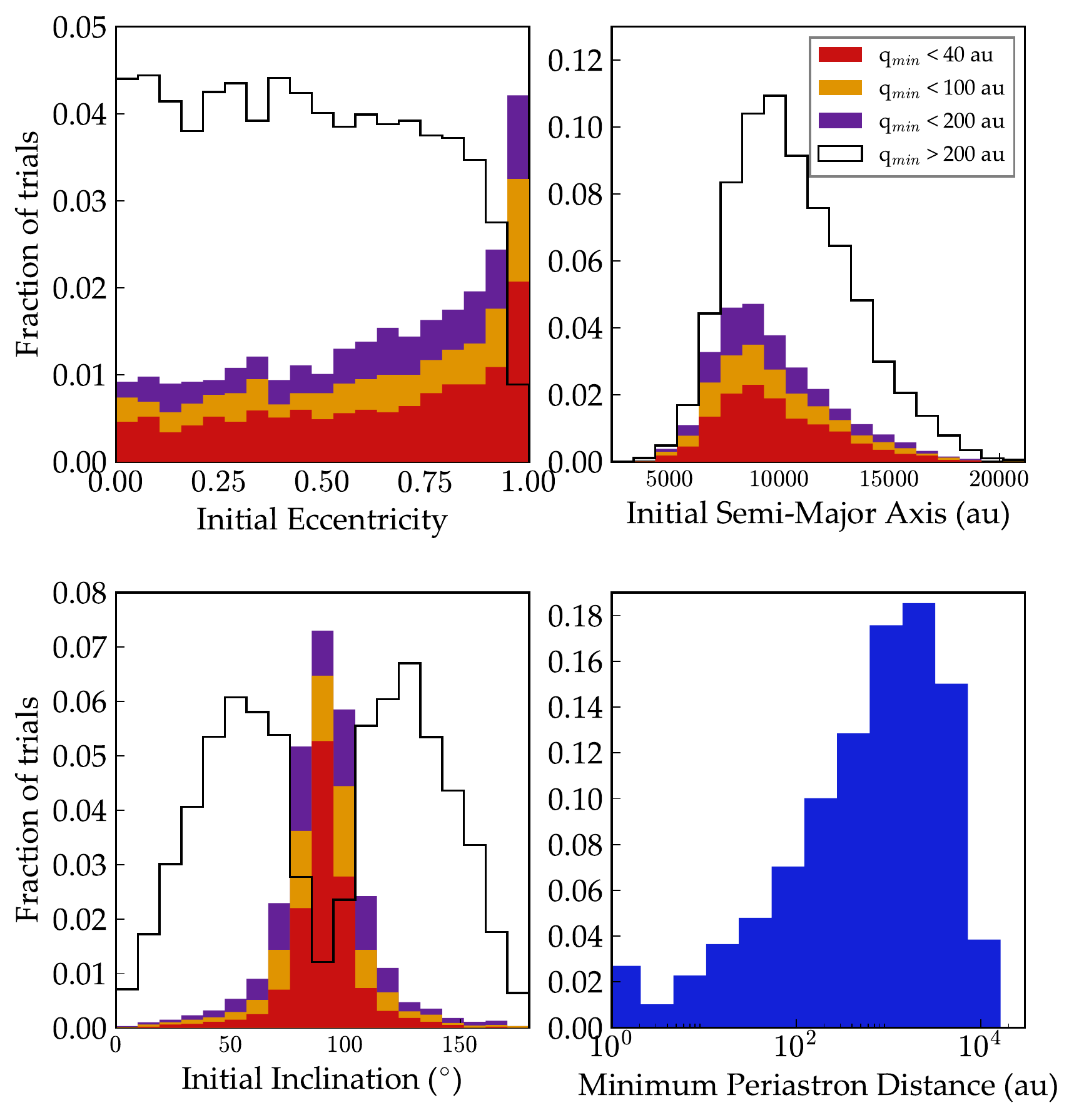}
\caption{Distribution of Proxima Centauri's minimum pericenter, $q$,
  without radial migration. Simulations in which $q$ passes below
  40 AU are shown in red. Stacked on top of that, simulations with
  $40<q<100$ AU are in orange, and $100<q<200$ in purple, i.e. red is a subset of orange is a subset of purple. The black line
  indicates simulations that never had passages within 200 AU.
  {\it Top left:} Initial
  eccentricity. {\it Top right:} Initial semi-major axis. {\it Bottom
    left:} Initial inclination relative to the galactic disk. {\it
    Bottom right:} Minimum periastron distance in all cases over the entire
  simulation. Generally, eccentricity and inclination determine the likelihood
  of close passages between Proxima and \acen, with high $e$ and $i \sim 90^{\circ}$
  (\ie low $\hat{Z}$-angular momentum) cases being the most likely to have
  such events.}
\label{fig:galacdist}
\end{figure*}

\begin{figure*}
\centering
\includegraphics[width=\textwidth]{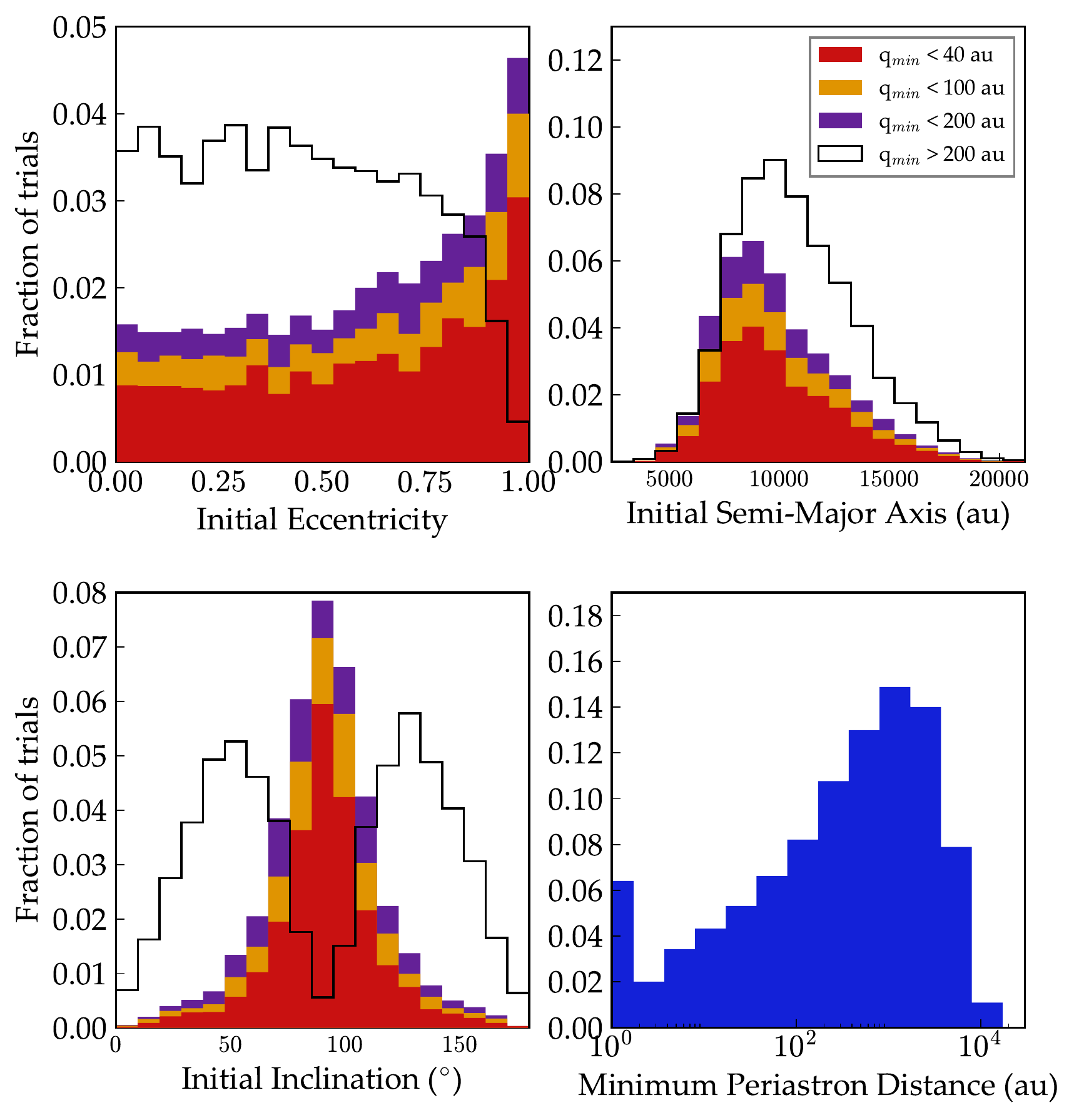}
\caption{Same as Fig.~\ref{fig:galacdist} but with radial migration.
  Systems which formed interior to 4.5 kpc from the galactic center
  have close passages with \acen~more frequently than
  those which were placed in the solar neighborhood from the
  beginning. }
\label{fig:galacdistmigr}
\end{figure*}

If Proxima is bound to \acen~A and B, its orbit will
be modified by the galactic tide and perturbations from passing
stars. We run two experiments to explore the effects of radial
migration: set \textbf{A} places the system in the solar neighborhood,
randomly selecting orbital parameters broadly consistent with the
observed positions, for 10,000 trials. In set \textbf{B}, we
take the same initial conditions and randomly select formation
distances over the range [1.5,4.5] kpc \citep{Loebman16} and migration
times over (1,5) Gyr after formation, after which the system is moved
to the solar neighborhood (8 kpc). In all cases, the initial
orbital elements for Proxima are randomly selected from
``Proxima-like'' conditions ({\it i.e.}, semi-major axes between $\sim$5000 and
$\sim$20000 AU, see Figs. \ref{fig:galacdist} and \ref{fig:galacdistmigr})
to simulate a myriad of possible histories
for the current system. Simulations are halted whenever Proxima becomes
gravitationally
unbound ($e > 1$)
or when it passes beyond 1 pc (at which point we consider it unbound).
Our model assuredly breaks down at separations of 1 AU (and probably larger,
see below)
and hence we also terminate simulations if that occurs.
In both sets, fewer than 1\% of the simulations are halted due to any of the
above conditions.

\begin{figure*}
\centering
\includegraphics[width=\textwidth]{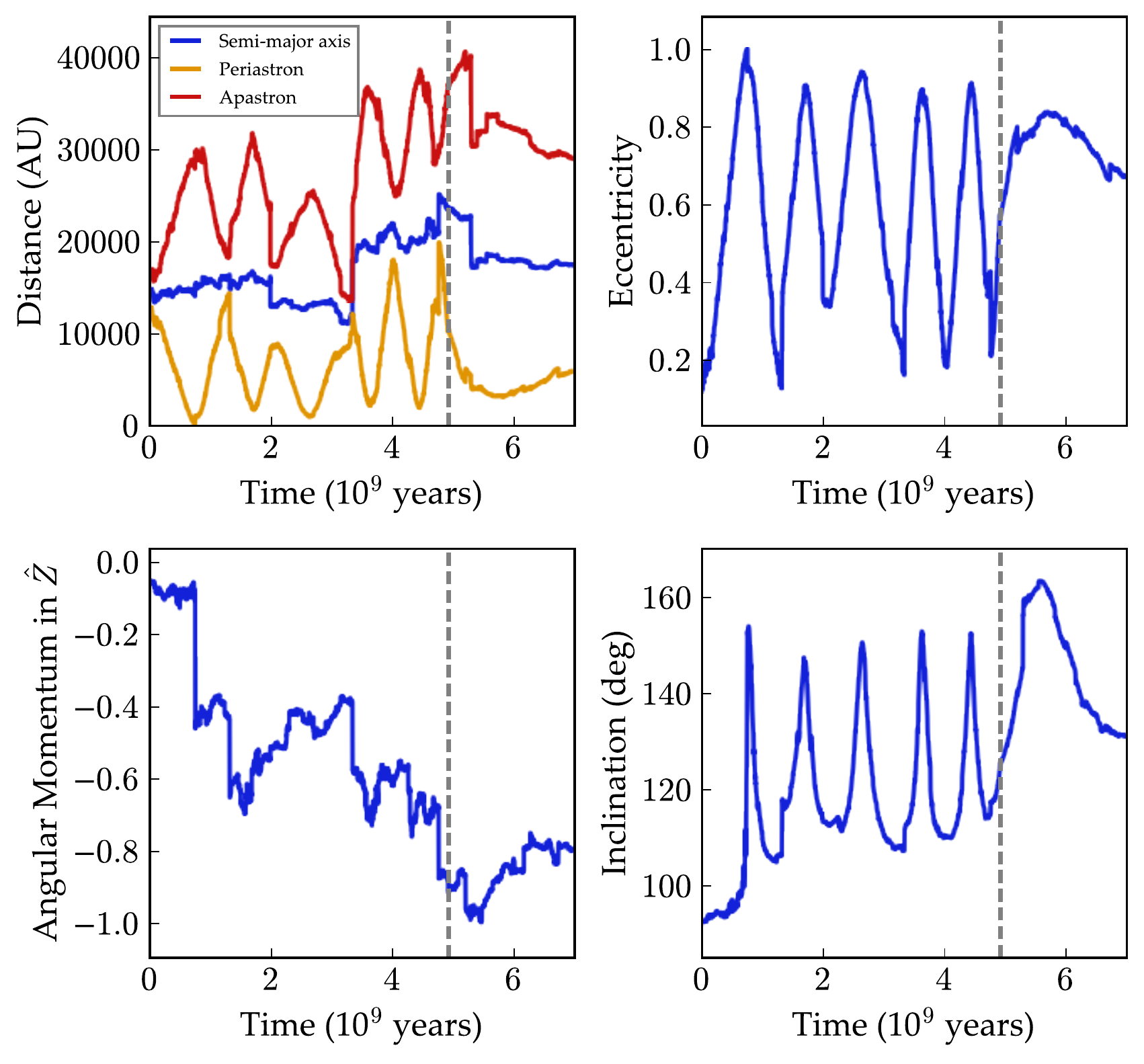}
\caption{An example of the orbital evolution of Proxima in the galactic
  simulations. The upper left panel shows the semi-major axis, periastron
  distance, and apastron distance, the upper right shows the eccentricity,
  the lower left shows the angular momentum in the $\hat{Z}-$direction,
  and the lower right shows the inclination with respect to plane of the
  galactic disk. The system was given a formation distance of $R = 3.63$
  kpc and the vertical dashed line shows the time of migration to 8 kpc.
  The angular momentum in $\hat{Z}$ (the action $J_z$) and semi-major axis
  are unchanged by
  galactic tides---eccentricity and inclination exchange angular momentum
  in such a way that these quantities are conserved---thus their evolution is
  purely
  due to stellar encounters. In this particular case, the eccentricity of Proxima
  grows such that its periastron dips within 40 AU of $\alpha$ Cen A and B.}
\label{fig:galevolution}
\end{figure*}

Figure \ref{fig:galevolution} shows an example of the evolution
of Proxima's orbit in set \textbf{B}. Pericenter gets close to \acen~a
number of times before radial migration occurs
(dashed line)---as discussed below, these close approaches are
potentially disruptive to a planetary system. Angular momentum in $\hat{Z}$ is exchanged
between the eccentricity and inclination because of the
galactic tides, while stellar encounters perturb all the orbital
elements, adding or removing energy and angular momentum
from the system.


We search both sets of simulations for the minimum pericenter
distance, $q_{min}$, that Proxima experiences over the course of
7 Gyr (the oldest of \acen's age estimates). A previous study
of planet stability in wide binaries \citep{Kaib13} found
that an extended system (like the solar system) orbiting a
solar mass star with a binary companion can be disrupted when
galactic influences drive the binary's pericenter to $50-250$ AU.
Extrapolating from
that study, which considered a larger mass host star than
Proxima and
smaller companion stellar masses ($\lesssim 1 M_{\odot}$) than
\acen, we anticipate
disruptions of an extended planetary system (if Proxima has or
ever had one) will occur when $q \lesssim 100-200$ AU. Bearing
that in mind, we separate our simulations into 5 groups:
those that had $q_{min}<40$ AU (\acen's orbit extends to $\sim 36$ AU), those with $40<q_{min}<100$ AU, those
with $100<q_{min}<200$ AU, those that never pass within $200$ AU, and
those that became gravitationally unbound. The numbers within
each category are shown in Table 1 for both
sets of simulations.

Figure \ref{fig:galacdist} shows the distribution of set
\textbf{A} simulations
within each category as a function of the initial semi-major axis,
eccentricity, and inclination. Also shown is the distribution of
minimum pericenter ($q_{min}$) distances within this set. Note that the
distributions for the intervals (40, 100) AU and (100, 200) AU
shown here are ``stacked'' on top of the previous categories
for ease of viewing.
Figure \ref{fig:galacdistmigr} shows
the same for set \textbf{B}. In the lower right panel, the
small spike at 1 AU is a result of our halting the simulations
that dip within 1 AU, which causes a pile up of simulations there.
Presumably, the tail of the distribution would continue on toward
smaller and smaller numbers if we did not halt the simulations at
that point.

As we expect, configurations that start with high eccentricity
or inclination close to $90^{\circ}$ (perpendicular to the
galactic midplane) are more likely to experience close encounters. This outcome is largely a
consequence of the galactic tide acting on a system with
low angular momentum in the $\pm\hat{Z}$ direction (perpendicular
to the galactic disk). Random encounters with passing stars
make the distribution messier---Proxima can be scattered into
or out of the ``close passage'' regime, which randomizes
the likelihood of close passages.

Close passages within 100 or 200 AU (orange and purple histograms)
are potentially destructive to an extended planetary system
orbiting Proxima (planets out to 20 to 30 AU), see next subsubsection.
Close passages
within $40$ AU (red) are potentially destructive to the
\emph{stellar} system itself, and thus may not be
representative of the real \acen~system. On the other hand, a close passage
may have resulted in the scattering the \acen~A+B orbit into its current high
eccentricity orbit. Regardless, our model does not account for the
three-body
physics of such an encounter, so robust conclusions should not be drawn from
our results.

\begin{figure}
\centering
\includegraphics[width=0.45\textwidth]{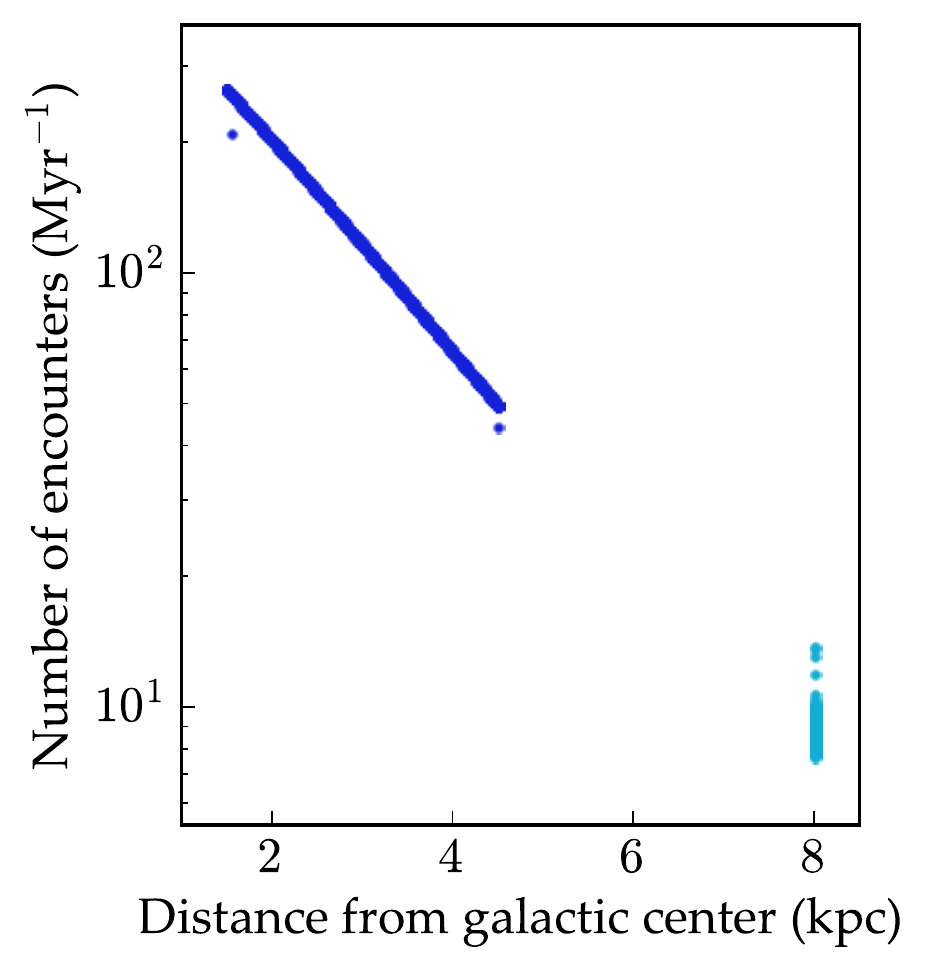}
\caption{Stellar encounter rates as a function of galactocentric distance.
  Dark blue points correspond to pre-migration encounter rates, light
  blue to post-migration. There is some
  scatter in the solar-neighborhood points because of the time dependence
  of the stellar velocity dispersion. At the tail end of the simulations, we
  match the encounter frequency of 10.5 Myr$^{-1}$ from previous
  studies \citep{Garciasanchez2001,Rickman2008}.}
\label{fig:encrates}
\end{figure}

In set \textbf{B}, in which \acen~forms closer to the denser galactic center and
undergoes radial migration, close passages are much more common
than in set \textbf{A}. Closer to the
galactic center, the system is subjected to stronger galactic
tides and more frequent stellar encounters, which make
close passages and complete disruption more likely.
Figure \ref{fig:encrates} shows how the encounter rate scales
with the galactocentric position.

\begin{table}[h]\label{tab:galacstat}
\centering
Table 1 -- Results of Galactic Simulations\\
\vspace{0.1cm}
\begin{tabular}{lcc}
\hline\hline \\[-1.5ex]
Minimum pericenter & Set A & Set B \\[0.5ex]
\hline \\ [-1.5ex]
(0, 40) AU & 1301 & 2321  \\
(40, 100) AU & 709 & 826  \\
(100, 200) AU & 756 & 790  \\
$>$ 200 AU & 7218 & 5981  \\
Unbound & 16 & 82 \\
\end{tabular}

\end{table}

\subsubsection{Consequences for Proxima's planetary system}

To investigate the impact of close passages on a planetary system
orbiting Proxima, we run short simulations in \texttt{HNBody},
wherein \acen~is placed in a high eccentricity orbit that brings
it close to Proxima (still treating \acen~as a single point mass).
We ran three sets of simulations, each with a different type of
planetary system: only Proxima b, Proxima b and the putative companion, Proxima b with the Solar
System's gas giants (see Table 2), and Proxima b with super-Earths in the same orbits as the Solar System's gas giants. In each set, we tested the effect of a close
passage at 40 AU, 100 AU, and 200 AU.

\begin{figure*}
\centering
\includegraphics[width=\textwidth]{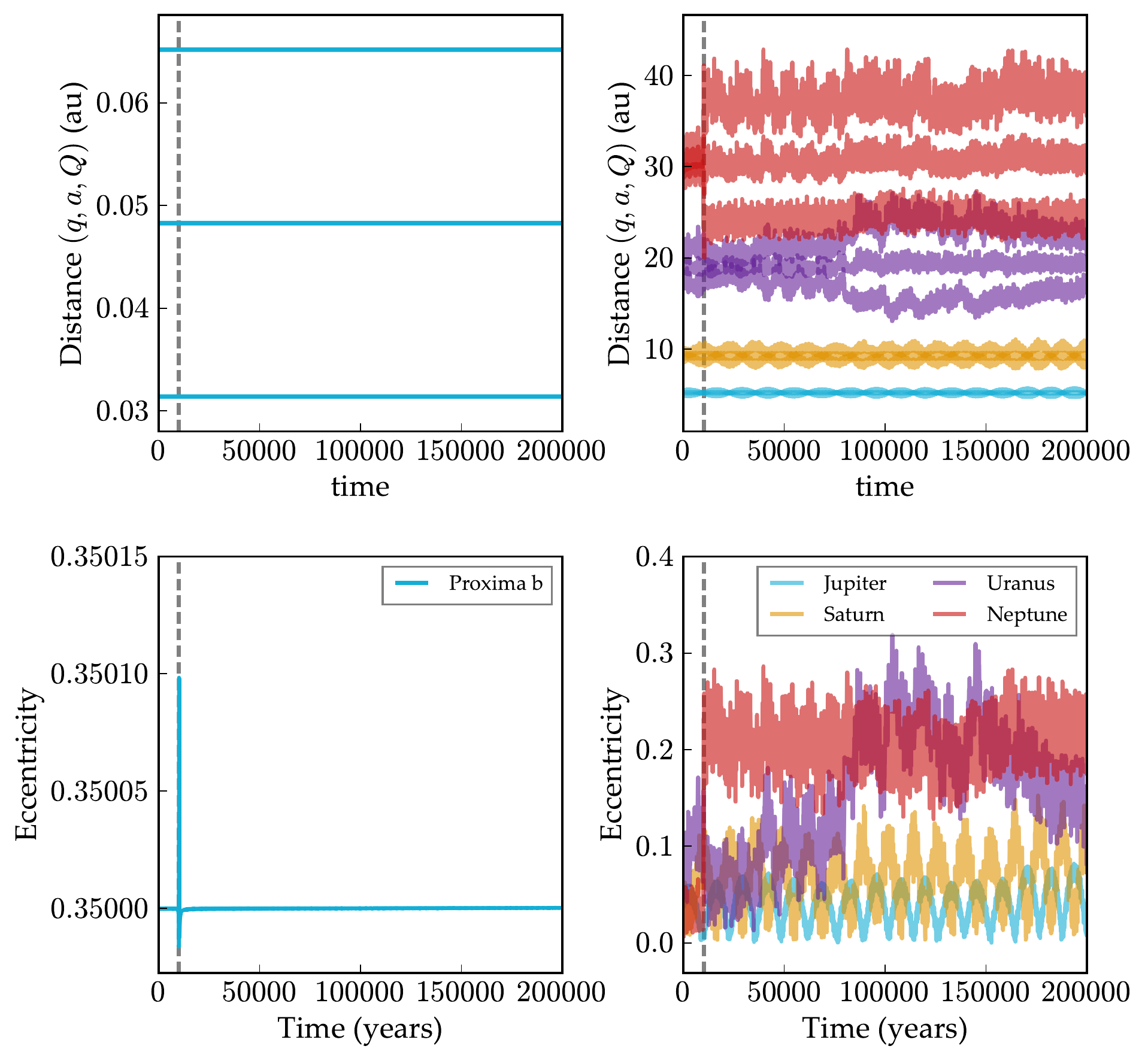}
\caption{A glimpse of the effects that close passages with \acen~can have
on Proxima's planetary system. Two cases are shown here: {\it Left:} Proxima b
solo, with an initial $e =0.35$ and \acen's pericenter occurring at 40 AU;
{\it Right:} An outer solar system analog orbiting Proxima, with
\acen's pericenter at 200 AU. Dashed gray lines
represent the time of \acen's pericenter passage.
The top panels show the location of
pericenter $q$, semi-major axis $a$, and apocenter $Q$ for each planet, seen as
three curves of the same color. In the Proxima b solo case, there is a spike of
small amplitude in the eccentricity, but after close passage the eccentricity
quickly settles to near its initial value. In the solar system analog case, the
eccentricities of the outer two planets are excited to the point that their
orbits cross---a highly unstable situation.}
\label{fig:planetpert}
\end{figure*}

In the first set, in which Proxima b is the only planet, nothing
dramatic occurs (see Table 3).
At low initial eccentricity ($e = 0.001$), the
perturbation from \acen's passage at 40 AU results in a change of $<10^{-6}$
in Proxima b's eccentricity. Increasing Proxima b's initial eccentricity
to the maximum allowed by the observations ($e= 0.35$), we do see a larger
deviation ($\Delta e = 10^{-4}$) but even in this case, the eccentricity
quickly settles down to its original value (see Fig. \ref{fig:planetpert},
left hand panels). In the second set, with Proxima b and a super-Earth
companion, the effects are also very minor. Even in the highest eccentricity
case (both planets have $e=0.35$ initially), the perturbation from
\acen~is completely drowned out by the perturbations between the two planets.

With the third set, we are extending the study by \cite{Kaib13} to a
lower mass host star and a higher mass perturber. We confirm their findings
that close passages are destructive to a system analogous to our outer
solar system. Even at a close passage of 200 AU, the eccentricities of the
Uranus and Neptune analogues are excited to the point that they cross orbits (Fig.
\ref{fig:planetpert}, right hand panels). Though this planetary system is not
completely destroyed over the course of this simulation (200,000 year),
it is almost certainly rendered unstable on longer time-spans. With
close passages of 40 and 100 AU, one planet (or more) is ejected almost immediately
after \acen's pericenter passage. If we perform the same experiment
with Neptune removed (so that the planets extend to 20 AU), the system
is destabilized at close passages of 40 or 100 AU, but not at 200 AU.

Giant planets may not, however, be common orbiting M dwarfs
\citep{Ida2005,Bonfils13}. To be certain that the instability mentioned in the
previous paragraph is not unique to planets as massive as our solar
system giants, we rerun the third set of simulations, but giving the
planets masses in the range 1 to 4 $M_{\oplus}$. In this situation, the
system is even \emph{more} prone to instability than the higher
mass cases, with close passages at 200 AU producing instabilities.

\begin{table}[h!]
\centering
Table 2 -- Hypothetical multiplanet systems of Proxima\\
\vspace{0.1cm}
\begin{tabular}{lccc}
\hline\hline \\[-1.5ex]
Set & Planet mass ($M_{\oplus}$) & a (AU) & e \\[0.5ex]
\hline \\ [-1.5ex]
PCb, e1 & 1.27 & 0.0482817 & 0.001 \\
\hline \\[-2ex]
PCb, e2 & 1.27 & 0.0482817 & 0.2 \\
\hline \\[-2ex]
PCb, e3 & 1.27 & 0.0482817 & 0.35 \\
\hline \\[-2ex]
PCb/c, e1 & 1.27 & 0.0482817 & 0.001 \\
    & 3.13 & 0.346 & 0.001 \\
\hline \\[-2ex]
PCb/c, e2 & 1.27 & 0.0482817 & 0.2 \\
    & 3.13 & 0.346 & 0.2 \\
\hline \\[-2ex]
PCb/c, e3 & 1.27 & 0.0482817 & 0.35 \\
    & 3.13 & 0.346 & 0.35 \\
\hline \\[-2ex]
JSUN & 317.79704651 & 5.20336 & 0.048393 \\
    & 95.15193166 & 9.53707 & 0.054151\\
    & 14.53439881 & 19.19126 & 0.047168 \\
    & 17.14527595 & 30.06896 & 0.008586 \\
\hline \\[-2ex]
JSU & 317.79704651 & 5.20336 & 0.048393 \\
    & 95.15193166 & 9.53707 & 0.054151\\
    & 14.53439881 & 19.19126 & 0.047168 \\
\hline \\[-2ex]
JSUNsm & 1.27 & 5.20336 & 0.048393 \\
    & 3.13 & 9.53707 & 0.054151\\
    & 1.016 & 19.19126 & 0.047168 \\
    & 2.087 & 30.06896 & 0.008586 \\
\hline \\[-2ex]
JSUsm & 1.27 & 5.20336 & 0.048393 \\
    & 3.13 & 9.53707 & 0.054151\\
    & 1.016 & 19.19126 & 0.047168 \\
\hline \\[-2ex]
\end{tabular}
\label{tab:galacpls}
\end{table}

\begin{table}[h!]
\centering
Table 3 -- Outcomes of Galactic Perturbations on Proxima planetary systems\\
\vspace{0.1cm}
\begin{tabular}{lcc}
\hline\hline \\[-1.5ex]
Set & Closest approach (AU) & Result \\[0.5ex]
\hline \\ [-1.5ex]
PCb, e1 & 200 & Negligible\\
        & 100 &Negligible \\
        & 40 &Negligible \\
PCb, e2 & 200 &Negligible \\
        & 100 & Negligible\\
        & 40 & Negligible\\
PCb, e3 & 200 & Negligible\\
        & 100 & Negligible\\
        & 40 &Negligible \\
PCb/c, e1 & 200 & Negligible\\
        & 100 & Negligible\\
        & 40 & Negligible\\
PCb/c, e2 & 200 & Negligible\\
        & 100 & Negligible\\
        & 40 & Negligible\\
PCb/c, e3 & 200 & Negligible\\
        & 100 & Negligible\\
        & 40 & Negligible\\
JSUN & 200 & Crossing orbits\\
    & 100 & Disruption\\
    & 40 & Disruption\\
JSU & 200 & Excited eccentricities\\
    & 100 & Disruption\\
    & 40 & Disruption\\
JSUNsm & 200 & Disruption\\
    & 100 & Disruption\\
    & 40 & Disruption\\
JSUsm & 200 & Disruption\\
    & 100 & Disruption\\
    & 40 & Disruption\\
\end{tabular}
\label{tab:galacplsresult}
\end{table}

We conclude that close passages
with \acen~might have truncated any planetary system or
planet forming disk extending beyond $\sim 10$ to $20$ AU from Proxima.
In other words, the extent and structure of Proxima's planetary system
is limited by the presence of these companion stars. Late (\ie in
the relatively recent past)
close passages could have destabilized a previously quiet
planetary system, leading to major alterations in Proxima b's orbit,
potentially even a late arrival to its present day location.
Future detections
of additional planets may be able to constrain Proxima's orbital
history. If Proxima has relatively few planets, or their orbits are
dynamically hot (\ie high eccentricity or inclination), that may
indicate that close encounters with \acen~have occurred.
Conversely, planets beyond $\sim 10$ or $20$ AU on
circular, coplanar orbits may indicate a relatively
peaceful history.

\subsection{Orbital/Rotational/Tidal Evolution}
\label{sec:results:orbital}

We begin exploring the dynamical properties of the {\it planetary} orbits and spins by
considering the tidal evolution of Proxima b if it is in isolation. In
this case, we need only apply \eqtide~to both Proxima and Proxima b and track
$a, e, P_{rot},$ and $\psi$. We find that if planet b has $Q=12$,
then an initially Earth-like rotation state becomes tidally locked in
$\sim10^4$ years, so it seems likely that if b formed near its
current location, then it reached a tidally locked state with
negligible obliquity almost immediately. Note that, contrary to popular belief, the term
``tidally locked'' does not necessarily mean that the planet is
synchronously rotating (see the end of this section).

\begin{figure}
\begin{center}
\includegraphics[width=0.45\textwidth]{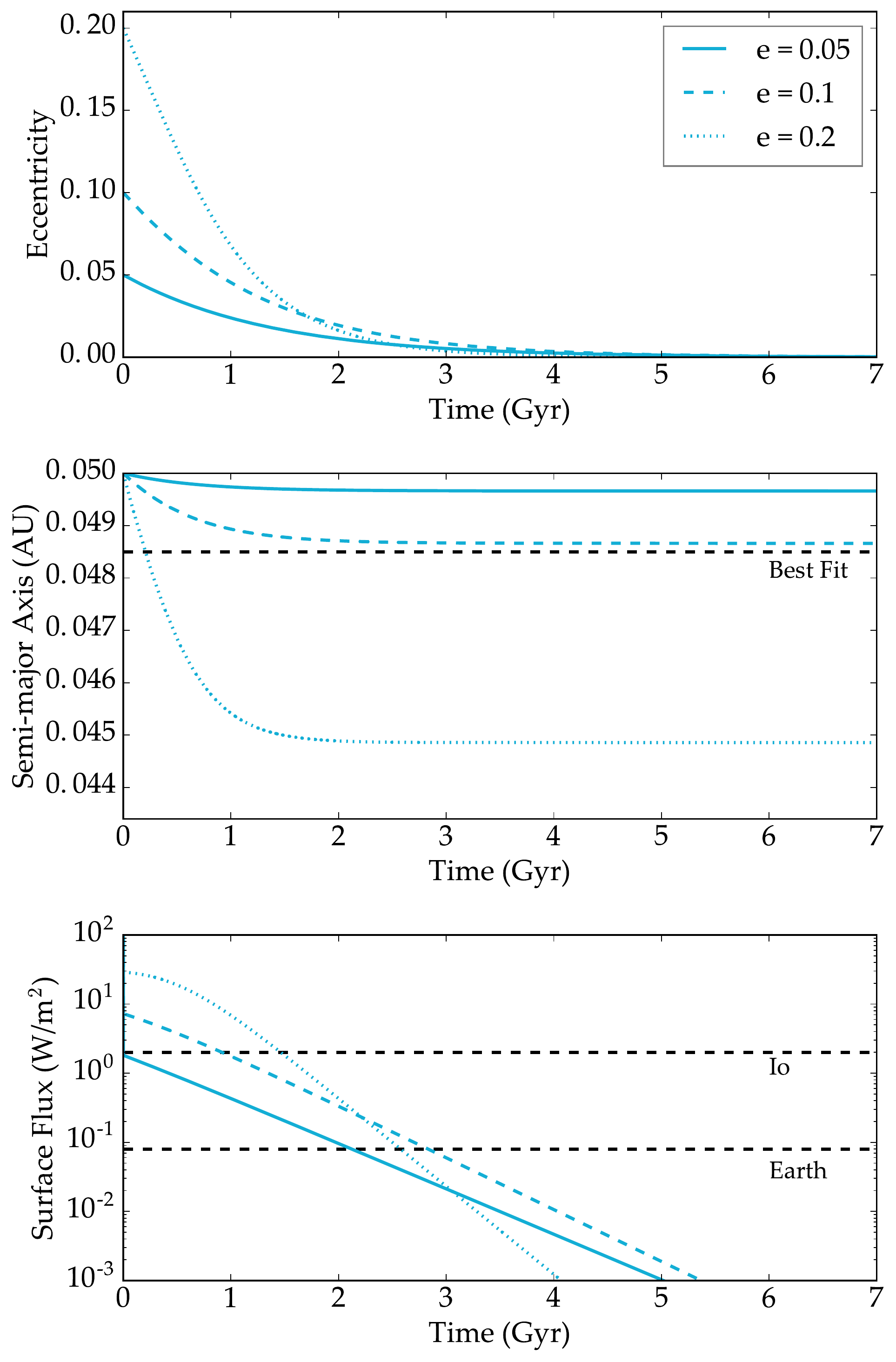}
\end{center}
\caption{Evolution of planet b's eccentricity (top), semi-major axis
  (middle), and tidal heating surface flux (bottom) assuming that
  initially $a~=~0.05$~AU and $e~=$~0.05 (dotted), 0.1 (solid) or 0.2
  (dashed). For reference the best fit semi-major axis and surface
  energy fluxes of Io and the modern Earth are shown by dashed black
  lines.}
\label{fig:eqtide}
\end{figure}

Unlike the rotational angular momentum, the orbit can evolve on long
timescales. In the top two panels of Fig.~\ref{fig:eqtide}, we
consider orbits that begin at $a=0.05$~AU and with different
eccentricities of 0.05 (dotted curves), 0.1 (solid curves) and 0.2
(dashed curves). In these cases $a$ and $e$ decrease and the amount of
inward migration depends on the initial eccentricity, which takes 2--3
Gyr to damp to $\sim0.01$. For initial eccentricities larger than
$\sim0.23$, the CPL model actually predicts eccentricity growth due
to angular momentum exchange between the star and planet
\citep{Barnes17}. This prediction is likely unphysical and due to the
low order of the CPL model; therefore we do not include evolutionary
tracks for higher eccentricities.

The equilibrium tide model posits that the lost rotational and orbital
energy is transformed into frictional heating inside the planet. The
bottom panel of Fig.~\ref{fig:eqtide} shows the average surface energy
flux as a function of time. We address the geophysical implications of
this tidal heating in $\S$~\ref{sec:results:internal:tides}. Note that if planet
b begins with a rotation period of 1 day and an obliquity of
$23.5^\circ$, then the initial surface energy flux due to tidal
heating is $\sim1$~kW/m$^{2}$.

Next, we consider the role of additional planets, specifically the
putative planet with a 215 day orbit \citep{AngladaEscude16}. For
these runs we now add the \distorb~and \distrot~modules and track the
orbital elements of both planets, the spins of the star and planet b,
and the dynamical ellipticity of planet b. A comprehensive exploration
of parameter space is beyond the scope of this study, so we consider
three end-member cases: a nearly coplanar, nearly circular system; a
system with high eccentricities and inclinations; and a system with high $e$ and low $i$. The initial orbital
elements and rotational properties of the bodies are listed in Table 4.

\begin{table*}[t]
\centering
Table 4 -- Initial conditions for two-planet Proxima systems\\
\vspace{0.1cm}
\begin{tabular}{lccccccccc}
\hline\hline \\[-1.5ex]
& $m$ ($M_{\oplus}$)  & $a_s$ (au) & $a_l$ (au) & $e$ & $i$ ($^{\circ}$)
 & $\omega$ ($^{\circ}$) & $\Omega$ ($^{\circ}$) & $\psi$ ($^{\circ}$) &
 $P_{rot}$ (days)\\[0.5ex]
\hline \\ [-1.5ex]
b & 1.27 & 0.0482817 & 0.05 & 0.001 & 0.001 & 248.87 & 20.68 & 23.5 & 1  \\
c & 3.13 & 0.346 & 0.346 & 0.001 & 0.001 & 336.71 & 20 & &  \\
\hline \\
b & 1.27 & 0.0482817 & 0.05 & 0.2 & 20 & 248.87 & 20.68 & 23.5 & 1  \\
c & 3.13 & 0.346 & 0.346 & 0.2 & 0.001 & 336.71 & 20 & &  \\
\end{tabular}
\label{tab:orbitic}
\end{table*}

\begin{figure*}
\begin{center}
\includegraphics[width=\textwidth]{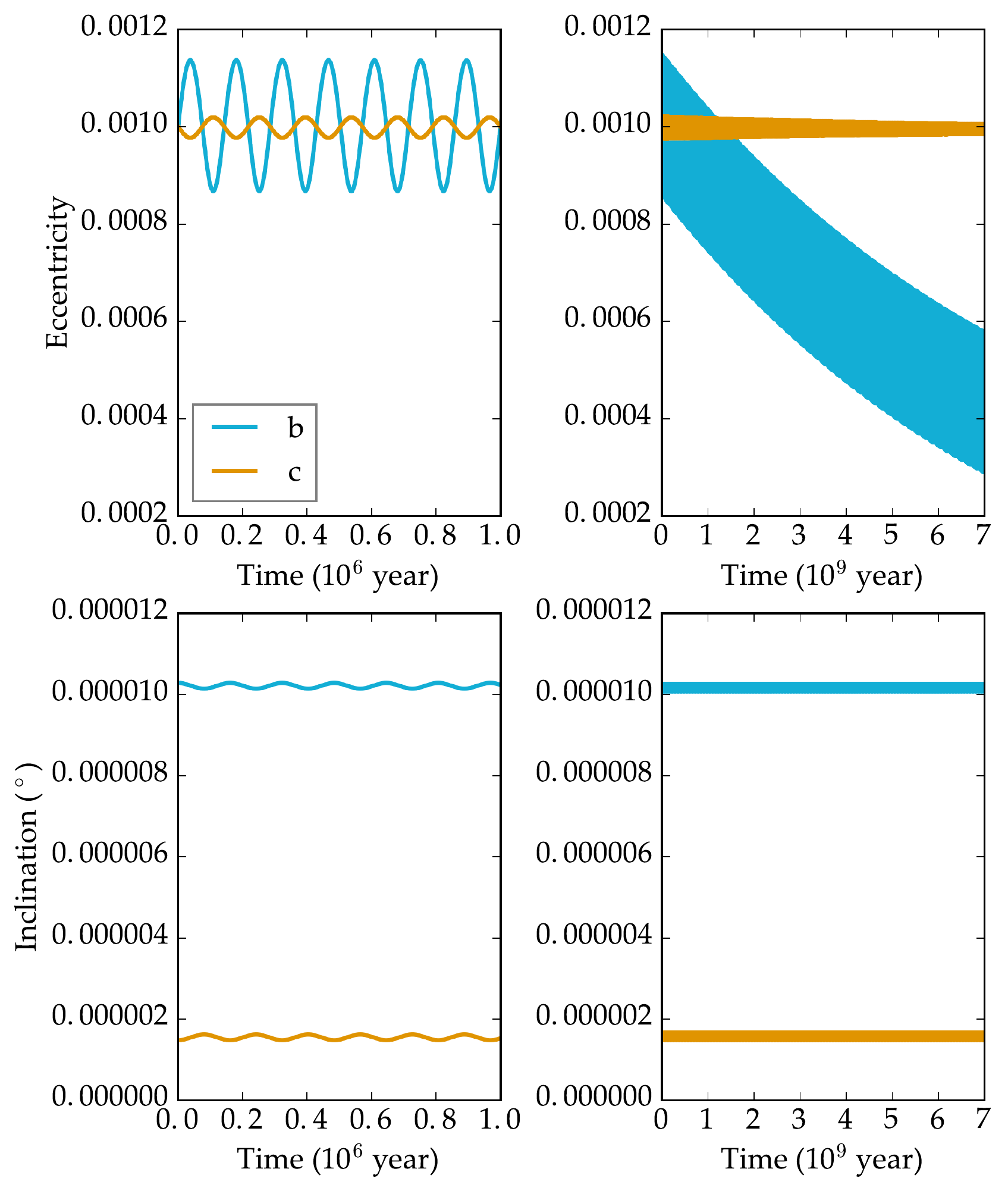}
\end{center}
\caption{Evolution of orbital elements if a putative planet c exists with an
orbital period of 215 days and both orbits are nearly circular and nearly
coplanar. {\it Top Row:} Eccentricity. {\it Bottom Row:} Inclination.}
\label{fig:MultiLow}
\end{figure*}

In Fig.~\ref{fig:MultiLow} we show the orbital evolution for the low
$e$ and $i$ case over short (left) and long (right) timescales. As
expected, the planets exchange angular momentum, but over the first
million years there is no apparent drift due to tidal effects. On
longer timescales, however, we see the eccentricity of $b$ slowly decay
due to tidal heating. Note the differences in timescale for the decay
between Figs.~\ref{fig:eqtide} and \ref{fig:MultiLow}. The
perturbations from a hypothetical ``planet c'' maintain significant
eccentricities for long periods of time.

\begin{figure*}
\begin{center}
\includegraphics[width=\textwidth]{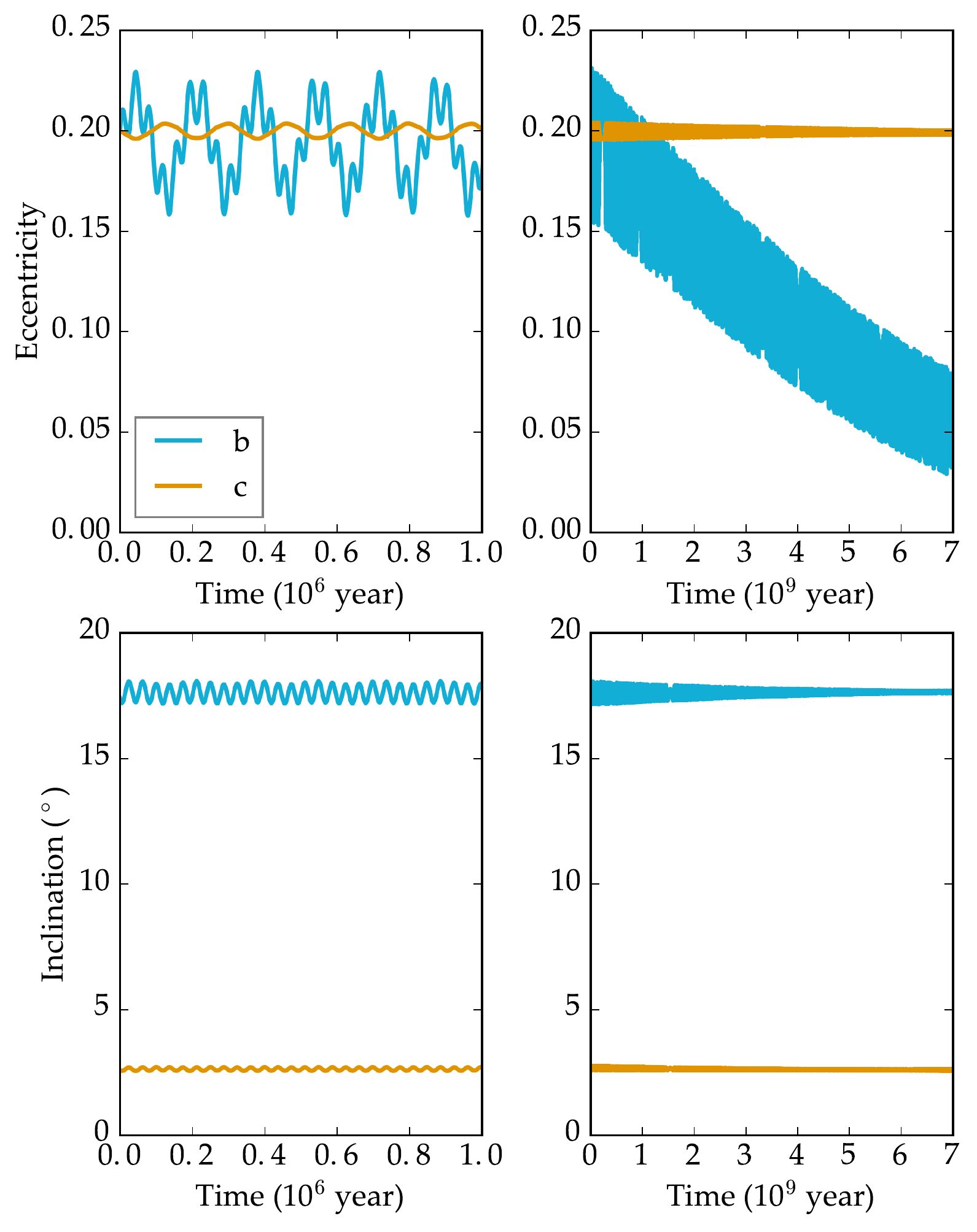}
\end{center}
\caption{Same as Fig.~\ref{fig:MultiLow}, but for the high $e,i$ case.}
\label{fig:MultiHigh}
\end{figure*}

In Fig.~\ref{fig:MultiHigh}, we plot the orbital evolution for the
high $e$ and $i$ case. The eccentricity and inclination oscillations
are longer, and the eccentricity cycles show several frequencies due
to the activation of coupling between $e$ and
$i$. As in the low $e$ and $i$ case, the eccentricity damps more
slowly than in the unperturbed case. Note as well that the inclination
oscillation amplitude decays with time.

\begin{figure*}
\begin{center}
\includegraphics[width=\textwidth]{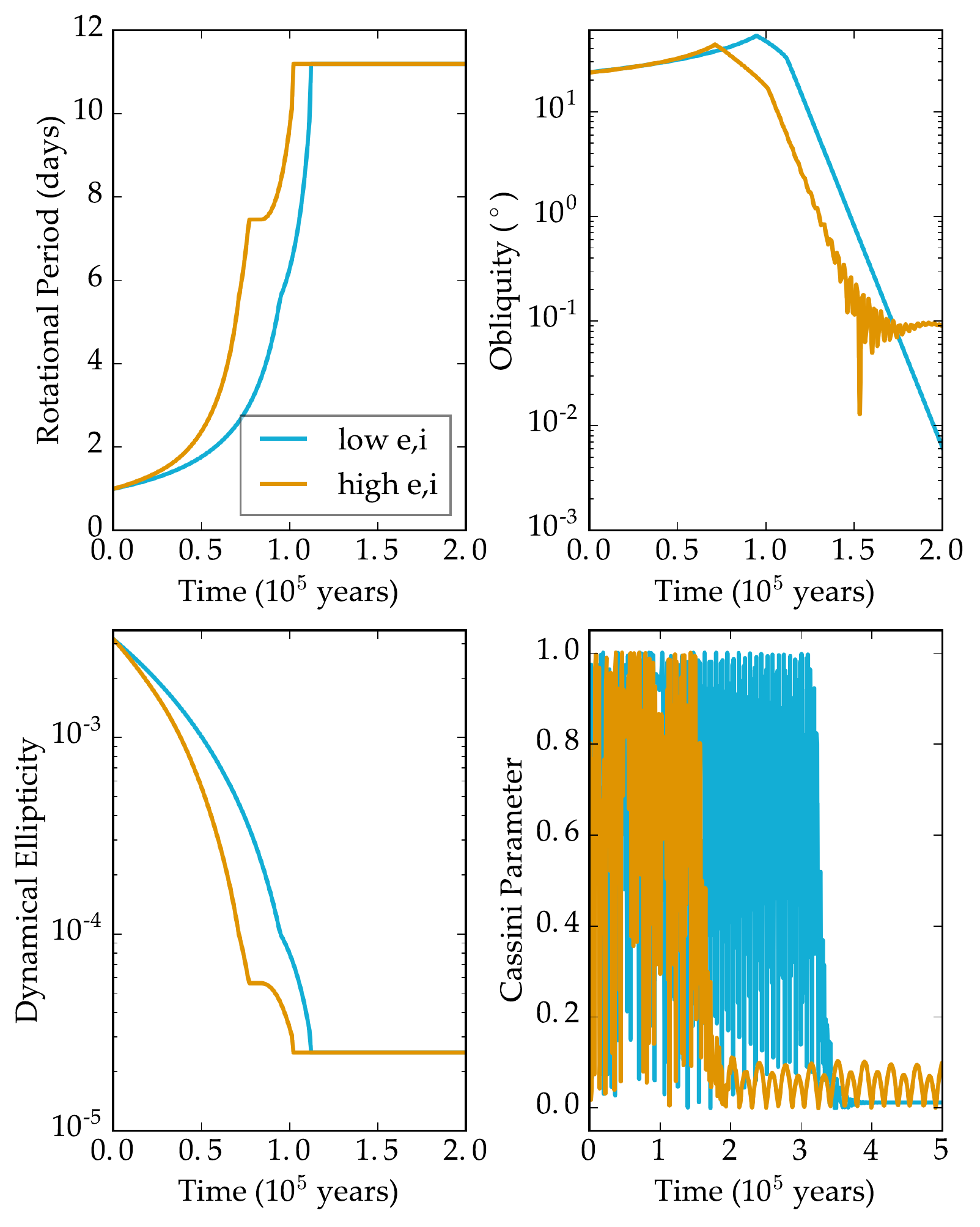}
\end{center}
\caption{Evolution of rotational properties of planet b for the
  low $e,i$ case
 (in blue), and high $e,i$ case (in orange). {\it Top left:} Rotation
 Period. {\it Top right:} Obliquity. {\it Bottom left:} Dynamical
 Ellipticity. {\it Bottom right:} Cassini Parameter.}
\label{fig:MultiSpins}
\end{figure*}

In Fig.~\ref{fig:MultiSpins}, we plot the evolution of the rotational
parameters for the two cases. In the top left panel, we show the
evolution of the rotational period. The rotation becomes tidally
locked very quickly (less than 1 Myr for all plausible values of $Q$
for an ocean-bearing world).  In the high $e,i$ case, the planet
briefly enters the 3:2 spin orbit frequency ratio (like the planet Mercury).
The obliquity initially grows due to conservation of angular momentum
\citep{Correia08}, but then damps down. For the high $e,i$ case, the
obliquity reaches an equilibrium value near $0.1^\circ$, while the low
$e,i$ case drops all the way to $10^{-8}$ degrees. The bottom left panel
shows the evolution of the dynamical ellipticity as predicted by the
formulae from \cite{Atobe2007}.  Realistically, the shape of the
planet should lag this shape by a timescale dependent on the planet's
rigidity, but we ignore that delay here. The lower right panel shows the
value of the Cassini parameter (see Equation \ref{eqn:cassini}) for
the two cases, both of which become locked near zero, indicating the
rotational and angular momentum have evolved into a Cassini state (in
this case, Cassini state 2), in which the spin and orbit vectors
of planet b are on opposite sides of the total angular momentum vector
of the planetary system.

\begin{figure*}[t]
\begin{center}
\includegraphics[width=\textwidth]{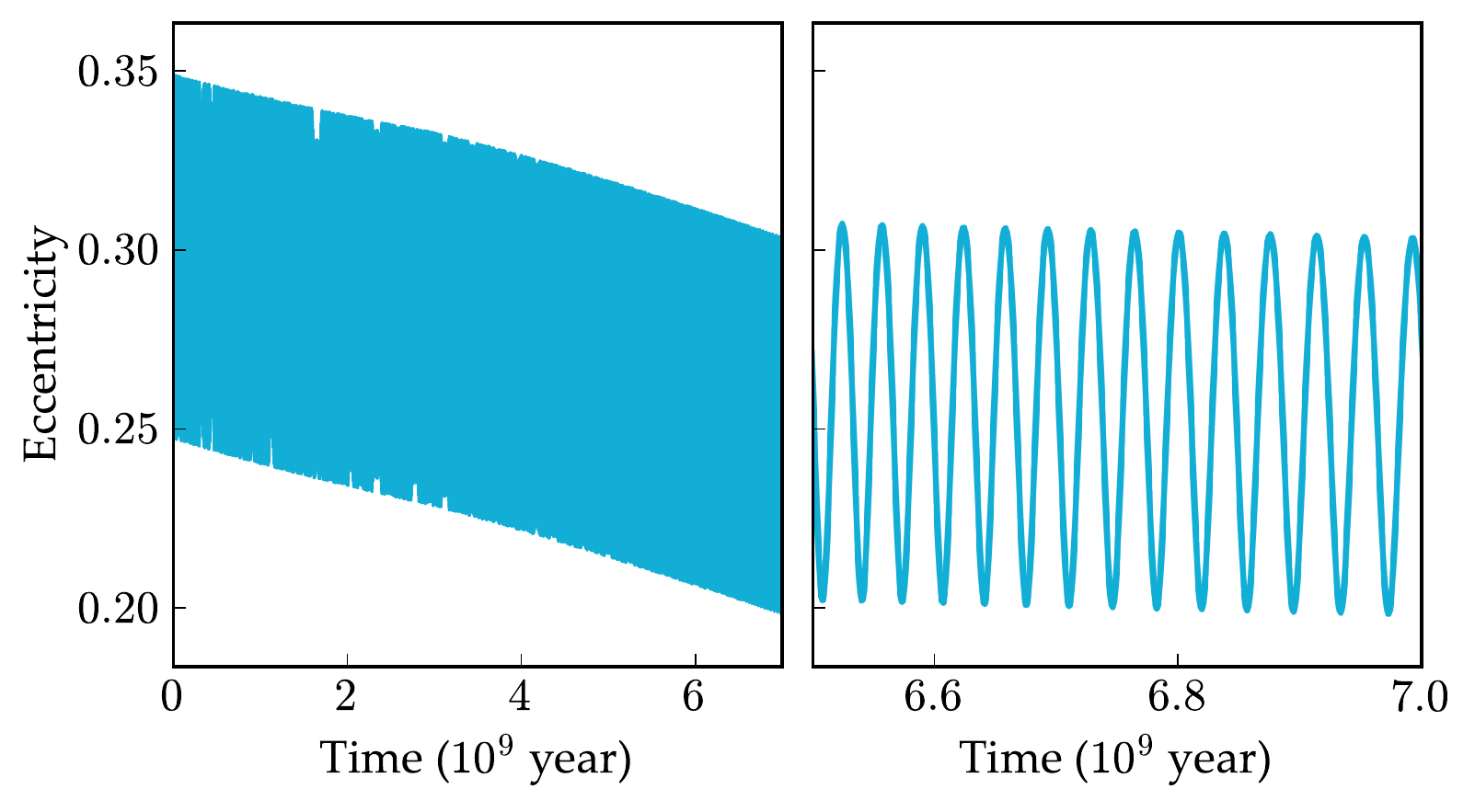}
\end{center}
\caption{Evolution of the eccentricity of planet b with a companion planet,
if the planets are coplanar and b's initial $e$ is 0.3. The left panel shows the full 7 Gyr evolution, while the right is zoom to show the last 500 Myr. A slow, large amplitude cycle persists after billions of years and the eccentricity remains above 0.2, indicating that spin-orbit resonances, i.e. super-synchronous rotation, is possible.}
\label{fig:HighEcc}
\end{figure*}

Figure \ref{fig:MultiSpins} hints at the possibility of non-synchronous
spin states for Proxima b, since the eccentricity is pumped to values large enough to force the rotation into a spin-orbit resonance. \cite{Ribas16} found a significant
probability that the planet is in a 3:2 spin orbit resonance. Our tidal
model does not take into account potential triaxiality
of the planet, and so we cannot  reproduce their results (\ie resonance trapping at
$e\gtrsim 0.1$). An expanded version of the model \citep{Rodriguez12}, shows
that the 3:2 state occurs generally for $e\gtrsim0.1$, consistent with the
\cite{Ribas16} result. In the case that Proxima has any additional planets,
it is entirely possible for planet b to maintain an eccentricity above
this value, even after 7 Gyrs. As an example, our third multiplanet case is like the configurations above, but coplanar and with a
slightly higher eccentricity for planet b ($e = 0.3$ initially). The
resulting eccentricity is shown in Figure \ref{fig:HighEcc}. The
eccentricity stays above 0.1 even after 7 Gyr of evolution. Though
the evolution is not fully self-consistent (it will be affected by
the planet's true spin state), it demonstrates that perturbations
can be strong enough to maintain a large eccentricity to force the planet into super-synchronous rotation.

\subsection{Stellar Evolution}
\label{sec:results:stellar}

\begin{figure}[ht]
\centering
\includegraphics[width=0.45\textwidth]{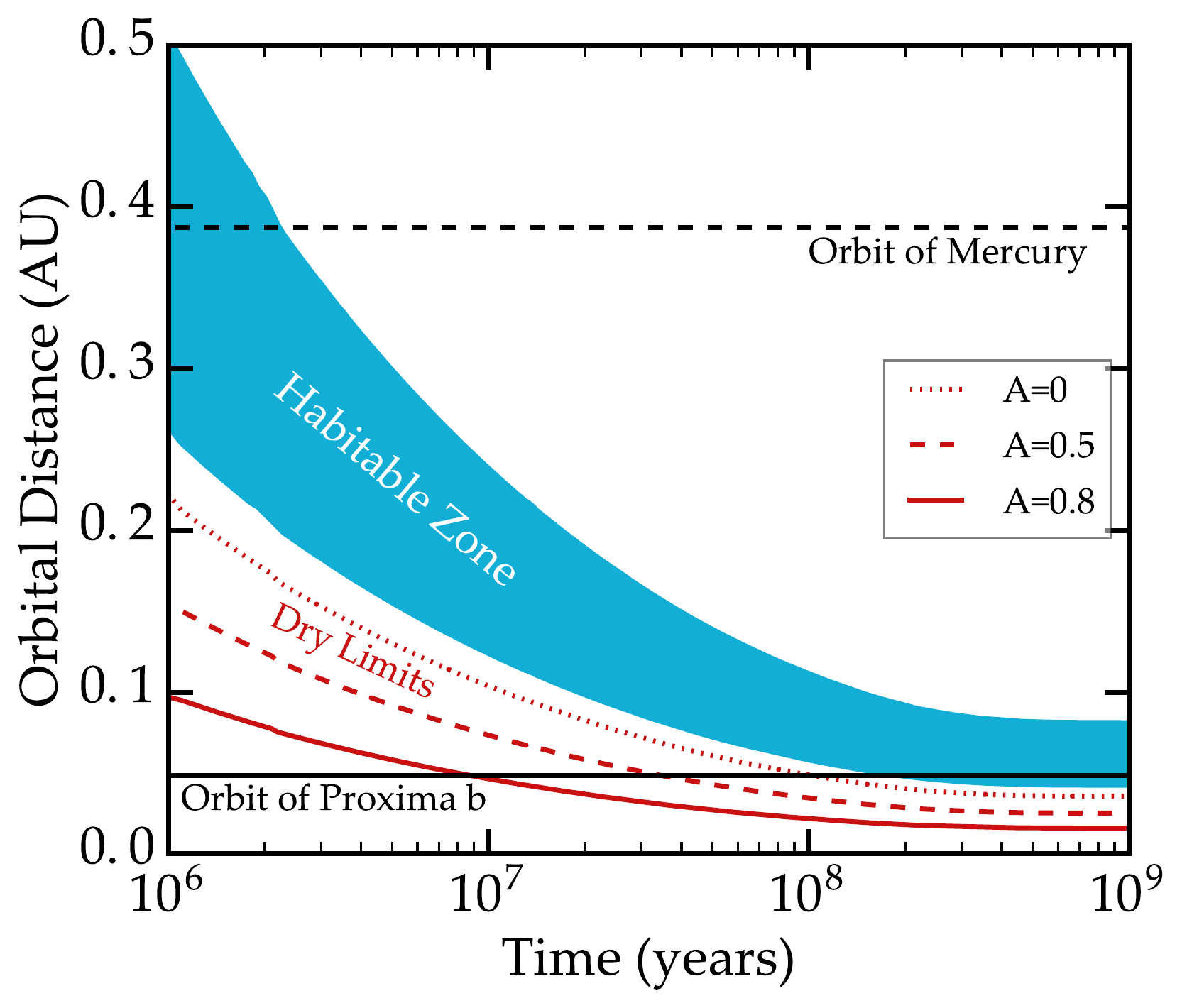}
\caption{Evolution of the HZ of Proxima Centauri assuming the fiducial stellar parameters (\S\ref{sec:models:stellar}), along with the orbits of Proxima Centauri b (solid line)
and Mercury (dashed line). The blue region is the conservative HZ of \cite{Kopparapu13}, while the red lines
are the HZ limits for dry planets of different albedos, $A$, from \cite{Abe11}.}
\label{fig:HZEvol}
\end{figure}

In Fig.~\ref{fig:HZEvol} we plot the evolution of the conservative HZ
limits of \cite{Kopparapu13} (blue region) as a function of time for our fiducial case (\S\ref{sec:models:stellar}); the
HZ is bounded by the runaway greenhouse limit on the side closest to
the star and by the maximum greenhouse limit on the opposite
side. The pre-MS luminosity evolution of
Proxima forces the HZ to slowly move inward for $\sim$ 1 Gyr, reaching the
current orbit of Proxima b after $\sim 169$ Myr, see $\S$~4.4.2.

The figure also shows the ``dry'' HZ limits of \cite{Abe11}, which
apply to planets with very limited surface water ($\lesssim 1\%$ of
the Earth's water inventory); these planets are significantly more
robust to an instellation-triggered runaway. Consequently, if Proxima
b's initial water content was very low, it would have spent
significantly less time in a runaway greenhouse. However, a dry
formation scenario for Proxima b does not help its present-day
habitability. As we show in \S\ref{sec:results:atmesc}, Proxima b
loses 1 ocean of water in $\lesssim 10$ Myr; if it formed with less water than that, it would be completely desiccated long before entering the \cite{Abe11} HZ. One can envision cases in which the planet forms dry but with a protective hydrogen envelope, or forms after $\sim$ 10 Myr, but such scenarios are unlikely \emph{a priori} and hence we do not consider them here.
In the atmospheric escape section below, we thus use the \cite{Kopparapu13} HZ limits to determine whether or not Proxima b is in a runaway greenhouse at any given time.

\vspace{1cm}
\subsection{Atmospheric Evolution}
\label{sec:results:atmesc}

\subsubsection{Fiducial cases}
\label{sec:results:atmesc:fiducial}
As we argued in \S\ref{sec:models:stellar}, the large uncertainties in the evolutionary history of the star make precise estimates of Proxima b's current atmospheric and water content impossible. In \S\ref{sec:results:atmesc:mcmc} below, we report the results of an ensemble of simulations that yield the posterior probability distribution for the planet's present-day water content given uncertainties on each of the parameters of the model. However, it is still instructive to consider the evolution of the system under our fiducial stellar parameters (\S\ref{sec:models:stellar}), which we briefly do here.

We consider two broad formation scenarios for Proxima b: one in which it formed with abundant water and a thin hydrogen envelope of up to 1\% by mass (due to either \emph{in situ} accretion or from planetary formation farther out followed by rapid disk-driven migration; see \cite{Luger15}), and one in which it formed with abundant water but no hydrogen. In both cases, we assume a fiducial planet mass of 1.27$\mathrm{M_\oplus}$.

\begin{figure*}[h!]
\centering
\includegraphics[width=\textwidth]{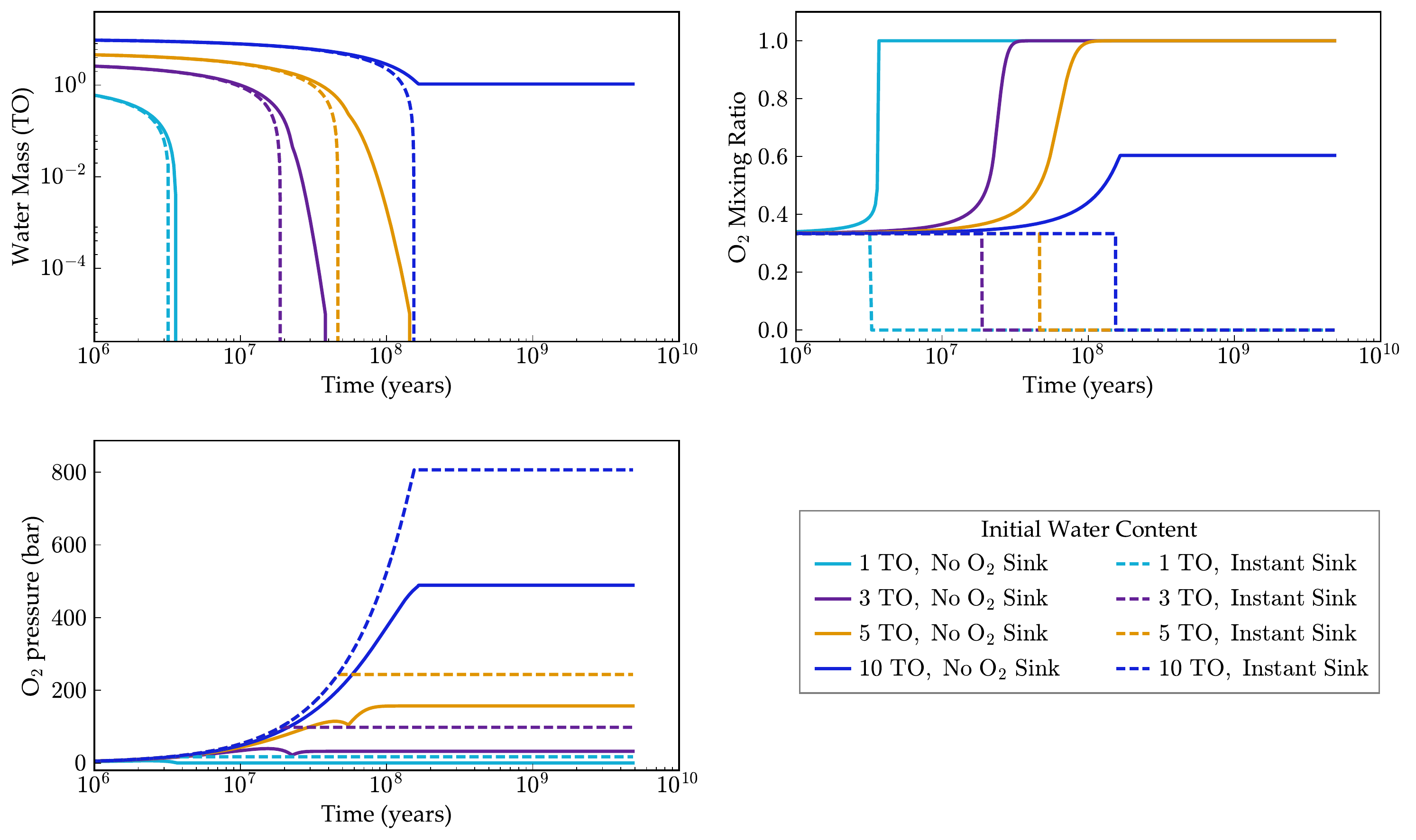}
\caption{Evolution of the water content and atmospheric O$_2$ pressure
  on Proxima b for the fiducial stellar parameters and different initial conditions. The initial water
  content is varied between 1 and 10 TO (various colors) for two
  different end-member scenarios: no O$_2$ surface sinks (solid lines)
  and instantaneous oxygen absorption at the surface (dashed
  lines). The planet mass is held constant at 1.27 M$_\oplus$ and the
  initial hydrogen envelope fraction is set to zero for all runs. In
  all but one of the runs, Proxima b is completely desiccated. For an
  initial water content of 10 TO and no surface sinks, the buildup of
  $\sim$ 500 bars of atmospheric O$_2$ slows the loss rate of H,
  preventing the last $\sim$ 1 TO of water from being lost. In the
  scenario that efficient oxygen sinks are present, the atmospheric
  O$_2$ mixing ratio never grows sufficiently to limit the escape of
  H, and desiccation occurs in all cases. Note that in this scenario,
  the curves in the ``O$_2$ pressure'' panel correspond to the
  equivalent O$_2$ pressure absorbed at the surface.\vspace{0.2in}}
\label{fig:atmesc:mirage}
\end{figure*}

In Fig.~\ref{fig:atmesc:mirage} we show the evolution of the latter type of planet, which formed with no significant primordial
envelope. We consider four different initial inventories of water: 1,
3, 5, and 10 terrestrial oceans (1 TO $\equiv 1.39\times 10^{24}$ g,
the total mass of surface water on Earth; see \cite{Kasting88}), but even larger inventories of water are theoretically possible. As
discussed in $\S$~\ref{sec:models:atmesc}, we also consider two end-member scenarios regarding the photolytically-produced O$_2$: no surface sinks (solid lines) and efficient surface sinks (dashed lines). In all cases but one, the planet is completely desiccated within the first 170 Myr, building up between tens and hundreds of bars of O$_2$ in
either its atmosphere or in the solid body. For an initial water
content of 10 TO and no surface sinks, O$_2$ builds up to high enough
levels to throttle the supply of H to the upper atmosphere and slow
the total escape rate. In this scenario, $\sim 1$ TO of water remains,
alongside a thick 500 bar O$_2$ atmosphere. If Proxima b formed with
less than ten times Earth's water content, and/or had a persistent
convecting, reducing magma ocean, it is desiccated today
according to our fiducial case.

\begin{figure*}[h!]
\centering
\includegraphics[width=\textwidth]{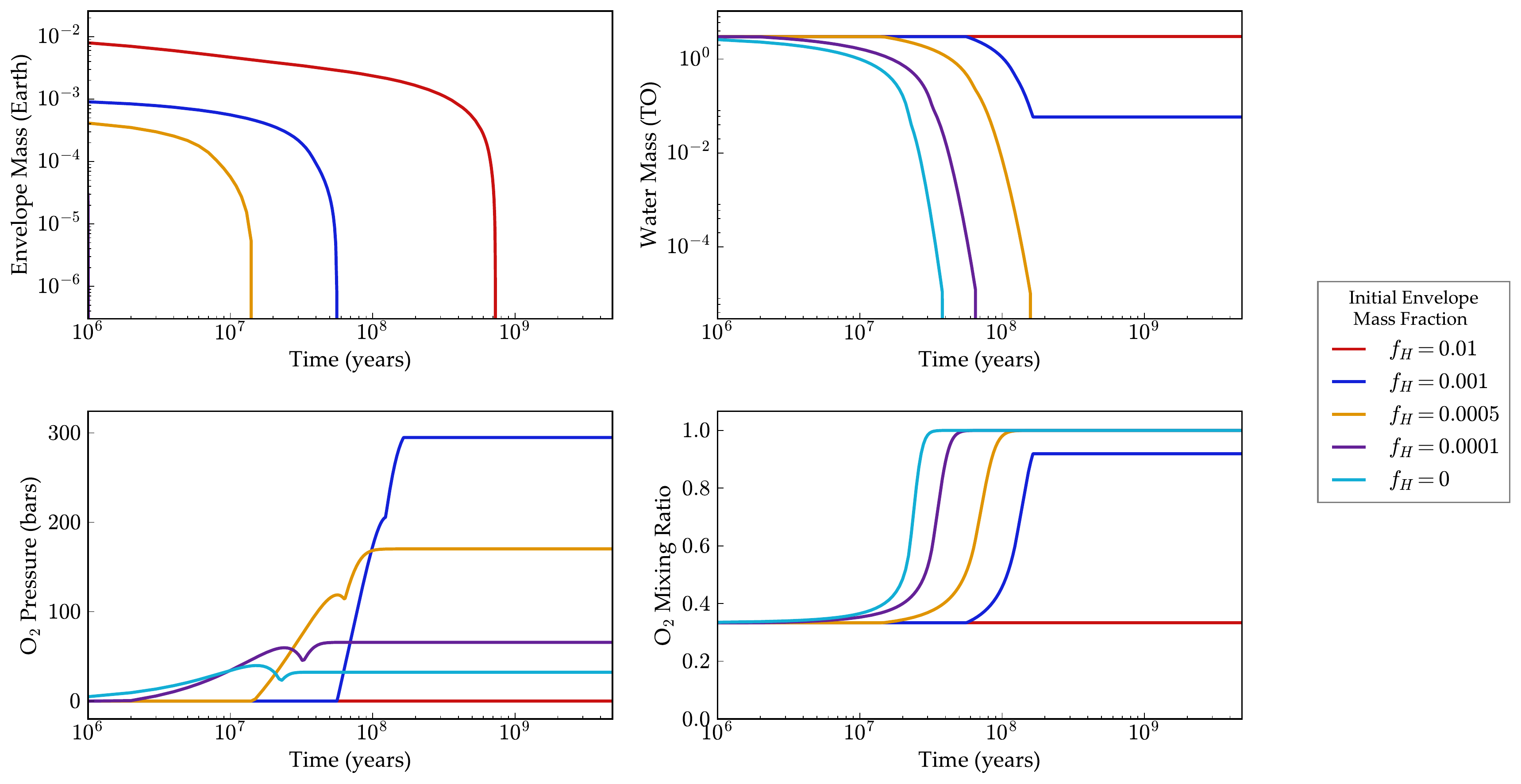}
\caption{Evolution of the water and O$_2$ contents assuming Proxima b
  formed with a hydrogen envelope and 3 TO. Line colors correspond to
  different initial envelope mass fractions $f_H$, ranging from 0.0001
  to 0.01. In all cases, the envelope escapes completely prior to 1
  Gyr. For $f_H \lesssim 0.001$, the H envelope escapes quickly
  enough to allow complete desiccation of the planet prior to its
  arrival in the HZ at $\sim$ 170 Myr. For $f_H \approx 0.01$, the
  envelope escapes a few hundred Myr \emph{after} the planet enters
  the HZ, preventing any water from being lost and O$_2$ from
  accumulating in the atmosphere.  This is the most favorable scenario
  for a potentially habitable Proxima b.\vspace{0.2in}}
\label{fig:atmesc:hec}
\end{figure*}

Next, in Fig.~\ref{fig:atmesc:hec}, we show the results assuming
Proxima b formed with a hydrogen envelope. We fix the initial water
content at 3 TO and consider initial envelope mass fractions $f_H$
ranging from $10^{-4}$ to $10^{-2}$. In all cases, the envelope
is lost completely within the first several hundred Myr. For $f_H
\lesssim 10^{-3}$, the envelope is lost early enough such that all
the water is still lost from the planet. For $f_H = 10^{-3}$, only
about 0.1 TO remain once the planet enters the HZ; only for $f_H \sim
10^{-2}$ does the presence of the envelope guard against all water
loss. In these calculations, we assumed inefficient surface sinks, so
the escape of water at late times was bottlenecked by the presence of
abundant O$_2$. Planets that form with hydrogen envelopes may have
quite reducing surfaces, which could absorb most of the O$_2$ and lead
to even higher total water loss. As before, a few tens to a few
hundreds of bars of O$_2$ remain in the atmosphere or in the solid
body at the end of the escape phase.

We note that we obtain slightly more hydrogen loss than
\cite{OwenMohanty16}, who find that planets more massive than $\sim
0.9 \mearth$ with $\sim 1\%$ hydrogen envelopes cannot fully lose
their envelopes around M dwarfs, due primarily to the transition from
hydrodynamic to ballistic escape at late times. However, their
calculations were performed for a $0.4 \msun$ M dwarf, whose pre-main
sequence phase lasts $\sim 200 \mathrm{Myr}$, five times shorter than
that for Proxima Centauri. Nevertheless, the discrepancy is small: we
find that for envelope fractions greater than 1\% or masses greater
than our fiducial value of $1.27 \mearth$, the envelope does not
completely escape, in which case Proxima b would likely be
uninhabitable.


\subsubsection{MCMC Simulations}
\label{sec:results:atmesc:mcmc}
In the previous section we showed that our model predicts that unless Proxima b formed with a substantial hydrogen envelope (mass fraction $f_H \gtrsim 0.001$) or more than 10 TO of water, it may be likely desiccated today under our fiducial stellar evolution tracks. However, the large uncertainties in the observed stellar properties and in our adopted model parameters translate to large uncertainties in these values. To robustly account for these uncertainties, in this section we perform
a suite of Markov Chain Monte Carlo (MCMC) runs. MCMC allows one to sample from multi-dimensional probability distributions that are difficult or impossible to obtain directly, which is the case for the ensemble of parameters that control the evolution of the planet surface water and oxygen content in \texttt{VPLANET}. Here we develop a framework for inferring the probability
distributions of these parameters conditioned on empirical data and our understanding
of the physical processes at play.

The input parameters to our model make up the state vector $\mathbf{x}$:
\begin{align}
\label{eq:mcmcx}
\mathbf{x} = \{f_\mathrm{sat}, t_\mathrm{sat}, \beta_\mathrm{xuv}, M_\star, t_\star, a, m\},
\end{align}
corresponding, respectively, to the XUV saturation fraction, the XUV saturation timescale,
the XUV power law exponent, the stellar mass, the stellar age, the semi-major axis of the planet, and the
mass of the planet. Given a value of $\mathbf{x}$, \texttt{VPLANET} computes the evolution of the system from
time $t = 0$ to $t = t_\star$, yielding the output vector $\mathbf{y}$:
\begin{align}
\label{eq:mcmcy}
\mathbf{y}(\mathbf{x}) = \{L_\star, L_\mathrm{xuv}, t_\mathrm{RG}, m_\mathrm{H}, m_\mathrm{H_2O}, P_\mathrm{O_2}\},
\end{align}
corresponding, respectively, to the stellar luminosity, the stellar XUV luminosity, the duration of the
runaway greenhouse phase, the mass of the
planet's hydrogen envelope, the mass of water remaining on its surface, and the amount of oxygen (expressed
as a partial pressure) retained
in either the atmosphere or the surface/mantle, all of which are evaluated at $t = t_\star$ (i.e., the present day).
Additional parameters that control the evolution of the
planet (initial water content, XUV absorption efficiency, etc.) are held fixed in individual runs; see below.

Our goal in this section is to derive posterior distributions for $\mathbf{y}$ (and in particular for $m_\mathrm{H_2O}$)
given prior information on both
$\mathbf{x}$ and $\mathbf{y}$. Some parameters---such as the present-day stellar luminosity---are well-constrained,
while others are less well-known and will thus be informed primarily by our choice of prior. This is the case for
the XUV saturation fraction, saturation timescale, and power law exponent, which have been studied in detail
for solar-like stars \citep{Ribas05} but are poorly constrained for M dwarfs \citep[see, e.g.,][]{LugerBarnes15}.
We therefore use flat-log priors for the saturation fraction and timescale, enforcing
$-5 \leq \log(f_\mathrm{sat}) \leq -2$ and $-0.3 \leq \log(t_\mathrm{sat} / \mathrm{Gyr}) \leq 1$. We use
a Gaussian prior for the XUV power law exponent, with a mean of 1.23, the value derived by \cite{Ribas05} for
solar-like stars: $\beta_\mathrm{xuv} \sim \mathcal{N}(-1.23, 0.1^2)$. We choose an ad hoc standard deviation
$\sigma = 0.1$ and verify \emph{a posteriori} that our results are not sensitive to this choice. As we show
below, $\beta_\mathrm{xuv}$ does not strongly correlate with the total water lost or total
amount of oxygen that builds up on the planet.

We also use a flat prior for the stellar mass ($0.1 \leq M_\star / \mathrm{M}_\oplus \leq 0.15$).
Although stronger constraints on the stellar mass exist \citep[e.g.,][]{Delfosse00,Segransan2003}, these are derived indirectly from mass-luminosity or mass-radius
relations, which are notoriously uncertain for low mass stars \citep[e.g.,][]{Boyajian12}. We thus
enforce a prior on the present-day luminosity to constrain the value of $M_\star$ via our stellar evolution model (see below).
We enforce a Gaussian prior on the stellar age $t_\star \sim \mathcal{N}(4.8, 1.4^2)$ Gyr based on the constraints discussed
in \S\ref{sec:obs:stellarsys} and \S\ref{sec:obs:inf}.

Our prior on the semi-major axis $a$ is a combination of a Gaussian prior on the orbital period,
$P \sim \mathcal{N}(11.186, 0.002^2)$ days \citep{AngladaEscude16}, and the stellar mass prior.
Finally, our prior on the planet mass $m$ combines the empirical minimum mass distribution,
$m\sin i \sim \mathcal{N}(1.27, 0.18^2)$ M$_\oplus$ \citep{AngladaEscude16}, and the {\it a priori} inclination distribution
for randomly aligned orbits,
$\sin i \sim \mathcal{U}(0, 1)$, where $\mathcal{U}$ is a uniform distribution \citep[e.g.,][]{Luger17}.

We further condition our model on measured values of the stellar luminosity $L_\star$ and
stellar XUV luminosity $L_\mathrm{xuv}$. We take
$L_\star \sim \mathcal{N}(1.65, 0.15^2) \times 10^{-3}$ L$_\odot$ \citep{Demory09} and
$\log L_\mathrm{xuv} \sim \mathcal{N}(-6.36, 0.3^2)$. We base the latter on \cite{Ribas16}, who compiled a comprehensive
list of measurements of the emission of Proxima Centauri in the wavelength range 0.6--118 nm. Summing the fluxes over
this range and neglecting the contribution of flares, we obtain an XUV flux at Proxima b of
$F_\mathrm{xuv} \approx 252\ \mathrm{erg\ cm^{-2}\ s^{-1}}$, corresponding to $\log(L_\mathrm{xuv}/L_\odot) = -6.36$ for $a = 0.0485$ AU. Given the
lack of uncertainties for many of the values compiled in \cite{Ribas16} and the fact that some of those estimates
are model extrapolations, it is difficult to establish a reliable error estimate for this value. We make the
{\it ad hoc}, but conservative, choice $\sigma = 0.3$ dex, noting that the three measurements that inform the X-ray luminosity
of the star in \cite{Ribas16} (which dominates its XUV emission) have a spread corresponding to $\sigma = 0.2$ dex.
However, more rigorous constraints on the XUV emission of Proxima with reliable uncertainties are direly needed
to obtain more reliable estimates of water loss from Proxima b.

Given these constraints, we wish to find the posterior distribution of each of the parameters in Equations~(\ref{eq:mcmcx})
and~(\ref{eq:mcmcy}). We thus define our likelihood function $\mathcal{L}$ for a given state vector $\mathbf{x}$ as
\begin{align}
\ln \mathcal{L}(\mathbf{x}) = &- \frac{1}{2}\left[\frac{(L_\mathrm{\star}(\mathbf{x}) - L_\mathrm{\star})^2}{\sigma_{L_\star}^2}
                              + \frac{(L_\mathrm{xuv}(\mathbf{x}) - L_\mathrm{xuv})^2}{\sigma_{L_\mathrm{xuv}}^2}\right]
                                \nonumber\\
                             &+ \ln \mathrm{Prior}(\mathbf{x}) + C,
\end{align}
where $L_\mathrm{\star}(\mathbf{x})$ and $L_\mathrm{xuv}(\mathbf{x})$ are, respectively, the model predictions for the present-day
stellar luminosity and stellar XUV luminosity given the state vector $\mathbf{x}$, $L_\mathrm{\star}$ and $L_\mathrm{xuv}$ are
their respective observed values, and $\sigma_{L_\star}^2$ and $\sigma_{L_\mathrm{xuv}}^2$ are the uncertainties on those observations. The
$\ln \mathrm{Prior}(\mathbf{x})$ term is the prior probability and $C$ is an arbitrary normalization constant. Expressed in this form,
the observed values of $L_\mathrm{\star}$ and $L_\mathrm{xuv}$ are our ``data,'' while the constraints on the other parameters
are ``priors,'' though the distinction is purely semantic.

\begin{figure*}[h!]
 \begin{center}
     \includegraphics[width=\textwidth]{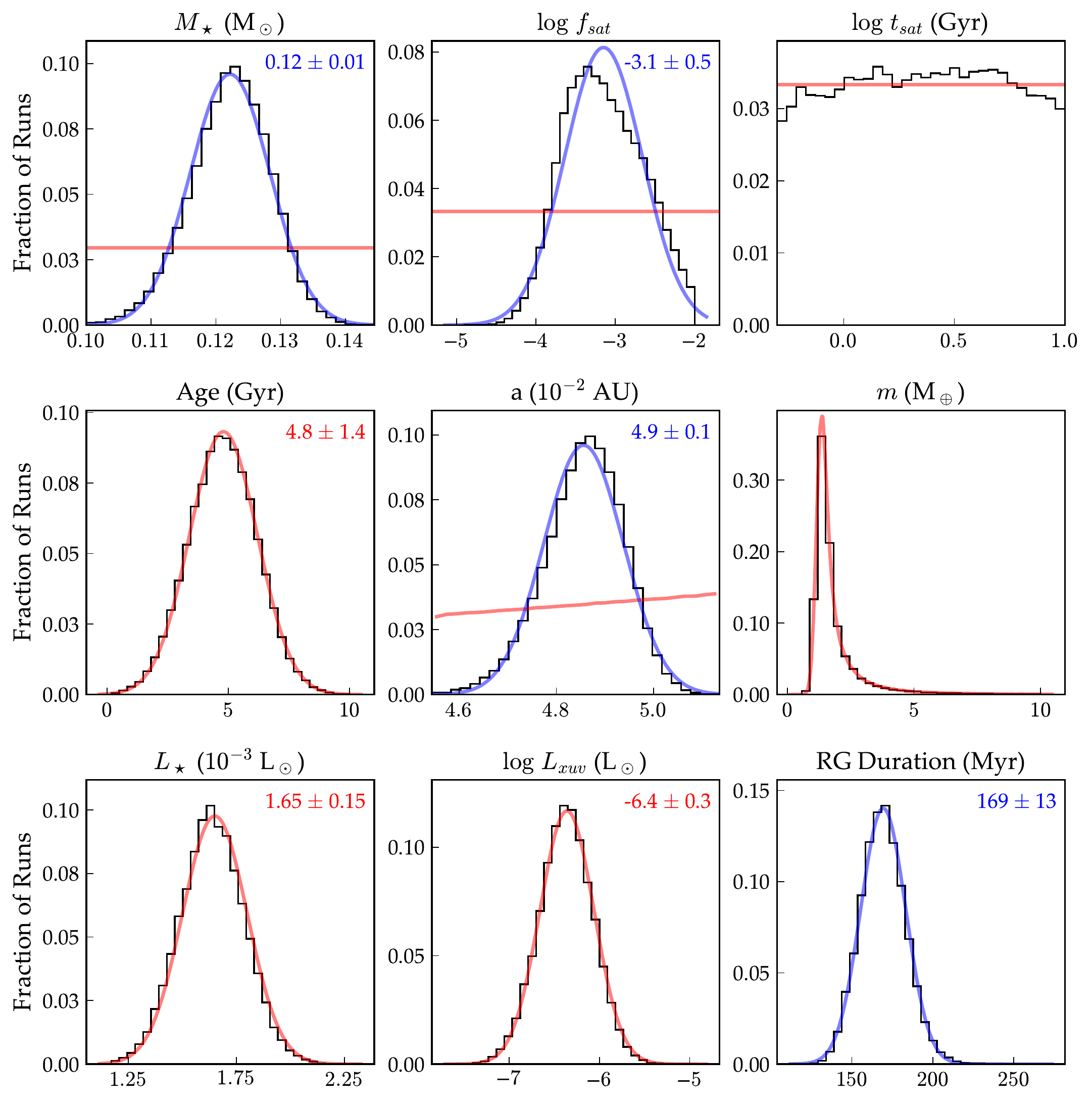}
      \caption{Posterior distributions for the various stellar parameters used in the model. The first eight
               parameters are model inputs, with their corresponding priors shown in red. The combination of
               these priors and the physical models in \texttt{VPLANET} constrain the stellar and planetary
               parameters shown in this section. Blue curves show Gaussian fits to the posterior distributions,
               with the mean and standard deviation indicated at the top right.
               The last panel shows the duration of the runaway greenhouse
               phase for Proxima Centauri b, one of the model outputs, which we find to be 169 $\pm$ 13 Myr.}
    \label{fig:star_posteriors}
 \end{center}
\end{figure*}

Given this likelihood function, we use the Python code package \texttt{emcee} \citep{ForemanMackey13} to obtain the posterior probability distributions for each of the parameters
of interest. We initialize each of the parameters in $\mathbf{x}$ by drawing from their respective prior distributions and run 40 parallel chains of 5,000
steps each, discarding the first 500 steps as burn-in. The marginalized posterior distributions for the stellar mass, saturation fraction,
saturation timescale, age, semi-major axis, planet mass, present-day stellar luminosity, present-day stellar XUV luminosity, and
duration of the runaway greenhouse are shown in Figure~\ref{fig:star_posteriors} as the black histograms. The red curves
indicate our priors/data, and the blue curves are Gaussian fits to the posteriors. The fit to the runaway greenhouse duration posterior yields $t_\mathrm{RG} = 169 \pm 13$ Myr.

By construction, the planet mass, stellar age, present-day stellar luminosity, and present-day stellar XUV luminosity posteriors reflect
their prior distributions. As mentioned above, the stellar mass posterior is entirely informed by the luminosity posterior via the
\cite{YonseiYale13} stellar evolution tracks. The stellar mass in turn constrains the semi-major axis (via the prior on the period and
Kepler's laws). The XUV saturation fraction is fairly well constrained by the present-day XUV luminosity; a log-normal fit
to its posterior yields $\log\ f_\mathrm{sat} = -3.1 \pm 0.5$, which is fully consistent with the observation that M dwarfs
saturate at or below $\log\ f_\mathrm{sat} \approx -3$ \citep{Jackson2012, Shkolnik2014}. The longer tail at high $f_\mathrm{sat}$ results from the fact that
this parameter is strongly correlated with the saturation timescale, $t_\mathrm{sat}$ (see Figure~\ref{fig:corner} below). If
saturation is short-lived, the initial saturation fraction must be higher to match the present-day XUV luminosity. Interestingly,
our runs do not provide any constraints on $t_\mathrm{sat}$, whose value is equally likely (in log space) across the range
$[0.5, 10]$ Gyr. Finally, the posterior for the XUV power law exponent $\beta_\mathrm{xuv}$ (not shown in the Figure) is
the same as the adopted prior, as the present data are insufficient to constrain it.

\begin{figure*}[h!]
 \begin{center}
     \includegraphics[width=\textwidth]{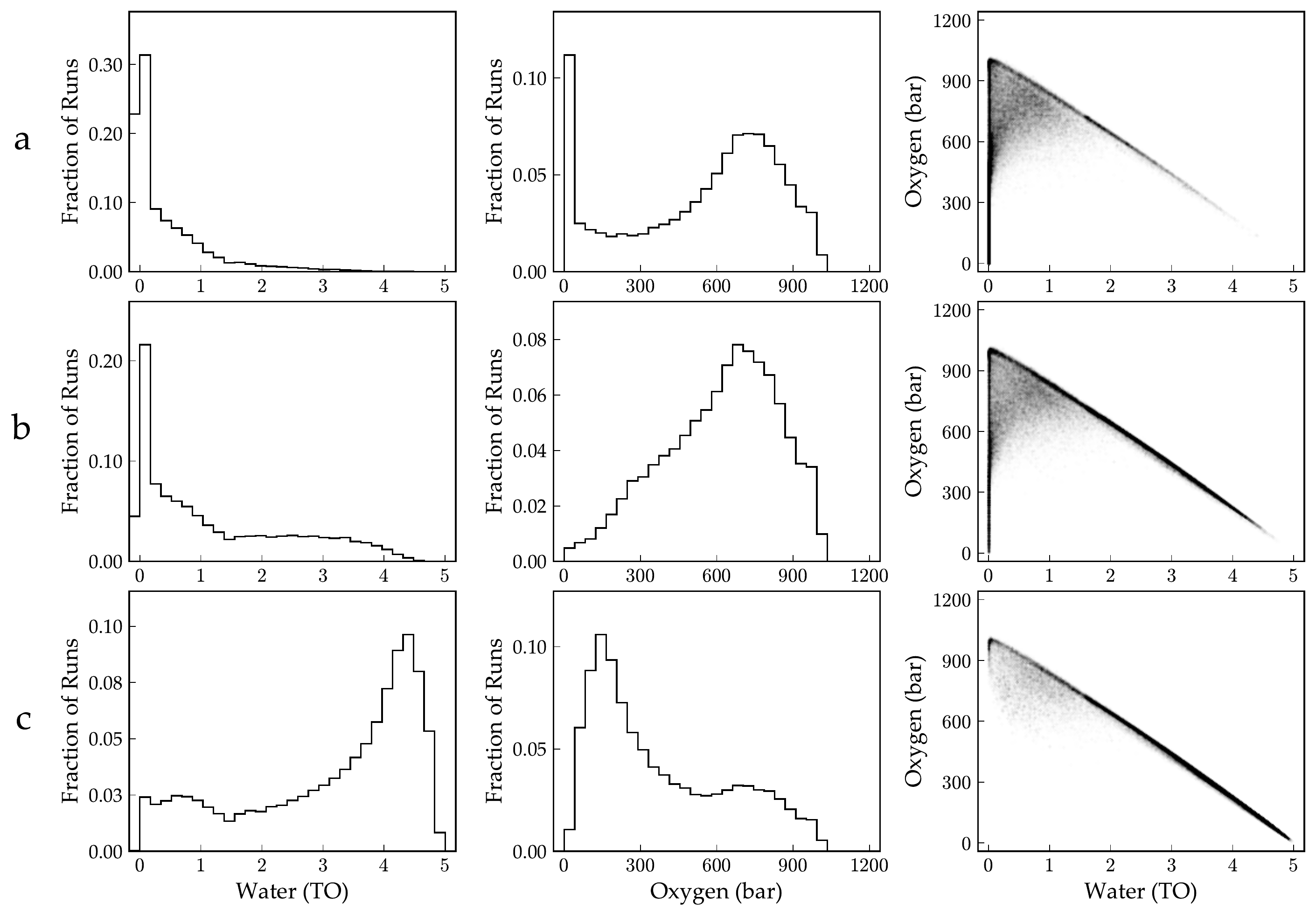}
      \caption{Marginalized posteriors for the present-day water content (left) and atmospheric oxygen
      pressure (center) on Proxima Cen b. The joint posteriors for these two parameters are shown at the
      right. \textbf{(a)} Posteriors for the default run ($m_\mathrm{H_2O}^0 = 5$ TO, $m_\mathrm{H}^0 = 0$,
      $\epsilon_\mathrm{xuv} = 0.15$, $\zeta_\mathrm{O_2} = 0$). \textbf{(b)} Same as (a), but for
      $\epsilon_\mathrm{xuv} = 0.05$. \textbf{(c)} Same as (a), but for
      $\epsilon_\mathrm{xuv} = 0.01$. For $\epsilon_\mathrm{xuv} \gtrsim 0.05$, the planet is desiccated or
      almost desiccated and builds up between 500 and 900 bars of O$_2$ in most runs. For $\epsilon_\mathrm{xuv} \sim 0.01$,
      the planet loses less water and builds up less O$_2$, though the loss of more than 1 TO is still likely.
      }
    \label{fig:planet_epsilon}
 \end{center}
\end{figure*}

\begin{figure*}[h!]
 \begin{center}
     \includegraphics[width=\textwidth]{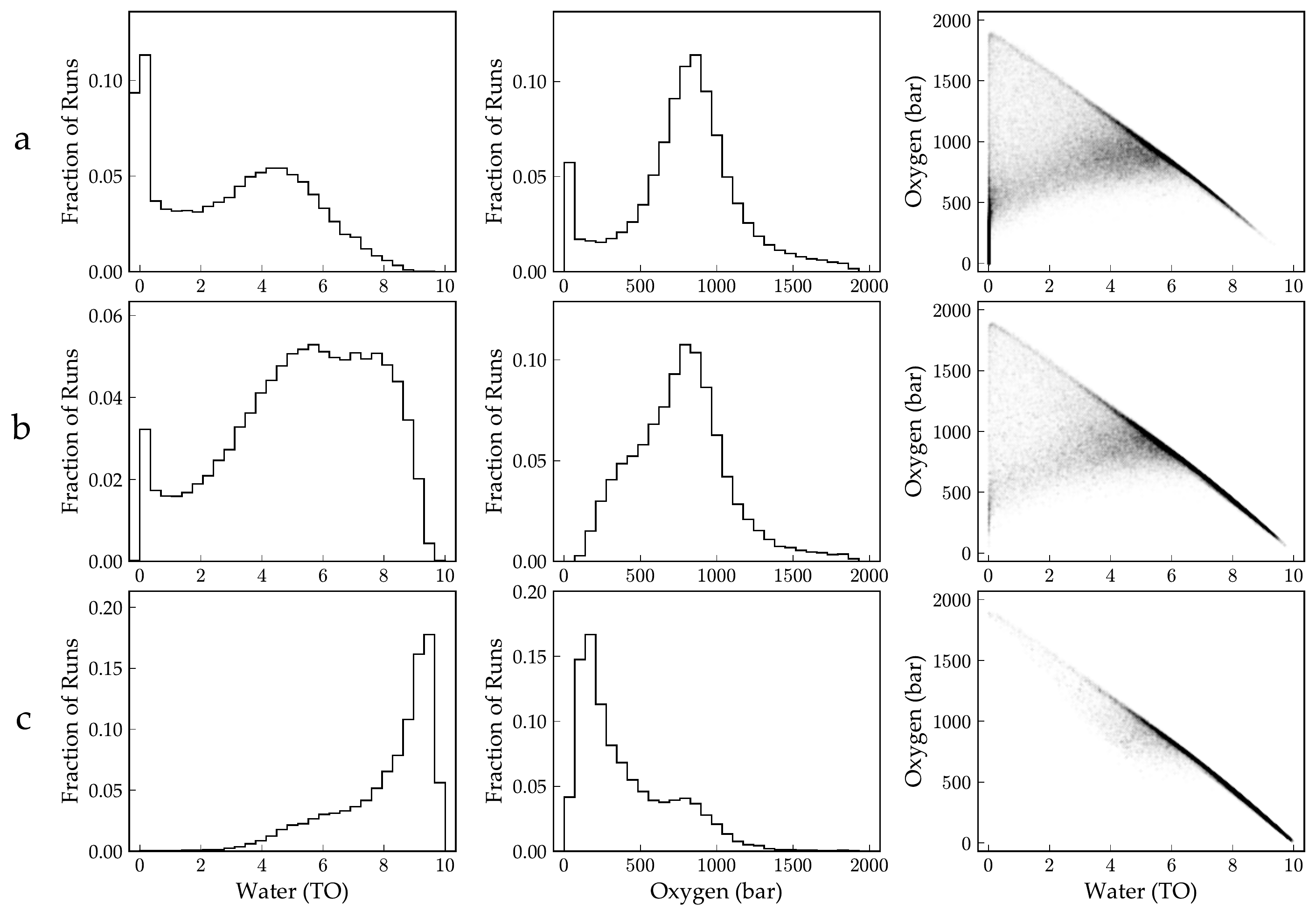}
      \caption{Similar to Figure~\ref{fig:planet_epsilon}, but for an initial water content
      $m_\mathrm{H_2O}^0 = 10$ TO. As before, the rows correspond to XUV escape efficiencies of
      0.15, 0.05, and 0.01 from top to bottom, respectively. For high XUV efficiency, Proxima
      Cen b loses more than 5 TO in most runs (and is desiccated in $\sim$ 20\% of runs). At lower
      efficiency, the planet loses less water. The amount of O$_2$ that builds up is similar
      to before, but a buildup of more than 1000 bars is now possible.}
    \label{fig:planet_epsilon10}
 \end{center}
\end{figure*}

The two quantities that are most relevant to habitability --- the final water content $m_\mathrm{H_2O}$ and final O$_2$ atmospheric
pressure $P_\mathrm{O_2}$ of Proxima b --- depend on four additional parameters we must specify: the initial water content $m_\mathrm{H_2O}^0$, the initial hydrogen
mass $m_\mathrm{H}^0$ (if the planet formed with a primordial envelope), the XUV escape efficiency $\epsilon_\mathrm{xuv}$, and
the O$_2$ uptake efficiency $\zeta_\mathrm{O_2}$ of the planet surface. In principle, planet formation models could provide
priors on $m_\mathrm{H_2O}^0$ and $m_\mathrm{H}^0$, but such models depend on additional parameters that are unknown or poorly
constrained. The same is true for the XUV escape efficiency, which can be modeled as in \cite{Ribas16}, and the rate of
absorption of O$_2$ at the surface, which can be computed as in \cite{Schaefer16}. However, given the large number of unknown
parameters needed to constrain these four parameters, for simplicity we perform independent MCMC runs for fixed combinations
of these parameters. This approach circumvents potential biases arising from incorrect priors on these parameters while still
highlighting how our results scale with different assumptions about their values. Note that the net water loss posteriors are qualitatively similar for $\epsilon_\mathrm{xuv} = 0.15$ and $\epsilon_\mathrm{xuv} = 0.05$, justifying our use of energy-limited escape.

In the runs discussed below, our default values are $m_\mathrm{H_2O}^0 = 5$ TO, $m_\mathrm{H}^0 = 0\ \mathrm{M_\oplus}$,
$\epsilon_\mathrm{xuv} = 0.15$, and $\zeta_\mathrm{O_2} = 0$, and we vary each of these parameters in turn.
Figure~\ref{fig:planet_epsilon} shows the marginalized posterior distributions for the present-day water content (left column)
and present-day O$_2$ atmospheric pressure (middle column), as well as a joint posterior for the two parameters (right column)
for three different values of $\epsilon_\mathrm{xuv}$: \textbf{(a)} 0.15, \textbf{(b)} 0.05, and \textbf{(c)} 0.01. In the first
two cases, the planet loses all or nearly all of the 5 TO it formed with, building up several hundred bars of O$_2$ (with
distributions peaking at about 700 bars and with a spread of several hundred bars). For $\epsilon_\mathrm{xuv} = 0.15$, about
10\% of runs result in no substantial oxygen remaining in the atmosphere; in these runs, the escape was so efficient as to
remove all of the O$_2$ along with the escaping H. In the final case, the amount of water lost is significantly smaller:
about 2 TO on average, with a peak in the distribution corresponding to a loss of about 0.8 TO. The amount of O$_2$
remaining is similarly smaller, but still exceeding 100 bars and with a similar spread as before. Finally, the joint posterior
plots emphasize how correlated the present-day water and oxygen content of Proxima b are. Since the rate at which
oxygen builds up in the atmosphere is initially constant \citep{LugerBarnes15}, and since the amount of water
lost scales with the duration of the escape period, there is a tight linear correlation between the two quantities
(lower right hand corner of the joint posterior plots). However, as the atmospheric mixing ratio of oxygen increases,
the rate at which hydrogen escapes---and thus the rate at which oxygen is produced---begins to decrease, leading to a break
in the linear relationship once $\sim$ 600--700 bars of oxygen build up and leading to the peak in the O$_2$ posteriors
at around that value.

Figure~\ref{fig:planet_epsilon10} is similar to Figure~\ref{fig:planet_epsilon}, but shows runs assuming Proxima Cen b
formed with 10 TO of water. As before, the rows correspond to different escape efficiencies (0.15, 0.05, 0.01, from top
to bottom). The amount of water lost increases in all cases, and for $\epsilon_\mathrm{xuv} = 0.15$ the planet is
desiccated or almost desiccated in about 20\% of runs. The amount of O$_2$ that builds up is similar to that in the
previous figure, but O$_2$ pressures exceeding 1000 bars are now possible in 20--30\% of cases for XUV efficiencies of 0.15
or 0.05.

\begin{figure*}[h!]
 \begin{center}
     \includegraphics[width=\textwidth]{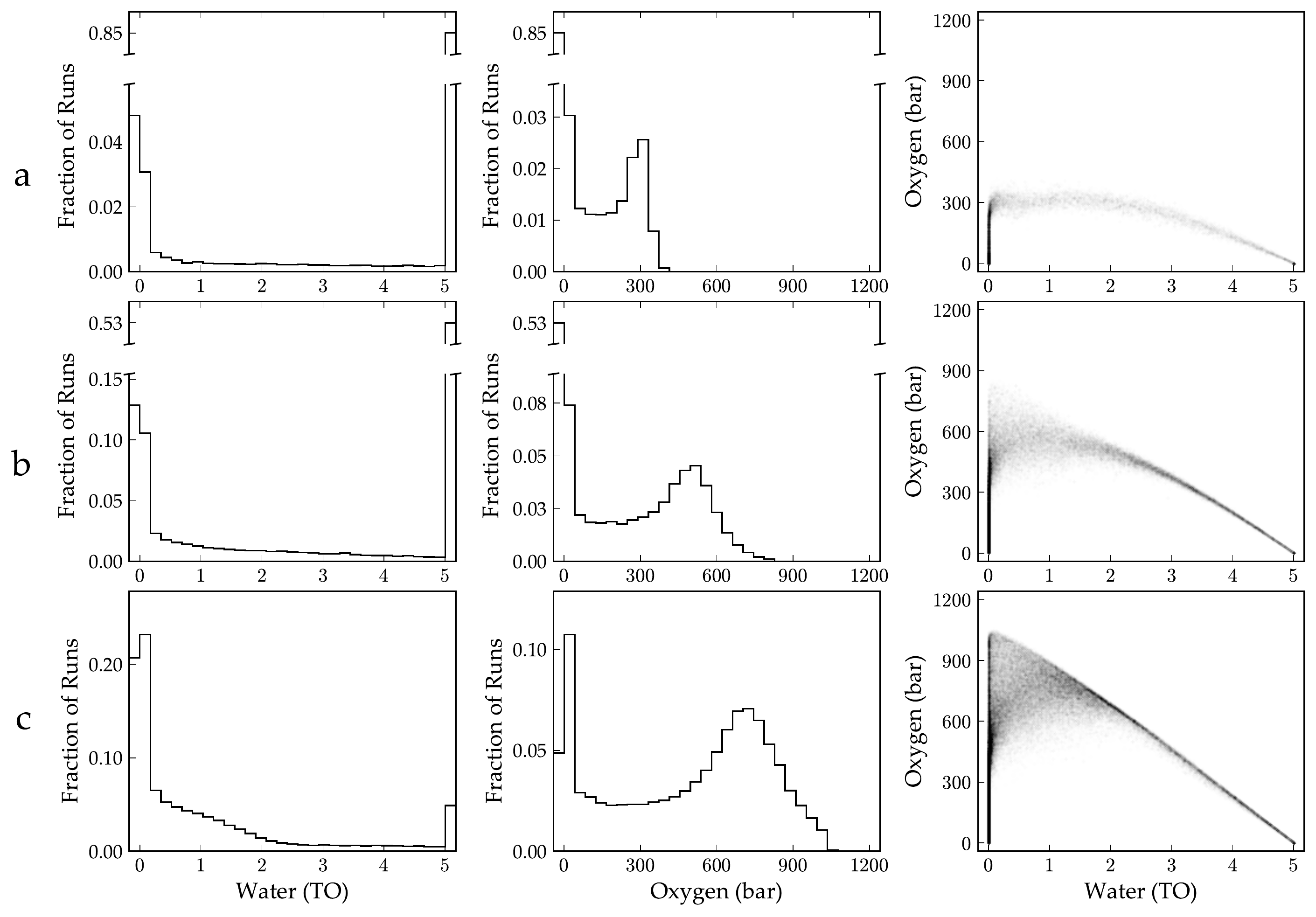}
      \caption{Similar to Figure~\ref{fig:planet_epsilon}, but this time varying the initial mass of the primordial
      hydrogen envelope of Proxima Cen b. Other parameters are set to their default values. The initial mass of hydrogen
      is $m_\mathrm{H}^0 =$ \textbf{(a)} $0.01\ \mathrm{M_\oplus}$, \textbf{(a)} $0.001\ \mathrm{M_\oplus}$, and
      \textbf{(a)} $10^{-4}~\mathrm{M_\oplus}$. Note the broken axes in the first two rows. In the first two cases,
      no water is lost in more than half of the runs; in such cases, a thin hydrogen envelope remains today. In the
      final case, most planets lost all their hydrogen and all their water. In order to prevent the runaway loss
      of its water, Proxima Cen b must have formed with more than 0.01\% of its mass in the form of a hydrogen envelope.
      }
    \label{fig:planet_hydrogen}
 \end{center}
\end{figure*}

It is interesting to note that in Figures~\ref{fig:planet_epsilon} and \ref{fig:planet_epsilon10} the posterior distributions for $\epsilon_\mathrm{xuv} = 0.15$ and $\epsilon_\mathrm{xuv} = 0.05$ are qualitatively similar. The median amount of water remaining in Figure~\ref{fig:planet_epsilon} is $\sim$ 0.1 and 0.3 TO for $\epsilon_\mathrm{xuv} = 0.15$ and 0.05, respectively, while the median oxygen pressure is $\sim$ 600 bar in both cases. In Figure~\ref{fig:planet_epsilon10}, the median water remaining is $\sim$ 3 and 5 TO for $\epsilon_\mathrm{xuv} = 0.15$ and 0.05 and the median oxygen pressure is again the same in both cases at $\sim$ 700 bar. This is because the total amount of water lost does \emph{not} scale linearly with $\epsilon_\mathrm{xuv}$, as the hydrodynamic drag on oxygen atoms acts as a negative feedback that stabilizes the net water loss rate. At high efficiency, the drag on the oxygen atoms is stronger and more of the energy goes into driving oxygen escape, which is \emph{less} efficient at depleting the water (since oxygen atoms are heavier). At low efficiency, most of the energy goes into driving hydrogen escape, which is \emph{more} efficient at depleting the water. As a result, our water loss amounts change little when $\epsilon_\mathrm{xuv}$ is varied by a factor as large as $\sim$ 3.

In Figure~\ref{fig:planet_hydrogen} we explore the effect of varying the initial hydrogen content of the planet.
From top to bottom, the rows correspond to initial hydrogen masses equal to 0.01, 0.001, and $10^{-4}$~$\mathrm{M_\oplus}$.
In the first two cases, the effect of the envelope is clear, as most planets lose no water and build up no oxygen.
These are mostly cases in which a portion of the hydrogen envelope remains at the present day. However, if the
initial hydrogen mass is on the order of $10^{-4}$~$\mathrm{M_\oplus}$ (corresponding to roughly 100 times Earth's
total atmospheric mass), the shielding effect of the envelope is
almost negligible; compare panel \textbf{(c)} to the top panel in Figure~\ref{fig:planet_epsilon} (the default run).
In this case, most of the water is lost to space in the majority of the runs.

\begin{figure*}[h!]
\begin{center}
    \includegraphics[width=\textwidth]{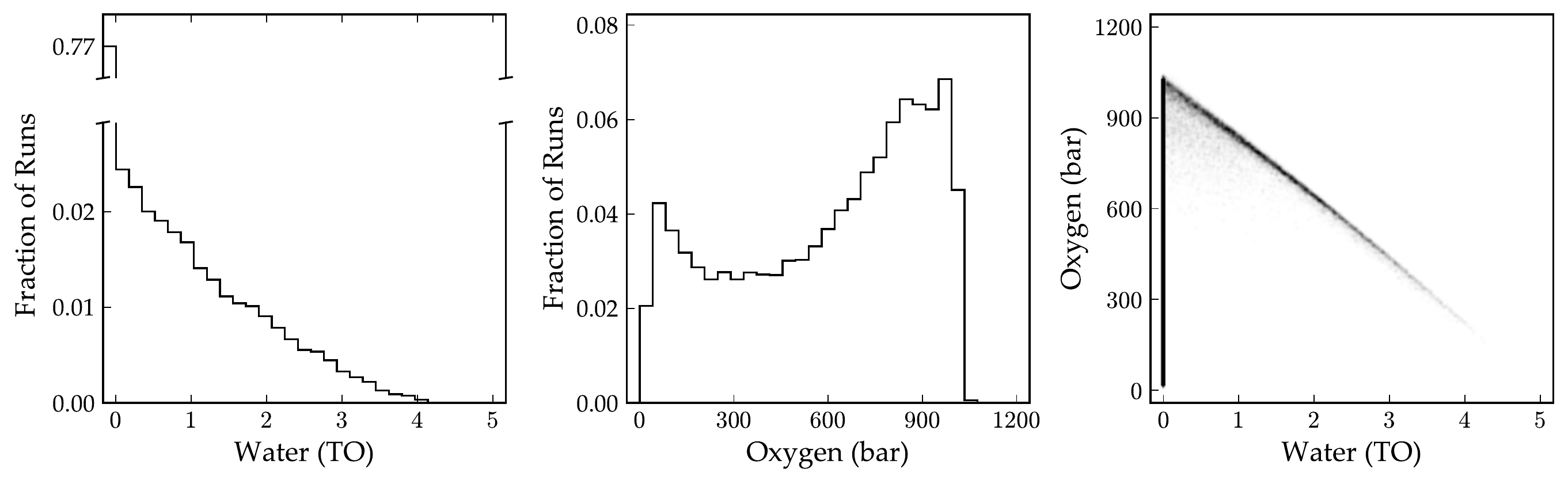}
     \caption{The same as panel (a) in Figure~\ref{fig:planet_epsilon}, but for efficient surface sinks
     ($\zeta_\mathrm{O_2} = 1$). The O$_2$ posterior now corresponds to the amount of oxygen (in bars) absorbed
     at the planet surface. The absence of atmospheric O$_2$ facilitates the loss of hydrogen, which must
     no longer diffuse through the O$_2$ to escape. In this case, nearly 80\% of runs result in complete desiccation
     (note the broken axis in the first panel). In all cases, Proxima b loses at least 1 TO.}
   \label{fig:planet_magma}
\end{center}
\end{figure*}

In Figure~\ref{fig:planet_magma} we show the posteriors assuming the O$_2$ uptake
efficiency of the surface $\zeta_\mathrm{O_2} = 1$, corresponding to instant O$_2$ removal by the surface. Compare
to the top panel of Figure~\ref{fig:planet_epsilon}. In this case, the O$_2$ posterior corresponds to the total amount
of oxygen absorbed by the surface, expressed in bars. While the total amount of oxygen retained by the planet
is similar, the fraction of runs in which the planet loses all of its water increases from $\sim 0.2\%$ to $\sim 0.8\%$.
This increase occurs because the buildup of atmospheric O$_2$ throttles the escape of hydrogen by decreasing its mixing
ratio in the upper atmosphere; when O$_2$ is quickly absorbed at the surface, hydrogen can escape more easily.

\begin{figure*}[h!]
 \begin{center}
     \includegraphics[width=\textwidth]{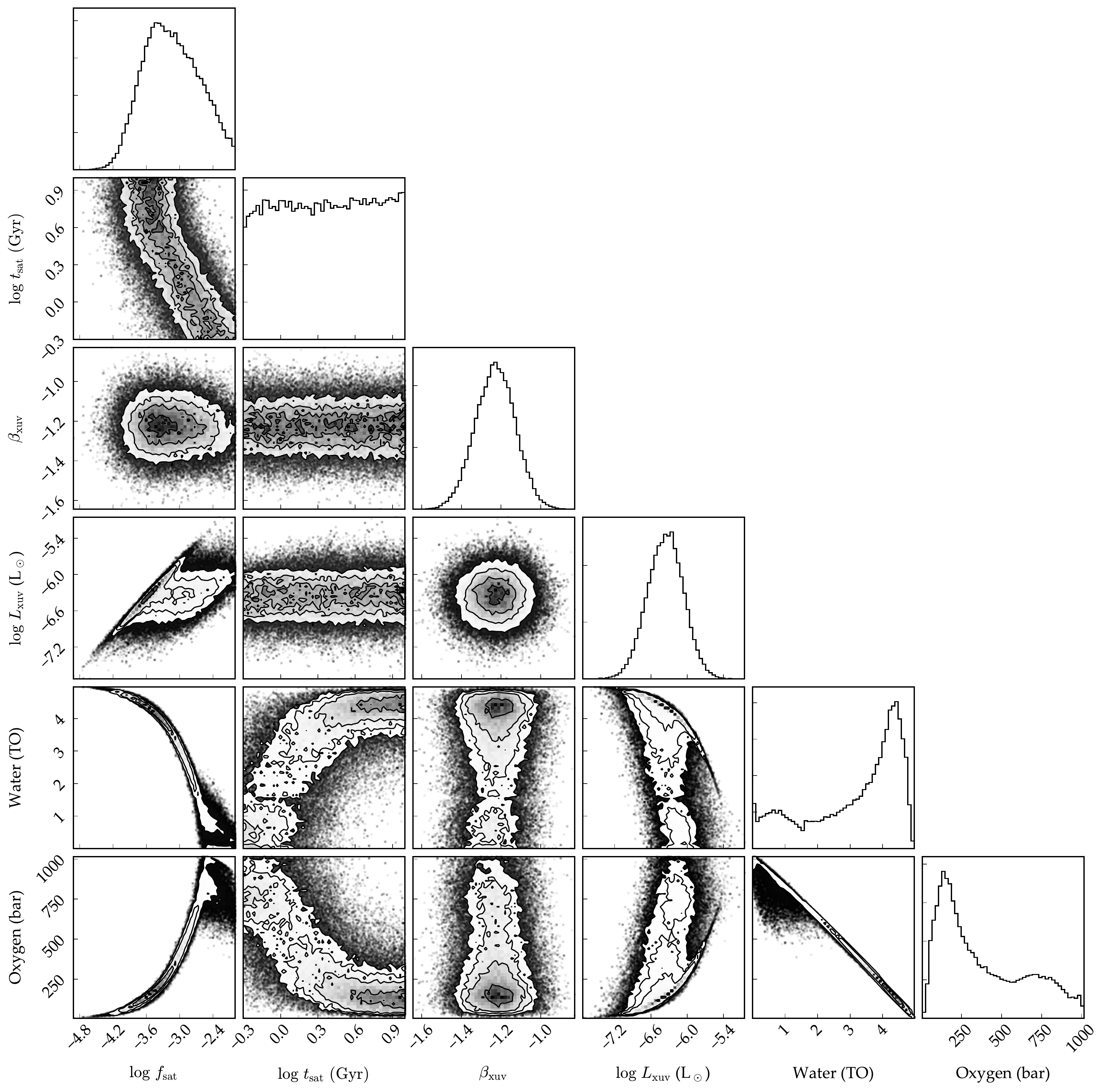}
      \caption{Joint posteriors of selected parameters for a run with $\epsilon_\mathrm{xuv} = 0.01$ [same as
       Figure~\ref{fig:planet_epsilon}(c)]. In addition to the correlation between the amount of water
       lost and the amount of O$_2$ that builds up, several strong correlations stand out.
       The strongest ones are between the XUV saturation fraction $f_\mathrm{sat}$ and
       the water content (negative) and O$_2$ pressure (positive). Since most of the water loss occurs
       in the first few 100 Myr, the value of $f_\mathrm{sat}$ is the single most important parameter
       controlling the present-day water and O$_2$ content of Proxima Cen b. The saturation timescale
       also correlates with the water and oxygen, but not as strongly; for $t_\mathrm{sat} \gtrsim 2$ Gyr,
       its exact value does not significantly affect the evolution of the planet. Shorter saturation
       timescales correlate with higher saturation fractions and therefore indirectly affect the evolution.
       Interestingly, the correlation between the XUV power law slope $\beta_\mathrm{xuv}$ and the water or
       O$_2$ content is negligible, since once saturation ends the water loss rate plummets --- the final
       water content depends almost entirely on the properties of the star early on.}
    \label{fig:corner}
 \end{center}
\end{figure*}

Finally, it is interesting to explore the various correlations between the parameters of the model. It is
clear from the previous figures that the amount of oxygen that builds up strongly correlates with the
amount of water lost from the planet, but additional correlations also exist. In Figure~\ref{fig:corner} we plot
the joint posteriors for the XUV saturation fraction, XUV saturation timescale, XUV power law exponent,
present-day XUV luminosity, present-day water content, and present-day O$_2$ content for a run with
$m_\mathrm{H_2O}^0 = 5$ TO, $m_\mathrm{H}^0 = 0$, $\epsilon_\mathrm{xuv} = 0.01$, and $\zeta_\mathrm{O_2} = 0$.
The marginalized posteriors are shown at the top. The strongest correlations are between the final
water and O$_2$ contents and the XUV saturation fraction (first column, bottom two panels). The higher
the XUV saturation fraction, the more water is lost and the more O$_2$ builds up. While this result
may be unsurprising, neither the saturation timescale (second column) nor the power law exponent
(third column) correlate as strongly
with the water and O$_2$ content. For saturation timescales longer than about 2 Gyr, the exact duration
of the saturation phase does not affect the evolution of the planet, since nearly all of the water loss
occurs in the first few 100 Myr. For the same reason, the value of the power law exponent does not
significantly correlate with the water or oxygen. On the other hand, the present-day XUV luminosity does correlate
with water loss, as it implies a higher XUV luminosity at early times.
An accurate determination of $f_\mathrm{sat}$ and more precise measurements of $L_\mathrm{xuv}$ are therefore
critical to determining the evolution of the water content of Proxima b.

\subsubsection{Summary of Atmospheric Evolution Simulations}
\label{sec:results:atmesc:summary}

The posterior distributions of Proxima b's present-day water and oxygen content support our findings in \S\ref{sec:results:atmesc:fiducial} based on the fiducial model parameters. Given our assumptions and evolutionary model, the planet spent $t_{RG} = 169 \pm 13$ Myr in a runaway greenhouse. During this time, we find that it may have lost on the order of 5 TO of water and built up more than 500 bars of O2 on average for a wide range of model assumptions. However, the posterior distributions for the water content (Figures~\ref{fig:planet_epsilon}--\ref{fig:planet_magma}, left columns) are distinctly non-Gaussian, with substantial peaks at or near a present-day water content of 0 for initial water contents up to 10 TO.

There are three broad scenarios in which Proxima b does not lose most of its initial water content. The first scenario is for initial water inventories of 10 TO (or higher). In these cases Proxima b may lose up to half its water and build up on the order of 500 bars of $O_2$, but may be habitable today.

The second scenario is for an XUV escape efficiency $\epsilon_\mathrm{xuv} \sim 0.01$, in which case only about 1 TO is lost on average. Such a low escape efficiency may have been possible during the first $\sim$ 10 Myr, when the XUV flux at Proxima b likely exceeded $10^{5}\ \mathrm{erg/cm^2/s}$ and recombination radiation contributed significantly to cooling the flow \citep{MurrayClay09}. In this case, hydrodynamic models predict efficiencies as low as 0.02--0.03 \citep{Bolmont16}. For reference, for Proxima b's present-day XUV flux, the model of \cite{Bolmont16} predicts $\epsilon_\mathrm{xuv} \approx 0.1$.

The final scenario in which Proxima b does not lose most of its water is if it formed with a substantial hydrogen envelope. For an initial envelope mass fraction $\gtrsim 10^{-4}$ (corresponding to a mass greater than 100 times Earth's total atmospheric mass), the envelope takes several hundred Myr to fully escape, shielding the surface water during the star's most active phase. This corresponds to the ``habitable evaporated core'' scenario of \cite{Luger15}. However, this scenario requires a certain amount of fine tuning, see $\S$~4.6. If Proxima b formed with more than 0.01 of its mass in hydrogen and/or if it is significantly more massive than $1.27 \mearth$, the envelope may not have completely escaped and the planet may not be habitable today.

\begin{sidewaysfigure}[p]
\centering
\subfigure{\label{fig:atmesc:synth:a}\imagebox{4in}{\includegraphics[width=5.5in]{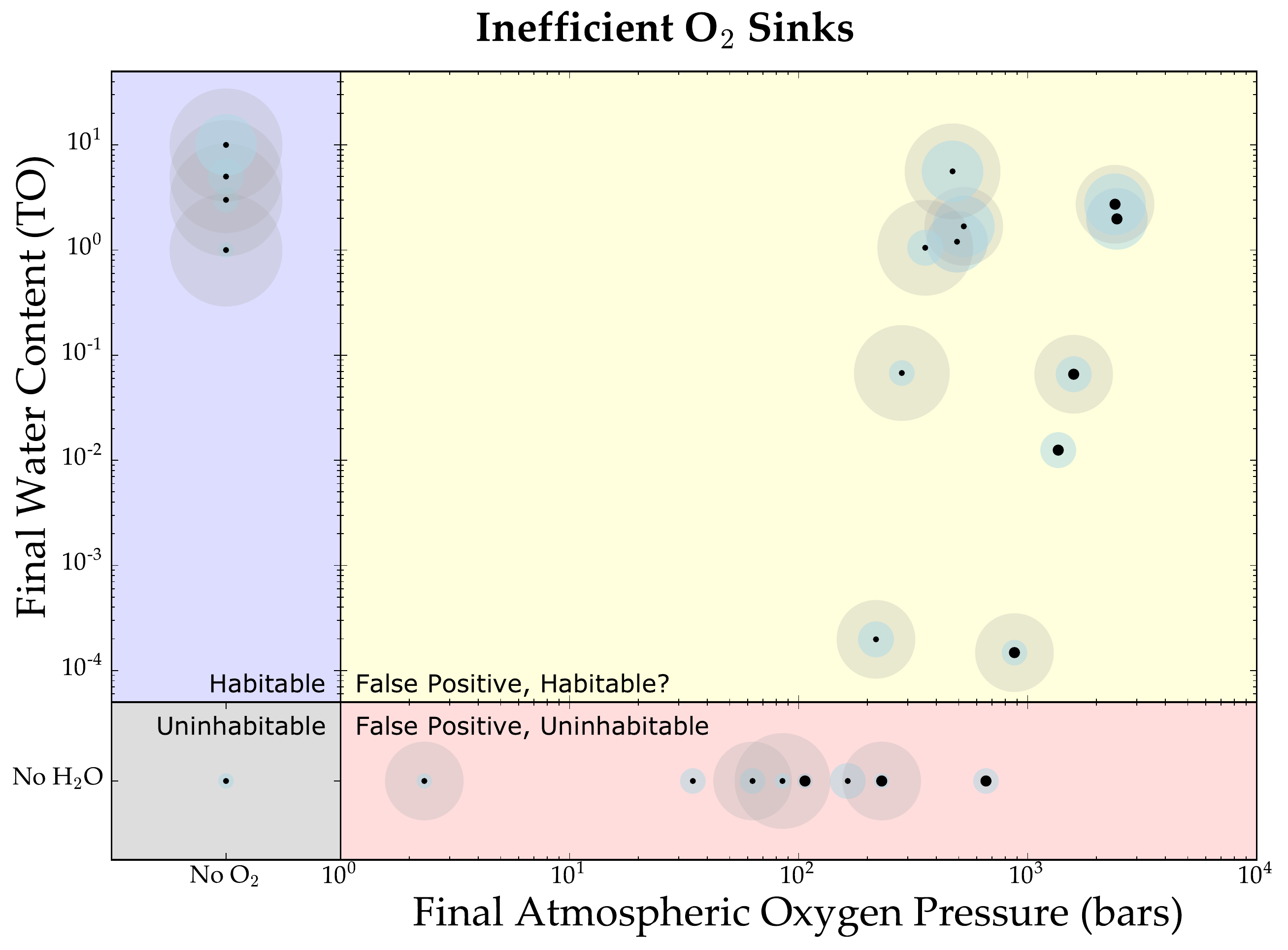}}}
\subfigure{\label{fig:atmesc:synth:c}\imagebox{4in}{\includegraphics[width=2.14in]{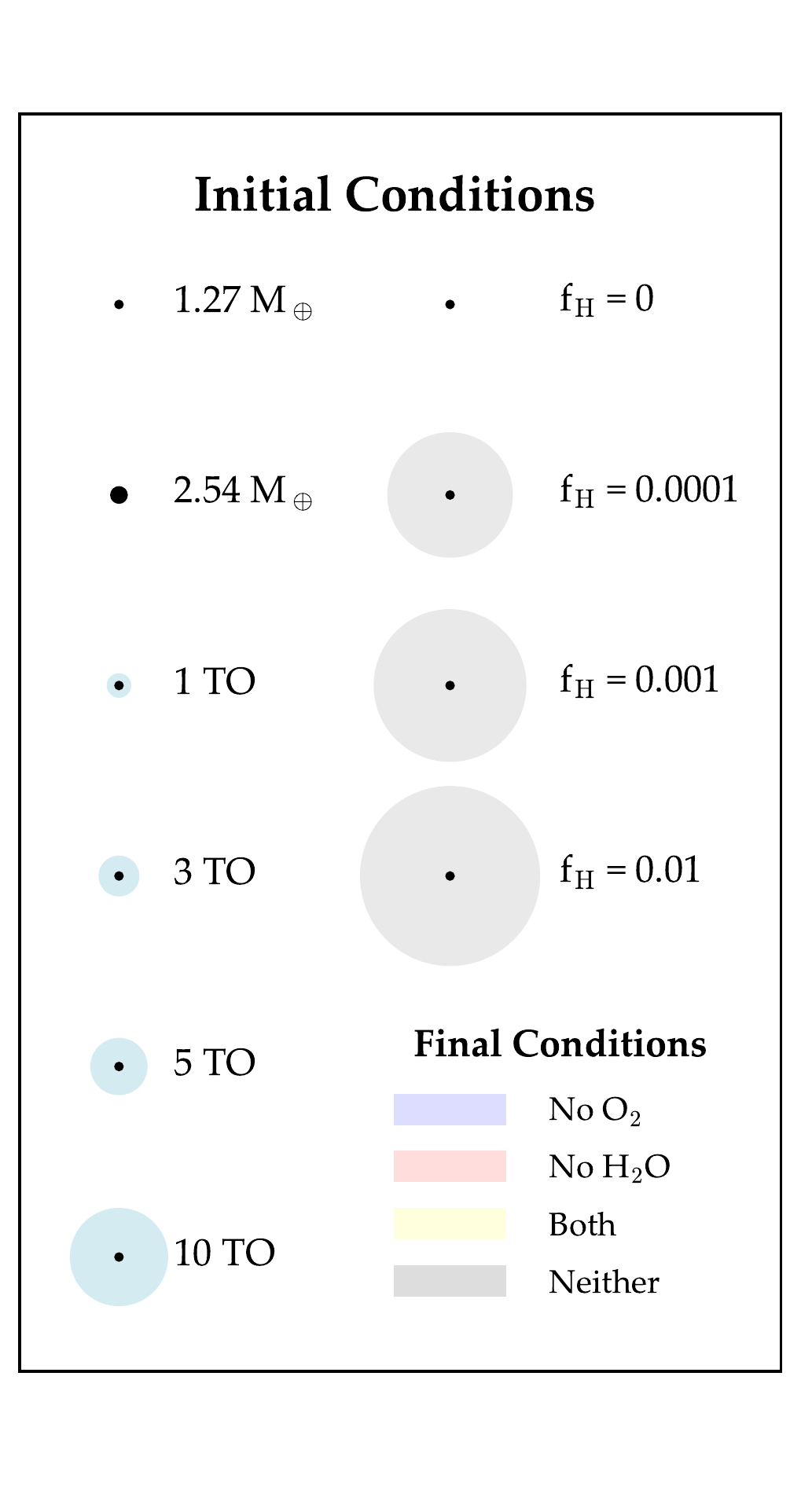}}}
\caption{Multiple evolutionary pathways for the water on Proxima
b. These plots show the final water and final atmospheric O$_2$ content of the planet for a suite of different initial conditions, assuming inefficient surface sinks for O$_2$. Different marker styles indicate different values of the planet mass, the initial water content, and the initial hydrogen envelope mass fraction $f_H$ (the final value of $f_H$ is zero for all planets shown here).  Each panel is divided into quadrants, corresponding to planets that at the end of the simulation have water but no O$_2$ (top left, blue), water and O$_2$ (top right, yellow), neither water nor O$_2$ (bottom left, gray), and O$_2$ but no water (bottom right, red). Habitable planets are those in the region shaded blue. Planets in the grey region are desiccated and therefore uninhabitable. Planets in the red region are likewise uninhabitable, but may have atmospheric O$_2$, which could be incorrectly attributed to biology. Finally, planets in the yellow region are habitable, since they have abundant surface water, but may also have substantial atmospheric O$_2$, which could be an impediment to the origin of life. These planets are also particularly problematic in the context of atmospheric characterization, as the presence of water and O$_2$ could fool observers into believing they are inhabited. \vspace{10cm}
}
\label{fig:atmesc:synthA}
\end{sidewaysfigure}

\begin{sidewaysfigure}[p]
\centering
\textcolor{white}{\rule{0.5\textheight}{0.4\textheight}}
\subfigure{\label{fig:atmesc:synth:b}\imagebox{4in}{\includegraphics[width=5.5in]{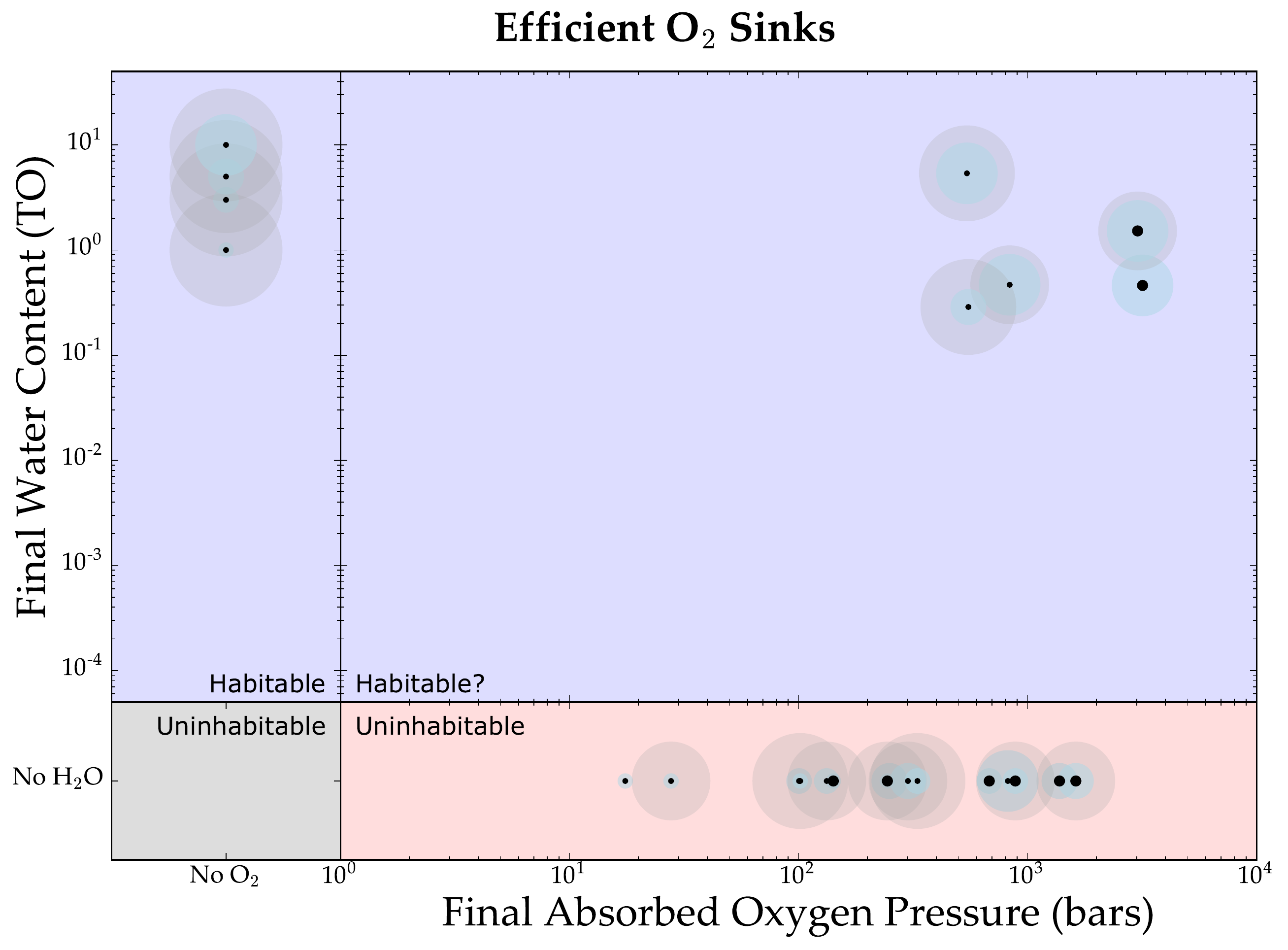}}}
\subfigure{\label{fig:atmesc:synth:d}\imagebox{4in}{\includegraphics[width=2.14in]{Figures/WaterLossSynth/synth_c.pdf}}}
\caption{Same as Fig.~\ref{fig:atmesc:synthB}, but assuming Proxima b has efficient O$_2$ sinks, preventing the buildup of significant oxygen in the atmosphere. The $x$ axis now shows the total amount of oxygen absorbed at the surface.}
\label{fig:atmesc:synthB}
\end{sidewaysfigure}

For reference, Figures~\ref{fig:atmesc:synthA} and \ref{fig:atmesc:synthB} show
a synthesis of runs of our model assuming the fiducial stellar parameters and
varying several of the planet properties. The two figures help identify cases
in which Proxima Cen b may presently be habitable.


\subsection{Internal Evolution}
\label{sec:results:internal}

\subsubsection{Role of Radiogenic Abundances}

Modeling the internal evolution of Proxima b is challenging due to
 the large number of unknowns about its composition,
structure, thermal state,
 atmosphere, and the evolution of its radiation environment.
In this section we first consider how
radiogenic abundances could affect its
evolution, followed by considerations of tidal heating.

As described in $\S$~\ref{sec:models:radheat} we consider four
possible abundance patterns for Proxima b: Earth-like, chondritic, 1
ppt $^{26}$Al, and inert (no radioactivity). In all cases we begin with a core
temperature of 6000~K and a mantle temperature of 3000~K.

In Fig.~\ref{fig:notides} we show the evolution of the radiogenic
power, mantle temperature, inner core radius, magnetic moment,
magnetopause radius and surface energy flux for the four cases. The
dashed black lines represent the modern Earth's value. In the top left panel we show the evolution of the total radiogenic
power produced in the core, mantle and surface. Initially, the
power from $^{26}$Al is over $2 \times 10^{18}$~W, but with a
half-life of 700,000 years, its contribution to the energy budget
drops to 0 within $10^7$ years. The Earth-like case is hidden behind
the $^{26}$Al curve except at $t=0$.

The mantle temperature is shown in the top right
panel.   As expected the model predicts a rapid cooling in response to the decay of $^{26}$Al, so that after 100 Myr the mantle temperatures are similar and the planet settles into an Earth-like evolution. Thus, if heating
from $^{26}$Al is just a passing energy source, it may not affect the
evolution. At this time, we merely point out that the
influence of $^{26}$Al could be significant for planets with formation times
of order 1~Myr, which is similar to $^{26}$Al's half-life. The inert case temperature drops quickly with no
radiogenic power in the interior, while the chondritic case shows increased
temperature for the entirety of the simulation for the mantle heat flow to accommodate the high heat source.

The cases with earliest inner core solidication (Figure \ref{fig:notides}, middle left panel) are the inert and chondritic
ones, which have the most and least total radiogenic power, respectively.
In the inert case the interior loses secular heat faster with no radiogenic heat source.  Counterintuitively the chrondritic case also cools faster than the nominal model, likely because the core cooling rate is roughly a constant fraction of the total surface heat flow, which is much higher in this case.  We note that changing the core composition could have a major impact on the inner core solidification rate and should be the subject of future work.

The middle right panel of Figure \ref{fig:notides} shows the evolution of the planet's magnetic moment for the
four different cases. Each case shows similar behavior --- a gradually
decaying field as the core cools, with a cusp when the inner core first nucleates.
The $^{26}$Al case shows the largest field due to an early super-heating of the core, despite being
otherwise very similar to the Earth case.

Finally, the bottom right panel shows the surface energy flux for each
case. Not surprisingly the chondritic case maintains the highest heat
fluxes, near 1 W/m$^2$, which is similar to Io's value of 2.5
W/m$^2$~\citep{Veeder94}.

\subsubsection{Evolution with Tidal Heating}
\label{sec:results:internal:tides}

\begin{figure*}[ht]
\begin{center}
\includegraphics[width=\textwidth]{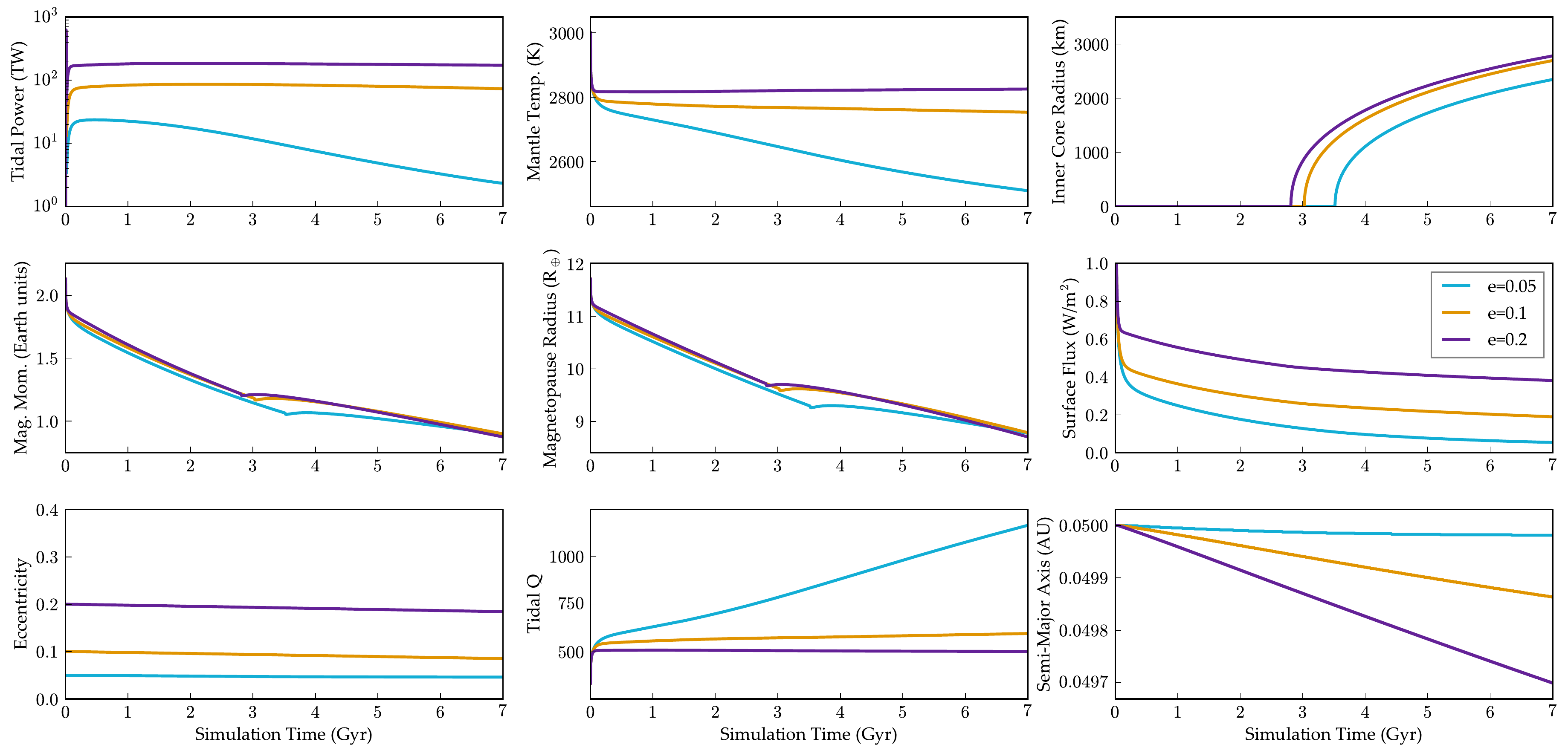}
\end{center}
\caption{Evolution of internal properties of planet b for three
  different initial eccentricities, as shown in the legend, and
  assuming the Earth-like levels of radiogenic isotopes. {\it Top
    left:} Power generated by tidal heating. {\it Top middle:} Mantle
  temperature. {\it Top right:} Radius of the inner core. {\it Middle
    left:} Magnetic moment. {\it Middle:} Magnetopause radius. {\it
    Middle right:} Surface energy flux. {\it Bottom left:} Orbital
  eccentricity. {\it Bottom middle:} Tidal $Q$. {\it Bottom right:}
  Semi-major axis.}
\label{fig:tides}
\end{figure*}

Next, we examine the role of tidal heating on the evolution of the planet's interior
and orbit. For simplicity we consider the ``Earth'' radiogenic case with
three initial eccentricities: 0.05, 0.1 and 0.2. In
Fig.~\ref{fig:tides} we show the evolution of 9 quantities as a
function of time.

The tidal power in these cases can be in excess of 100 TW, or twice the total
power of the modern Earth. Similar to \cite{DriscollBarnes15}, we find
that the planet's tidal
properties evolve in response to the thermal state of the
interior, avoiding unrealistically large tidal powers predicted by
simpler equilibrium tide models;
 see above. Tidal heating
increases mantle temperature,
lowering mantle viscosity,
which raises the tidal $Q$ after an initial peak during mantle solidification. A
 ``tidal steady state'', as defined in
\cite{DriscollBarnes15}, is possible
where the surface heat flow balances tidal dissipation in the mantle so that the
planet cools very slowly.  This quasi-steady state relies on the negative feedback between mantle
temperature and tidal dissipation and the positive feedback between temperature and heat flow,
so that a decrease in temperature causes increased tidal heating, pushing the temperature back up.

\begin{figure*}[h!]
\begin{center}
\includegraphics[width=\textwidth]{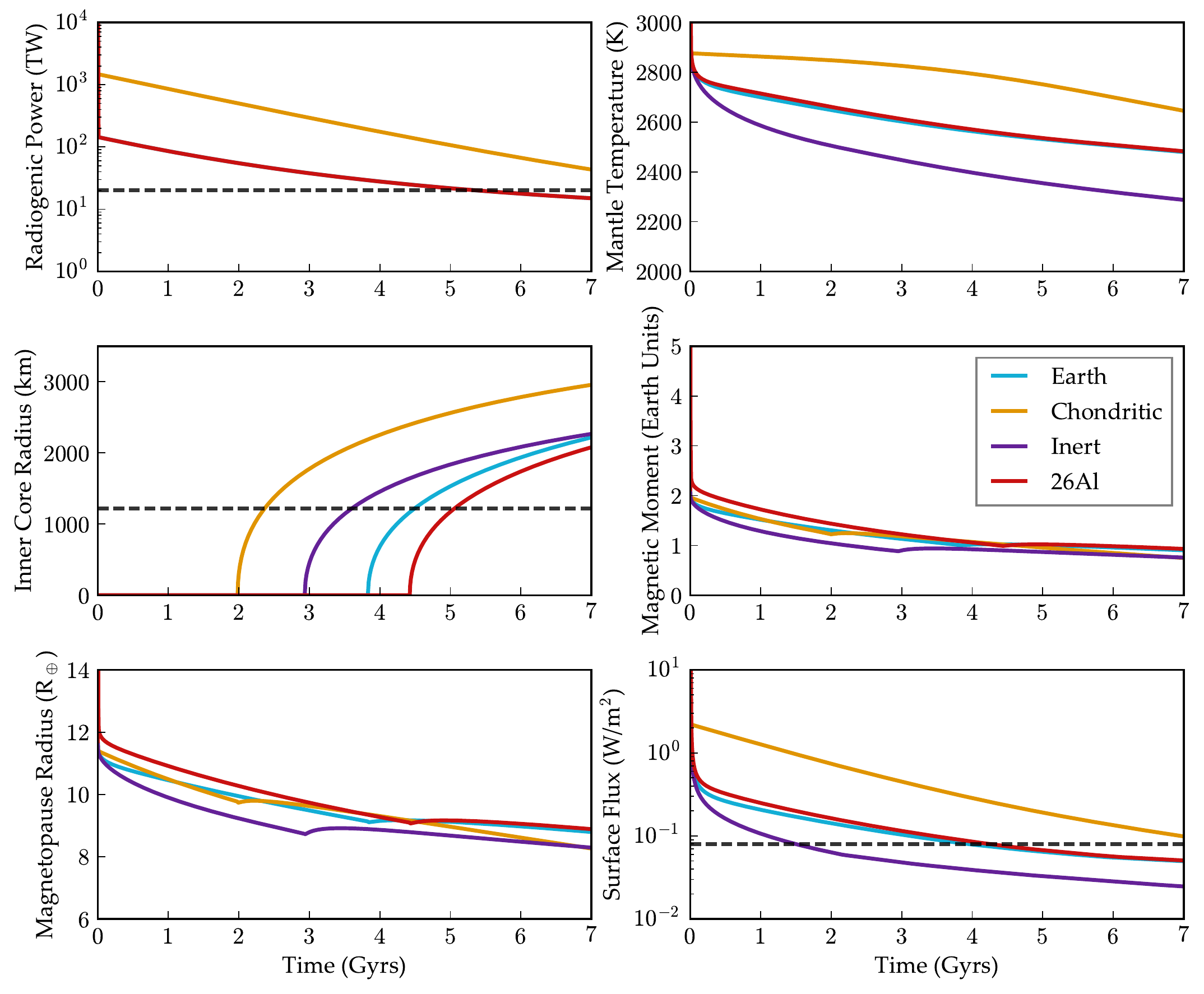}
\end{center}
\caption{Evolution of internal properties of planet b for different
  assumptions of radiogenic inventory: Earth-like in blue, chondritic
  in orange, 1 part per trillion $^{26}$Al in red, and inert in
  purple.
  Values for the modern Earth are shown with the dashed black
  line. {\it Top left:} Radiogenic power. The Earth curve is behind
  the $^{26}$Al curve except for time = 0. {\it Top right:} Mantle
  Temperature.
  {\it Middle left}: Size of the solid
  inner core. {\it Middle right}: Magnetic moment.
  {\it Bottom left:} Magnetopause radius assuming the solar wind pressure at Proxima b is
  0.2 times that at Earth. {\it Bottom right:} Surface energy
  flux.}
\label{fig:notides}
\end{figure*}

The mantle cooling rate is more sensitive to tidal dissipation than the core because dissipation
in the model occurs only in the mantle.  The core cooling rate does change somewhat with tidal
dissipation, but its effect on the magnetic moment is muted because the magnetic moment depends
on core convective heat flux to the $1/3$ power.
 Note, however, that
the inner core grows earlier and more rapidly than in the
Earth case of Fig.~\ref{fig:notides} because a hotter mantle is less viscous, thinning the thermal boundary layer above the core-mantle boundary layer, allowing the core to cool slightly faster \citep{DriscollBarnes15}.
Although none of our cases achieves a fully
solid core, which would quench the dynamo, they are approaching that limit, and given
the uncertainty in both the composition and structure of Proxima b,
it is possible that the
core has already solidified, preventing a core dynamo, and exposing the atmosphere to
stellar flares.

\clearpage

\clearpage

\subsection{Habitable Evaporated Core Scenarios}
\label{sec:results:tidal_hec}

Since a possible path towards habitability for Proxima b is the
``habitable evaporated core'' scenario of \citet{Luger15}, we seek to
model how the presence of a hydrogen envelope and surface
oceans undergoing atmospheric escape impact the tidal and orbital evolution of Proxima b.
To do so, we couple the atmospheric escape physics of \atmesc, tidal evolution
using \eqtide, the Earth-calibrated geophysical interior models of
\radheat and \thermint and the stellar evolution of \stellar.

We model the combined tidal contributions of the envelope, oceans, and
solid interior via the following equation
\begin{equation}
\label{eqn:Q_hec}
-Im(k_2) = -Im(k_2)_{interior} + \frac{ k_{2_{ocean}}}{Q_{ocean}} +  
\frac{ k_{2_{envelope}}}{Q_{envelope}},
\end{equation}
where $Im(k_2)$ is the imaginary part of the Love number, $k_2$ is the Love number of order 2, and $Q$ is the tidal quality factor
\citep[see][]{Barnes13,DriscollBarnes15}. In Eq.~(\ref{eqn:Q_hec}) we remove terms if there is no
component to contribute to Proxima b's net tidal interaction, \ie no ocean or no envelope.

In the general case when a hydrogen envelope is present, we only
consider the coupled tidal effects of the interior and the envelope as any
water is likely to be supercritical due to the high pressure exerted
by the envelope.  In our model, we account for this case by neglecting the ocean term
from Eq.~(\ref{eqn:Q_hec}) when the mass of the hydrogen envelope is non-zero.  When an envelope is not present,
we consider the tidal contribution of surface oceans only if the planet is
not in the runaway greenhouse limit since otherwise all water would be present
in the atmosphere as steam.  To determine if the planet is in the runaway greenhouse limit, we check to see if the flux
the planet receives is greater than or equal to the mass-dependent runaway greenhouse limit of \citet{Kopparapu14}
appropriate for Proxima Centauri.  In our model when the planet is in the runaway greenhouse limit, we
neglect the ocean term from Eq.~(\ref{eqn:Q_hec}) and the planet's $Im(k_2)$
is set by the solid interior term in Eq.~(\ref{eqn:Q_hec}).

\begin{figure*}[h!]
\centering
\includegraphics[width=\textwidth]{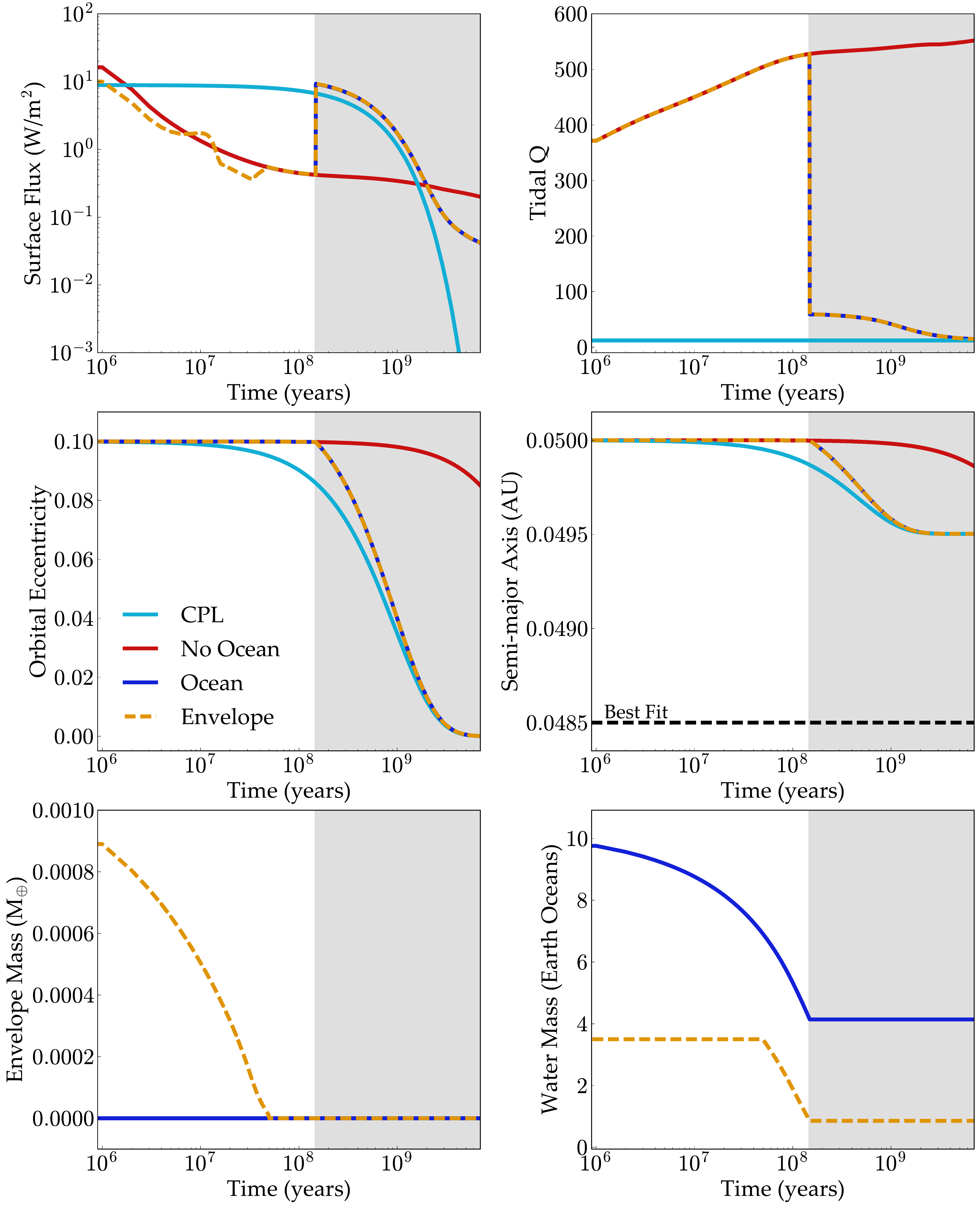}
\caption{Evolution of the orbital, tidal and atmospheric properties of
  Proxima b for the ``CPL" case in light blue, ``No Ocean" case in
  red, ``Ocean" case in dark blue, and the ``Envelope" case in orange
  with the dashed line for clarity.  The grey shaded region indicates
  when the planet is in the HZ. {\it Top left:} Surface Flux. {\it Top
    right:} Tidal Q. {\it Middle left:} Orbital Eccentricity. {\it
    Middle right:} Semi-major Axis. {\it Bottom left:} Envelope
  Mass. {\it Bottom right:} Surface Water Mass.}
\label{fig:tidal_hec}
\end{figure*}

We simulate four cases that bracket the potential
past tidal evolution of Proxima b.  The first case, ``CPL,'' is a simple application
of the constant phase lag tidal model that probes how Proxima b's orbit would evolve due to tides
if it was an Earth-like planet with persistent surface oceans.  This case assumes
constant tidal $Q = 12$, consistent with observations of
Earth today \citep{Dickey94,Williams78,Yoder95}.  This low tidal $Q$
is due to efficient energy dissipation by oceans and leads to rapid tidal evolution, while still acknowledging that the tidal $Q$ of the surface oceans may depend on the
total ocean mass and the presence and topology of shallow seas, which should be explored in
future studies. The ``No Ocean'' case assumes a dry, rocky planet in which
the tidal interaction is dominated by the planet's interior as determined by \radheat and \thermint.
The ``Ocean" case couples the rocky planet from the ``No Ocean'' case with an ocean containing an
initial water inventory of 10 TO with $Q_{ocean} = 12$ \citep{Dickey94,Williams78,Yoder95}.
We chose an initial water inventory of 10 TO
to ensure that water loss due to atmospheric escape would not desiccate the planet such that the liquid
surface oceans would exist after the runaway greenhouse phase allowing both the oceans and interior to impact the planet's
tidal evolution. Finally, the ``Envelope'' case examines the tidal history of the
``habitable evaporated core" scenario first explored by \citet{Luger15}, and described in detail for Proxima b above.  For this simulation, we
couple the interior evolution from the ``No Ocean" case, liquid surface oceans, and a hydrogen envelope that has an
initial mass fraction of $0.001$ of the planet's total mass with $Q_{envelope} = 10^4$.
This choice of the tidal Q for the gaseous envelope is consistent with measurements of Neptune's
tidal Q \citep{ZhangHamilton08}.  The ``Envelope'' case starts with 3.5 TO of liquid
surface oceans to demonstrate the envelope's ability to shield surface water from
atmospheric escape (see $\S$~\ref{sec:results:atmesc}).  We set $k_{2_{ocean}} = 0.3$ and
$k_{2_{envelope}} = 0.01$ to let the thermal
interior and oceans determine $k_2$ as these components likely dominate the tidal dissipation in the
planet.  All four simulations include the full stellar evolution of the star
as determined by the \stellar module.  Simulations which include liquid surface oceans and/or a gaseous
envelope use the atmospheric escape module \atmesc.  Atmospheric escape calculations assume
$\epsilon_\mathrm{xuv} = 0.15$ as per the fiducial case of $\S$~\ref{sec:results:atmesc}.
The results of the simulations are shown in Figure \ref{fig:tidal_hec}.

The ``No Ocean'' case, with tidal dissipation occurying primarily in the mantle, reaches a tidal $Q$ of ${\sim}500$
and undergoes minimal tidally-driven orbital evolution until after $\sim$1 Gyr. Initially, the bulk of the surface flux stems from
rotational tidal energy dissipation which lessens as the planet
approaches a tidally locked state at around 15 Myr. Early on in the
``Ocean'' case, the planet is in a runaway greenhouse phase in which all
the water is locked up in the atmosphere and subject to escape of
hydrogen and oxygen via photolysis, decreasing the water mass.  Its
tidal evolution mirrors the ``No Ocean'' case as the tidal dissipation takes place in the mantle.
After the stellar luminosity decreases to its main sequence value,
the planet enters the HZ at $\sim$150 Myr and the remaining water condenses to the
surface. The presence of surface oceans after the runaway greenhouse phase for the
``Ocean" and ``Envelope" cases dramatically decreases the
tidal $Q$, leading to rapid orbital circularization and a substantial
surface flux increase via tidal energy dissipation.

Early on, the ``Envelope'' case has a slightly different surface flux and tidal $Q$
than the ``Ocean" and ``No Ocean" cases as the envelope contributes
minimally to the initial tidal evolution and as the planet's radius
evolves as the envelope experiences atmospheric escape.  In the ``Envelope" case, stellar XUV flux completely strips the
hydrogen envelope after about $5 \times 10^7$ years, causing the mantle
to again dictate the tidal interaction.  The hydrogen envelope
shields the planet's water before it is completely stripped away, allowing enough to survive
subsequent photolysis. About 1 TO of the initial 3.5 TO
remain once the planet enters the HZ, assuming the planet has
inefficient oxygen sinks (see $\S$~\ref{sec:results:atmesc}).  With the envelope gone,
surface water dominates the tidal interaction and the planet's evolution mirrors the ``Ocean'' case.

The coupling of tidal evolution, interior geophysics, surface oceans, atmospheric escape,
and stellar evolution permit a complicated evolutionary history for Proxima b which depend
immensely on its properties at formation.  Future planet formation studies that examine initial volatiles, analogous to \citet{Raymond04,Raymond07},  are critical to constrain potential evolutionary histories for this system.

\section{Discussion}\label{sec:disc}

In the previous section we outlined numerous influences on the
history of Proxima b. We considered processes spanning the planet's
core to the Milky Way galaxy and find that effects at all these scales
could be important in the history of our closest exoplanet. In this
section, we summarize the results in terms of the potential
atmospheres that Proxima b might have, which are considered in
detail in \cite{Meadows18}. Then we examine the likelihood that it is
currently habitable.

\subsection{Atmospheric States}
\label{sec:results:atmstates}

\subsubsection{Earth-Like}
\label{sec:results:atmstates:earthlike}

We find that some scenarios produce a planet today that has surface water and rotates super-synchronously, i.e. is ``Earth-like.'' In particular, the ``habitable evaporated core'' scenario
\citep{Luger15} is promising as it can mitigate both the
high-luminosity pre-MS phase and any devastating tidal
heating that may occur during orbital circularization from a primordial non-circular orbit after orbital destabilization \citep{Barnes13,DriscollBarnes15}. Even if the planet became
desiccated during the pre-MS phase, impact from water-rich bodies
could simultaneously blow off the CO$_2$ and/or O$_2$ atmosphere while
delivering water. This scenario would require a specific set of events
to occur, but we note that close passages between Proxima and \acen~A
and B could destabilize any putative ``exo-Oort Clouds'' that could
have existed around the stars. The observed dust belt \citep{Anglada17} may be a remnant of such an encounter. Current numerical models do not permit
a robust calculation of the probability of such an event, but we cannot exclude it at this time.

Tidal damping of the spin is rapid, but there are at least four possibilities as to how the planet may currently rotate super-synchronously, all of which rely on Proxima b orbiting with non-zero eccentricity today. For $e>0$, torques on the planet tend to increase the rotation period into either a ``pseudo-equilibrium'' value \citep{Goldreich66,ZanazziLai17}, or into a spin-orbit resonance \citep{Rodriguez12,Ribas16}. The first possibility is that, if the system is on the younger side of the age distribution, and the initial eccentricity was very large, then
the eccentricity might not have damped away, see Fig.~\ref{fig:eqtide}. Second, the results of Section \ref{sec:results:galactic} demonstrate that a close encounter between Proxima and
\acen~A and B can destabilize an extended planetary system (Fig.
\ref{fig:planetpert}) and scatter Proxima b into a close, high eccentricity orbit. Since Proxima's orbit evolves stochastically, and we cannot constrain the timing of the galactic migration, this instability could have occurred in the recent past, leaving Proxima b on an eccentric orbit today. Third, the planetary system could have experienced a late stage instability, independent of \acen~A and B. The so-called Nice instability in our Solar System occurred $\sim 700$~Myr after it formed \citep{Gomes05}, and evidence is mounting that compact planetary systems are marginally stable and instabilities can occur Gyr after formation \citep{VolkGladman15}.  Fourth, perturbations from other planets may delay circularization sufficiently that the orbit of Proxima b is currently large enough for capture into a spin-orbit resonance. We conclude that it is plausible that the rotation period of Proxima b is non-synchronous.

The obliquity of Proxima b is likely small, even in a Cassini state
due to the influence of companion
planets, due to tidal damping. Thus, we do not expect the
obliquity to significantly impact the climate in the form of seasons or heat
transport. The approximately zero obliquity of Proxima b is one feature that will produce a different climate than Earth's.

\subsubsection{Habitable and Synchronously Rotating}

If the eccentricity is modest today, then Proxima b could be rotating synchronously. Although this rotation state has previously been considered dangerous for habitability \citep{Mumford09,Dole64,Kasting93}, these planets may still be habitable \citep{Joshi97,Pierrehumbert11,Yang13,Shields16}. If the planet is on a circular orbit and $\sin(\psi)=0$, then the substellar point is constant in latitude and longitude and every location on the planet is either in permanent starlight or darkness. In this case, the daylight side may experience near-constant cloud-cover as the high temperature at the sub-stellar point constantly evaporates water \citep{Yang13}. However, that result is for longer rotation periods that lead to weak a Coriolis force. Proxima b, with an orbital period of 11.2 days, may have a strong enough Coriolis force to shear the substellar clouds and weaken the cloud-albedo feedback. On the other hand, if the planet is covered in a global ocean and the greenhouse forcing is weak, then the planet may be in an ``eyeball state'' in which the planet is covered in ice except for a circular patch of open ocean at the substellar point \citep{Pierrehumbert11}.

If $e>0$, then Proxima b will librate once per orbit as the ratio between the (constant) rotational frequency and the (oscillating) orbital frequency changes from pericenter to apocenter and back \citep{Makarov16}. In this case, regions near the terminator can experience a day/night cycle, in which the star rises and sets from the same point in the horizon once per orbit. We are unaware of any research that explores the climate of a librating exoplanet, but our results suggest this state is the most likely for Proxima b and therefore such a simulation could be very enlightening. We hypothesize that for low eccentricity (low amplitude libration) the cloud feedback of \citet{Yang13} would still operate, but that for larger $e$, the apparent movement of the substellar point could significantly impact the feedback, possibly destroying it. For reference, for $e=0.05$, the librational amplitude is $\sim 5^\circ$.

\subsubsection{Super-Io}

Although Proxima b may have liquid water and be Earth-like, the planetary interior may be significantly hotter than Earth due to tidal heating or increased radiogenic abundances. Non-circular orbits, non-zero obliquities, and non-synchronous rotation all contribute to tidal heating, and hence even if the planet has water and super-synchronous rotation, it is probably not an ``Earth twin'' as far as the internal properties are concerned. \cite{Barnes09} dubbed hot and very volcanically active terrestrial planets ``super-Ios,'' see also \cite{Jackson08c}. If the planet formed in colder region, it may be enhanced in $^{40}$K relative to Earth and its interior may be very hot regardless of tidal heating. Although not modeled here, outgassing rates on Proxima b are probably higher than Earth's due to tidal heating and/or larger radiogenic heating, and potentially altering photochemistry and atmospheric structure. With no constraints on the composition of the solid planet, the thermal evolution and outgassing are poorly constrained, but the planet's interior may be so hot that the surface is similar to Io's. If the planet's composition is similar to Io and volcanoes are constantly erupting, sulfur species may be present and remotely detectable in the atmospheres \citep{Jackson08c,Misra15}.

\subsubsection{Habitable and Dry}

As shown in Fig.~\ref{fig:HZEvol}, a dry planet would enter a
potentially habitable state earlier than an Earth-like planet, and
hence it may be that Proxima b is a habitable world, but with little
liquid water \citep{Abe11}. This could occur if the planet formed {\it in situ} and
the gas disk was able to shield the planet from water loss. The
``habitable evaporated core'' scenario is also a possibility if the core
was dry and a thin layer of hydrogen could be blown away. On the other
hand, a planet without much water does not need to spend much time in
a runaway greenhouse to become desiccated. Based on the modeling in $\S$~\ref{sec:results}, we conclude that this possibility is very unlikely, but we cannot exclude it.

\subsubsection{Venus-Like}
\label{sec:results:atmstates:venuslike}

Regardless of whether Proxima b spent significant time in a runaway
greenhouse prior to the arrival of the HZ, it could be in a runaway
greenhouse state like Venus. If it formed {\it in situ}, then this
possibility is more likely because its first $\sim$170 Myr were spent
interior to the HZ and hence it may have developed a dense CO$_2$
atmosphere as has occurred on Venus. In the companion paper \citep{Meadows18}, we show that this case is indeed
uninhabitable as the surface temperature reaches 640~K and the planet is too hot for liquid water. Our companion paper, \citep{Meadows18}, shows that the higher surface temperature results from greenhouse warming by
CO$_2$ alone.

Even if the planet avoided desiccation during the pre-MS stage, it is
reasonable to assume that a Venus-like atmosphere is still possible. CO$_2$
is a very abundant molecule in planetary atmospheres, and given its
strong ability to heat the surface, high molecular weight, and strong
chemical bonds, it may be able to accumulate in the atmospheres of planets in the HZ to large enough
levels to trigger the runaway greenhouse.

A final possibility, mentioned above, is that past tidal heating drove
the planet into a runaway greenhouse \citep{Barnes13}. If the planet
was ever in a high eccentricity state ($e~>~0.35$) then the surface
energy flux from the interior could have reached the critical limit of
$\sim$300 W/m$^2$ \citep{Kasting93,Abe93,Goldblatt15}. Such high
surface fluxes may be short-lived if the heating can only come from
the ocean \citep{DriscollBarnes15}.

\subsubsection{Mini-Neptune}
\label{sec:results:atmstates:neptunelike}

Proxima b may have formed with sufficient hydrogen that some has been
retained despite all the high energy processes that can remove
it. This possibility is especially likely if it formed beyond the snow
line and migrated in. Similarly to \cite{OwenMohanty16}, we find that
if Proxima b formed with $\gtrsim$ 1\% of its mass in the form of a
hydrogen envelope (and/or is significantly more massive than 1.27$\mathrm{M_\oplus}$), it could still possess some hydrogen, in which
case the surface may be too hot and/or the surface pressure is too
high for habitability.  If Proxima b has a mass much larger than the minimum, its surface gravity could retain H in spite of high XUV fluxes \citep{Luger15,OwenMohanty16,LehmerCatling17}. Future measurements of Proxima b's radius can
inform its present-day composition and thus settle this issue. However, we also note that our model does not include several important processes, such as flaring and coronal mass ejections, that remove additional mass than our model. A proper assessment of the longevity of a hydrogen envelope requires a better understanding of Proxima's activity since its birth, as well as sophisticated space weather models \citep{Garraffo16,Airapetian17}.

\subsubsection{Abiotic Oxygen Atmosphere}
\label{sec:results:atmstates:o2atmos}

If Proxima b formed with one or more TO of water, photolysis followed
by hydrogen escape during the stellar pre-MS phase could have led to
the buildup of substantial O$_2$ in the atmosphere. Although oxygen is
highly reactive, thousands of bars of oxygen can be liberated through
this mechanism \citep{LugerBarnes15} and hence all sinks for it may
become saturated \citep{Schaefer16}. In principle, thousands of bars
of oxygen could remain in the atmosphere, but this figure is most
likely lower, as much of the oxygen will be consumed in the process of
oxidizing the surface. In this scenario, the planet may have an atmopshere dominated by oxygen molecules.

In many of the scenarios in which Proxima b develops an O$_2$-rich
atmosphere, it also retains at least some of its initial water.
After the end of the Pre-MS phase, the H$_2$O and CO$_2$
greenhouse warming could be sufficient to prevent water from
accumulating on the surface, and hence it could have significant
abundance in the stratosphere. The residence time of water is therefore of crucial importance, and it may be short due to flaring, coronal mass ejections, and other forms of stellar activity, especially while the system is young. Thus, even though the planet may have had H$_2$O after the HZ arrived, water may have remained in the stratosphere and been photolyzed by high energy radiation and the H atoms lost to space, ultimately producing a dry planet. We do not model this possibility here, but such a scenario was found to be plausible in \cite{Ribas16}.

\subsubsection{Water and Oxygen, but Uninhabitable?}

The joint oxygen/water posterior in Figure \ref{fig:corner} reveals an interesting possibility in which large amounts of oxygen are built up
by the pre-MS runaway, but not all the water is lost. In this case Proxima b may be uninhabitable,
given that little free energy may be available at the surface for
early organisms to exploit. Life on Earth is thought to have
emerged in an extremely reducing environment \citep{Oparin24,
  Haldane29}, with access to large energy gradients to fuel early
metabolisms; such a reducing environment may not have ever been present on Proxima b. While the simultaneous detection of
water and oxygen has traditionally been envisaged as an ideal
combination for life detection \citep{DesMarais2002,Meadows17}, in the case of
Proxima b, it is insufficient to prove that life is present. More information will be necessary to confirm the presence of life on Proxima b \citep{Meadows18}.

If the greenhouse gases are at low enough levels in the
atmosphere, then it may be possible for liquid water to accumulate on
the planetary surface and this planet would meet our
definition of ``habitable.'' However, as argued above, such a planet
would likely be incapable of abiogenesis.  Thus, the detection of large amounts of atmospheric oxygen, e.g. through $O_4$ bands, as well as the presence of surface liquid by other
means, \eg glint \citep{Robinson10}, would still not be sufficient evidence
that the planet is habitable \citep{Schwieterman16,Meadows17,Meadows18}.

\subsubsection{No Atmosphere and No Water}
\label{sec:results:atmstates:noatmos}

Since Proxima b is subjected to repeated flaring events and other
activity \citep{Walker81,Davenport16}, the atmosphere may be
constantly or permanently destroyed. Such a process is difficult to
envision, as it would require all the volatiles in the mantle to have
been degassed and blown away. However, if the planet was tidally and/or radiogenically
heated for a long time, mantle convection may have been vigorous enough to completely devolatilize the planet through outgassing. This possibility is most likely if the planet
is of order 7 Gyr old and if the core has solidified, quenching the
magnetic dynamo \citep{DriscollBarnes15}. Another possibility is that
a recent stellar eruption has temporarily stripped away the
atmosphere, which will later reform by outgassing.

\subsubsection{Sub-surface Liquid Water Layer}

A final possibility is that Proxima b, receiving only 65\% of Earth's
insolation, may have an ice-covered surface, but with a liquid water
mantle, similar to Europa. For such a planet, the water is heated by
the energy from accretion, radiogenic sources and/or tidal
heating. Water ice is much more absorptive at the longer wavelengths of
light that Proxima emits and so it may be difficult to ice over the
planet \citep{Pierrehumbert11,JoshiHaberle12,Shields13,Checlair17}, especially since it probably
spent hundreds of Myr in a runaway greenhouse. But if a reflective
haze and/or cloud layer could form, it could reflect away the light
before it reaches the surface \citep[\eg][]{Arney16}. This possibility
is most likely if the planet is rotating syper-synchronously, otherwise the ``eyeball state'' could develop \citep{Pierrehumbert11}.

Unlike the icy worlds in our Solar System, Proxima b has sufficient gravity that atmospheric molecules can remain bound to the planet, unless stellar activity removes them. If the subsurface layer is inhabited, then biomolecular gases may float to the ice layer and slowly find their way to the atmosphere. If they remain in the atmosphere for long periods of time, they could constitute a biosignature. If the planet has no atmosphere, then reflectance spectroscopy could reveal organic compounds on the surface, similar to those seen on Europa \citep[\eg][]{Noll95}.

\clearpage
\subsection{Is Proxima b Habitable?}
\label{sec:results:habitable}

Planetary habitability is a complicated feature to model
quantitatively and Proxima b is no exception. We do know that the
planet has sufficient energy to support life, almost assuredly has significant abundances of
bioessential elements since they are some of the most common in the galaxy, and is old enough for life to have gained a
foothold. The biggest
questions are if it is terrestrial, if it possesses vast reservoirs
of liquid water, and if it quickly formed a highly oxidized atmosphere. Without tighter constraints on the initial volatile inventory, it is impossible to determine the
probability that it does support liquid water, so we cannot answer the
eponymous question. As always, more data are needed.

However, our analysis does provide some important information on where
to focus future efforts. As liquid water is vital, it is paramount to
determine the pathways that allow the planet to have accreted and
retained the water. Even if the planet forms with water, our
investigations have shown that it will not necessarily be retained. If
it formed {\it in situ} or arrived in the HZ at the time of the
dispersal of the gas disk, then Proxima b had to endure $\sim$ 170
Myr in a runaway greenhouse state; see $\S$~\ref{sec:results:atmesc}.

Even if the planet arrived in its orbit late, perhaps following an
orbital instability, the water may have to survive a ``tidal
greenhouse'' in which tidal heating drives water loss, see
$\S$~\ref{sec:results:internal} and \cite{Barnes13}. Such high tidal
heating rates may require very large eccentricity and/or abnormally
low $Q$ values, but the former is certainly possible during
planet-planet scattering events \citep{Chatterjee08}, or perhaps by
Kozai-like oscillation driven by perturbations from the \acen~A and B
pair if the orbit of Proxima Centauri was much smaller in the past
\citep{DesideraBarbieri07}. If an additional planet in the system is
massive, on an eccentric orbit, and/or on a highly inclined orbit,
then it, too, may induce perturbations that maintain eccentricities
\citep{TakedaRasio05}, possibly in the range of a tidal greenhouse. A
planet in a Cassini state may receive additional tidal heating due to non-zero obliquity \citep{Heller11},
further increasing the risk of a tidal greenhouse. As shown in
$\S$~\ref{sec:results:internal}, the eccentricity of b will damp, but
the timescale can be very long. Large eccentricities are not
well-modeled by equilibrium tide theory, even with a proper accounting
of geophysical features as in the \thermint~module, so it is difficult
at present to assess the role of tidal heating in water retention.

Another possible route to water loss is through temporary or permanent
atmosphere erosion by flares and coronal mass ejections. These events could blast away the atmosphere completely,
in which case liquid water on the surface is not stable. Should the
atmosphere reform, the water may return to the liquid state, but it is
certainly plausible that some events are powerful enough to remove the
water in one event, or, more likely, repeated bombardments would
slowly remove the atmosphere \citep{Cohen15}.  Our analysis doesn't
provide a direct measurement of this phenomenon, but we note that if the dynamo is quenched, perhaps because the core has completely nucleated, charged
particles can reach the surface. Even if the planet does have a magnetic field,
\cite{Vidotto13} find that planets around typical M dwarfs may have
their magnetopause distances driven to the planet surface by the
star's magnetic field. However, it is not clear that a magnetic field
is always beneficial for life, as it also increases the
cross-sectional area of the planet for charged particles and funnels
the energy into the magnetic poles, possibly increasing mass loss.

These processes are all clear dangers for the habitability of Proxima
b. Yet, we are also able to identify pathways that produce decidedly
Earth-like versions of planet b. As shown in Fig.~\ref{fig:tidal_hec},
if the planet formed with 0.1\% of its mass in a hydrogen envelope, 3.5
Earth oceans of water, then the combined effects of the stellar
evolution, envelope evolution and atmospheric escape, tidal evolution, orbital evolution,
and geophysical evolution predict a planet with 1 Earth ocean of
surface water, no hydrogen envelope, no abiotic oxygen build-up, and a semi-major axis within the
observed uncertainties. If this scenario is true, then our model predicts Proxima b could be a true Earth analog today.

Proxima b may or may not be habitable. While we are only able to
identify a narrow range of pathways that permit habitability, we must
bear in mind that our model, while including phenomena over sizescales
of meters to kpc, is simple and does not include many potential
feedbacks. The geochemistry of exoplanets is a gaping hole in
scientific knowledge, and one can easily imagine how other systems may
maintain liquid water with geochemical cycles not present in our Solar
System.

\section{Conclusions}\label{sec:concl}
We have performed a detailed analysis of the evolution of the
Alpha Centauri triple star system with a specific focus on Proxima
Centauri b's habitability. We find that many disparate factors are
important, including the stellar system's orbit in the galaxy
($\S$~\ref{sec:results:galactic}), the orbital and rotational
evolution of the planets ($\S$~\ref{sec:results:orbital}), the stellar
evolution ($\S$~\ref{sec:results:stellar}), the atmospheric evolution
($\S$~\ref{sec:results:atmesc}), and the geophysical evolution
($\S$~\ref{sec:results:internal}). We find that many evolutionary
pathways are permitted by the data and hence the planet may currently
exist in one of many possible states.

We conclude that Proxima b may be habitable, and identify the
retention of water as the biggest obstacle for Proxima b to support
life. Water loss may occur through multiple channels operating in
tandem or in isolation, including desiccation during the Pre-MS,
excessive tidal heating, or atmospheric destruction by flares and
coronal mass ejections. We find the most likely pathway for
habitability is if planet b formed with a thin hydrogen envelope of
order $10^{-4}$ to $10^{-2}~\mearth$ which was eroded by the early XUV evolution of
the host star; see $\S$~\ref{sec:results:stellar} and
\cite{Luger15}. In that case, Proxima b is a ``habitable evaporated
core'' and has followed a very different trajectory than Earth did on
their paths to liquid surface water. Our conclusions regarding water loss, tidal effects, and potential habitability are broadly consistent with the independent and simultaneous analysis in \citep{Ribas16}.

Regardless of Proxima's habitability, it offers scientists an
unprecedented window into the nature of terrestrial planets. At only
1.3 pc distance, we will be able to study this planet in detail with
future missions, should they be designed appropriately; see
\cite{Meadows18}. If Proxima b is uninhabitable, we may be able to
determine how that happened and how Earth avoided the same fate. At a
minimum, the discovery of Proxima Centauri b, as well as more recent discoveries such as the TRAPPIST-1 planets and LHS~1140~b \citep{Gillon17,Dittman17}, have ushered in a new era
of comparative planetology and analyses such as this one can provide a foundation for intepreting observational data, such as in \citep{Meadows18}.

The research described here shows that numerous possibilities exist for Proxima b's current state. These hypotheses can be tested with future telescopes and spacecraft, such as the TMT and LUVOIR. In the second paper in this series \citep{Meadows18}, the general features of the atmospheres we predict are transformed into self-consistent atmospheric models and the observable features are computed. The generation of evolutionary pathways provides a foundation to interpret future observations, and the results of the second paper find that it is indeed possible to distinguish most of the planetary states we predict.

\vspace{1cm} We thank G. Anglada-Escud{\' e} for sharing the results,
and for leading the Pale Red Dot campaign. This work was supported by
the NASA Astrobiology Institute's Virtual Planetary Laboratory under
Cooperative Agreement number NNA13AA93A.  David Fleming is supported
by an NSF IGERT DGE-1258485 fellowship. This work was performed in part under contract with the California
Institute of Technology (Caltech)/Jet Propulsion Laboratory (JPL)
funded by NASA through the Sagan Fellowship Program executed by the
NASA Exoplanet Science Institute. N.A. Kaib acknowledges support from HST-AR-13898. The results reported herein benefited from the authors’ affiliation with the NASA's Nexus for Exoplanet System Science (NExSS) research coordination network sponsored by NASA's Science Mission Directorate.

\bibliography{bib}

\begin{thebibliography}{223}
\expandafter\ifx\csname natexlab\endcsname\relax\def\natexlab#1{#1}\fi

\bibitem[{{Abe}(1993)}]{Abe93}
{Abe}, Y. 1993, Lithos, 30, 223

\bibitem[{{Abe} {et~al.}(2011){Abe}, {Abe-Ouchi}, {Sleep}, \& {Zahnle}}]{Abe11}
{Abe}, Y., {Abe-Ouchi}, A., {Sleep}, N.~H., \& {Zahnle}, K.~J. 2011,
  Astrobiology, 11, 443

\bibitem[{{Aguilar} \& {White}(1985)}]{Aguilar1985}
{Aguilar}, L.~A. \& {White}, S.~D.~M. 1985, \apj, 295, 374

\bibitem[{{Airapetian} {et~al.}(2017){Airapetian}, {Glocer}, {Khazanov},
  {Loyd}, {France}, {Sojka}, {Danchi}, \& {Liemohn}}]{Airapetian17}
{Airapetian}, V.~S., {Glocer}, A., {Khazanov}, G.~V., {Loyd}, R.~O.~P.,
  {France}, K., {Sojka}, J., {Danchi}, W.~C., \& {Liemohn}, M.~W. 2017, \apjl,
  836, L3

\bibitem[{{Aksnes} \& {Franklin}(2001)}]{AksnesFranklin01}
{Aksnes}, K. \& {Franklin}, F.~A. 2001, \aj, 122, 2734

\bibitem[{{Alibert} \& {Benz}(2017)}]{AlibertBenz17}
{Alibert}, Y. \& {Benz}, W. 2017, \aap, 598, L5

\bibitem[{{Allen} \& {Herrera}(1998)}]{AllenHerrera98}
{Allen}, C. \& {Herrera}, M.~A. 1998, Rev.~Mexicana Astron. Astrofis., 34, 37

\bibitem[{{Allende Prieto} {et~al.}(2004){Allende Prieto}, {Barklem},
  {Lambert}, \& {Cunha}}]{AllendePrieto04}
{Allende Prieto}, C., {Barklem}, P.~S., {Lambert}, D.~L., \& {Cunha}, K. 2004,
  \aap, 420, 183

\bibitem[{{Anders} \& {Grevesse}(1989)}]{AndersGrevesse89}
{Anders}, E. \& {Grevesse}, N. 1989, \gca, 53, 197

\bibitem[{{Anglada} {et~al.}(2017){Anglada}, {Amado}, {Ortiz}, {G{\'o}mez},
  {Mac{\'{\i}}as}, {Alberdi}, {Osorio}, {G{\'o}mez}, {de Gregorio-Monsalvo},
  {P{\'e}rez-Torres}, {Anglada-Escud{\'e}}, {Berdi{\~n}as}, {Jenkins},
  {Jimenez-Serra}, {Lara}, {L{\'o}pez-Gonz{\'a}lez}, {L{\'o}pez-Puertas},
  {Morales}, {Ribas}, {Richards}, {Rodr{\'{\i}}guez-L{\'o}pez}, \&
  {Rodriguez}}]{Anglada17}
{Anglada}, G., {Amado}, P.~J., {Ortiz}, J.~L., {G{\'o}mez}, J.~F.,
  {Mac{\'{\i}}as}, E., {Alberdi}, A., {Osorio}, M., {G{\'o}mez}, J.~L., {de
  Gregorio-Monsalvo}, I., {P{\'e}rez-Torres}, M.~A., {Anglada-Escud{\'e}}, G.,
  {Berdi{\~n}as}, Z.~M., {Jenkins}, J.~S., {Jimenez-Serra}, I., {Lara}, L.~M.,
  {L{\'o}pez-Gonz{\'a}lez}, M.~J., {L{\'o}pez-Puertas}, M., {Morales}, N.,
  {Ribas}, I., {Richards}, A.~M.~S., {Rodr{\'{\i}}guez-L{\'o}pez}, C., \&
  {Rodriguez}, E. 2017, \apjl, 850, L6

\bibitem[{{Anglada-Escud{\'e}} {et~al.}(2016){Anglada-Escud{\'e}}, {Amado},
  {Barnes}, {Berdi{\~n}as}, {Butler}, {Coleman}, {de La Cueva}, {Dreizler},
  {Endl}, {Giesers}, {Jeffers}, {Jenkins}, {Jones}, {Kiraga}, {K{\"u}rster},
  {L{\'o}pez-Gonz{\'a}lez}, {Marvin}, {Morales}, {Morin}, {Nelson}, {Ortiz},
  {Ofir}, {Paardekooper}, {Reiners}, {Rodr{\'{\i}}guez},
  {Rodr{\'{\i}}guez-L{\'o}pez}, {Sarmiento}, {Strachan}, {Tsapras}, {Tuomi}, \&
  {Zechmeister}}]{AngladaEscude16}
{Anglada-Escud{\'e}}, G., {Amado}, P.~J., {Barnes}, J., {Berdi{\~n}as}, Z.~M.,
  {Butler}, R.~P., {Coleman}, G.~A.~L., {de La Cueva}, I., {Dreizler}, S.,
  {Endl}, M., {Giesers}, B., {Jeffers}, S.~V., {Jenkins}, J.~S., {Jones},
  H.~R.~A., {Kiraga}, M., {K{\"u}rster}, M., {L{\'o}pez-Gonz{\'a}lez}, M.~J.,
  {Marvin}, C.~J., {Morales}, N., {Morin}, J., {Nelson}, R.~P., {Ortiz}, J.~L.,
  {Ofir}, A., {Paardekooper}, S.-J., {Reiners}, A., {Rodr{\'{\i}}guez}, E.,
  {Rodr{\'{\i}}guez-L{\'o}pez}, C., {Sarmiento}, L.~F., {Strachan}, J.~P.,
  {Tsapras}, Y., {Tuomi}, M., \& {Zechmeister}, M. 2016, \nat, 536, 437

\bibitem[{{Anosova} {et~al.}(1994){Anosova}, {Orlov}, \& {Pavlova}}]{Anosova94}
{Anosova}, J., {Orlov}, V.~V., \& {Pavlova}, N.~A. 1994, \aap, 292, 115

\bibitem[{{Araki} {et~al.}(2005){Araki}, {Enomoto}, {Furuno}, {Gando},
  {Ichimura}, {Ikeda}, {Inoue}, {Kishimoto}, {Koga}, {Koseki}, {Maeda},
  {Mitsui}, {Motoki}, {Nakajima}, {Ogawa}, {Ogawa}, {Owada}, {Ricol},
  {Shimizu}, {Shirai}, {Suekane}, {Suzuki}, {Tada}, {Takeuchi}, {Tamae},
  {Tsuda}, {Watanabe}, {Busenitz}, {Classen}, {Djurcic}, {Keefer}, {Leonard},
  {Piepke}, {Yakushev}, {Berger}, {Chan}, {Decowski}, {Dwyer}, {Freedman},
  {Fujikawa}, {Goldman}, {Gray}, {Heeger}, {Hsu}, {Lesko}, {Luk}, {Murayama},
  {O'Donnell}, {Poon}, {Steiner}, {Winslow}, {Mauger}, {McKeown}, {Vogel},
  {Lane}, {Miletic}, {Guillian}, {Learned}, {Maricic}, {Matsuno}, {Pakvasa},
  {Horton-Smith}, {Dazeley}, {Hatakeyama}, {Rojas}, {Svoboda}, {Dieterle},
  {Detwiler}, {Gratta}, {Ishii}, {Tolich}, {Uchida}, {Batygov}, {Bugg},
  {Efremenko}, {Kamyshkov}, {Kozlov}, {Nakamura}, {Karwowski}, {Markoff},
  {Nakamura}, {Rohm}, {Tornow}, {Wendell}, {Chen}, {Wang}, \&
  {Piquemal}}]{Araki05}
{Araki}, T., {Enomoto}, S., {Furuno}, K., {Gando}, Y., {Ichimura}, K., {Ikeda},
  H., {Inoue}, K., {Kishimoto}, Y., {Koga}, M., {Koseki}, Y., {Maeda}, T.,
  {Mitsui}, T., {Motoki}, M., {Nakajima}, K., {Ogawa}, H., {Ogawa}, M.,
  {Owada}, K., {Ricol}, J.-S., {Shimizu}, I., {Shirai}, J., {Suekane}, F.,
  {Suzuki}, A., {Tada}, K., {Takeuchi}, S., {Tamae}, K., {Tsuda}, Y.,
  {Watanabe}, H., {Busenitz}, J., {Classen}, T., {Djurcic}, Z., {Keefer}, G.,
  {Leonard}, D., {Piepke}, A., {Yakushev}, E., {Berger}, B.~E., {Chan}, Y.~D.,
  {Decowski}, M.~P., {Dwyer}, D.~A., {Freedman}, S.~J., {Fujikawa}, B.~K.,
  {Goldman}, J., {Gray}, F., {Heeger}, K.~M., {Hsu}, L., {Lesko}, K.~T., {Luk},
  K.-B., {Murayama}, H., {O'Donnell}, T., {Poon}, A.~W.~P., {Steiner}, H.~M.,
  {Winslow}, L.~A., {Mauger}, C., {McKeown}, R.~D., {Vogel}, P., {Lane}, C.~E.,
  {Miletic}, T., {Guillian}, G., {Learned}, J.~G., {Maricic}, J., {Matsuno},
  S., {Pakvasa}, S., {Horton-Smith}, G.~A., {Dazeley}, S., {Hatakeyama}, S.,
  {Rojas}, A., {Svoboda}, R., {Dieterle}, B.~D., {Detwiler}, J., {Gratta}, G.,
  {Ishii}, K., {Tolich}, N., {Uchida}, Y., {Batygov}, M., {Bugg}, W.,
  {Efremenko}, Y., {Kamyshkov}, Y., {Kozlov}, A., {Nakamura}, Y., {Karwowski},
  H.~J., {Markoff}, D.~M., {Nakamura}, K., {Rohm}, R.~M., {Tornow}, W.,
  {Wendell}, R., {Chen}, M.-J., {Wang}, Y.-F., \& {Piquemal}, F. 2005, \nat,
  436, 499

\bibitem[{{Arevalo} {et~al.}(2009){Arevalo}, {McDonough}, \&
  {Luong}}]{Arevalo09}
{Arevalo}, R., {McDonough}, W.~F., \& {Luong}, M. 2009, Earth and Planetary
  Science Letters, 278, 361

\bibitem[{{Armstrong} {et~al.}(2014){Armstrong}, {Barnes}, {Domagal-Goldman},
  {Breiner}, {Quinn}, \& {Meadows}}]{Armstrong14}
{Armstrong}, J.~C., {Barnes}, R., {Domagal-Goldman}, S., {Breiner}, J.,
  {Quinn}, T.~R., \& {Meadows}, V.~S. 2014, Astrobiology, 14, 277

\bibitem[{{Arney}(2016)}]{Arney16}
{Arney}, G. 2016, AsBio, accepted

\bibitem[{{Atobe} \& {Ida}(2007)}]{Atobe2007}
{Atobe}, K. \& {Ida}, S. 2007, \icarus, 188, 1

\bibitem[{{Atri}(2017)}]{Atri17}
{Atri}, D. 2017, \mnras, 465, L34

\bibitem[{{Baraffe} {et~al.}(1998){Baraffe}, {Chabrier}, {Allard}, \&
  {Hauschildt}}]{Baraffe98}
{Baraffe}, I., {Chabrier}, G., {Allard}, F., \& {Hauschildt}, P.~H. 1998, A\&A,
  337, 403

\bibitem[{{Baraffe} {et~al.}(2015){Baraffe}, {Homeier}, {Allard}, \&
  {Chabrier}}]{Baraffe15}
{Baraffe}, I., {Homeier}, D., {Allard}, F., \& {Chabrier}, G. 2015, \aap, 577,
  A42

\bibitem[{{Barnes} {et~al.}(2014){Barnes}, {Jenkins}, {Jones}, {Jeffers},
  {Rojo}, {Arriagada}, {Jord{\'a}n}, {Minniti}, {Tuomi}, {Pinfield}, \&
  {Anglada-Escud{\'e}}}]{Barnes14}
{Barnes}, J.~R., {Jenkins}, J.~S., {Jones}, H.~R.~A., {Jeffers}, S.~V., {Rojo},
  P., {Arriagada}, P., {Jord{\'a}n}, A., {Minniti}, D., {Tuomi}, M.,
  {Pinfield}, D., \& {Anglada-Escud{\'e}}, G. 2014, \mnras, 439, 3094

\bibitem[{{Barnes}(2017)}]{Barnes17}
{Barnes}, R. 2017, Celestial Mechanics and Dynamical Astronomy, 129, 509

\bibitem[{{Barnes} {et~al.}(2011){Barnes}, {Greenberg}, {Quinn}, {McArthur}, \&
  {Benedict}}]{Barnes11}
{Barnes}, R., {Greenberg}, R., {Quinn}, T.~R., {McArthur}, B.~E., \&
  {Benedict}, G.~F. 2011, \apj, 726, 71

\bibitem[{{Barnes} {et~al.}(2009){Barnes}, {Jackson}, {Raymond}, {West}, \&
  {Greenberg}}]{Barnes09}
{Barnes}, R., {Jackson}, B., {Raymond}, S.~N., {West}, A.~A., \& {Greenberg},
  R. 2009, \apj, 695, 1006

\bibitem[{{Barnes} {et~al.}(2013){Barnes}, {Mullins}, {Goldblatt}, {Meadows},
  {Kasting}, \& {Heller}}]{Barnes13}
{Barnes}, R., {Mullins}, K., {Goldblatt}, C., {Meadows}, V.~S., {Kasting},
  J.~F., \& {Heller}, R. 2013, Astrobiology, 13, 225

\bibitem[{{Barnes} {et~al.}(2008){Barnes}, {Raymond}, {Jackson}, \&
  {Greenberg}}]{Barnes08}
{Barnes}, R., {Raymond}, S.~N., {Jackson}, B., \& {Greenberg}, R. 2008,
  Astrobiology, 8, 557

\bibitem[{{Bazot} {et~al.}(2016){Bazot}, {Christensen-Dalsgaard}, {Gizon}, \&
  {Benomar}}]{Bazot16}
{Bazot}, M., {Christensen-Dalsgaard}, J., {Gizon}, L., \& {Benomar}, O. 2016,
  \mnras, 460, 1254

\bibitem[{{Benedict} {et~al.}(1998){Benedict}, {McArthur}, {Nelan}, {Story},
  {Whipple}, {Shelus}, {Jefferys}, {Hemenway}, {Franz}, {Wasserman},
  {Duncombe}, {van Altena}, \& {Fredrick}}]{Benedict98}
{Benedict}, G.~F., {McArthur}, B., {Nelan}, E., {Story}, D., {Whipple}, A.~L.,
  {Shelus}, P.~J., {Jefferys}, W.~H., {Hemenway}, P.~D., {Franz}, O.~G.,
  {Wasserman}, L.~H., {Duncombe}, R.~L., {van Altena}, W., \& {Fredrick}, L.~W.
  1998, \aj, 116, 429

\bibitem[{{Benedict} {et~al.}(1993){Benedict}, {Nelan}, {McArthur}, {Story},
  {van Altena}, {Ting-Gao}, {Jefferys}, {Hemenway}, {Shelus}, {Whipple},
  {Franz}, {Fredrick}, \& {Duncombe}}]{Benedict93}
{Benedict}, G.~F., {Nelan}, E., {McArthur}, B., {Story}, D., {van Altena}, W.,
  {Ting-Gao}, Y., {Jefferys}, W.~H., {Hemenway}, P.~D., {Shelus}, P.~J.,
  {Whipple}, A.~L., {Franz}, O.~G., {Fredrick}, L.~W., \& {Duncombe}, R.~L.
  1993, \pasp, 105, 487

\bibitem[{{Bixel} \& {Apai}(2017)}]{BixelApai17}
{Bixel}, A. \& {Apai}, D. 2017, \apjl, 836, L31

\bibitem[{{Bolmont} {et~al.}(2017){Bolmont}, {Selsis}, {Owen}, {Ribas},
  {Raymond}, {Leconte}, \& {Gillon}}]{Bolmont16}
{Bolmont}, E., {Selsis}, F., {Owen}, J.~E., {Ribas}, I., {Raymond}, S.~N.,
  {Leconte}, J., \& {Gillon}, M. 2017, \mnras, 464, 3728

\bibitem[{{Bonfils} {et~al.}(2013){Bonfils}, {Delfosse}, {Udry}, {Forveille},
  {Mayor}, {Perrier}, {Bouchy}, {Gillon}, {Lovis}, {Pepe}, {Queloz}, {Santos},
  {S{\'e}gransan}, \& {Bertaux}}]{Bonfils13}
{Bonfils}, X., {Delfosse}, X., {Udry}, S., {Forveille}, T., {Mayor}, M.,
  {Perrier}, C., {Bouchy}, F., {Gillon}, M., {Lovis}, C., {Pepe}, F., {Queloz},
  D., {Santos}, N.~C., {S{\'e}gransan}, D., \& {Bertaux}, J.-L. 2013, \aap,
  549, A109

\bibitem[{{Bouchy} \& {Carrier}(2001)}]{Bouchy01}
{Bouchy}, F. \& {Carrier}, F. 2001, \aap, 374, L5

\bibitem[{{Bouchy} \& {Carrier}(2002)}]{Bouchy02}
---. 2002, \aap, 390, 205

\bibitem[{{Boyajian} {et~al.}(2012){Boyajian}, {von Braun}, {van Belle},
  {McAlister}, {ten Brummelaar}, {Kane}, {Muirhead}, {Jones}, {White},
  {Schaefer}, {Ciardi}, {Henry}, {L{\'o}pez-Morales}, {Ridgway}, {Gies}, {Jao},
  {Rojas-Ayala}, {Parks}, {Sturmann}, {Sturmann}, {Turner}, {Farrington},
  {Goldfinger}, \& {Berger}}]{Boyajian12}
{Boyajian}, T.~S., {von Braun}, K., {van Belle}, G., {McAlister}, H.~A., {ten
  Brummelaar}, T.~A., {Kane}, S.~R., {Muirhead}, P.~S., {Jones}, J., {White},
  R., {Schaefer}, G., {Ciardi}, D., {Henry}, T., {L{\'o}pez-Morales}, M.,
  {Ridgway}, S., {Gies}, D., {Jao}, W.-C., {Rojas-Ayala}, B., {Parks}, J.~R.,
  {Sturmann}, L., {Sturmann}, J., {Turner}, N.~H., {Farrington}, C.,
  {Goldfinger}, P.~J., \& {Berger}, D.~H. 2012, \apj, 757, 112

\bibitem[{{Brasser} {et~al.}(2014){Brasser}, {Ida}, \& {Kokubo}}]{Brasser2014}
{Brasser}, R., {Ida}, S., \& {Kokubo}, E. 2014, \mnras, 440, 3685

\bibitem[{{Breiter} \& {Vokrouhlick{\'y}}(2015)}]{Breiter2015}
{Breiter}, S. \& {Vokrouhlick{\'y}}, D. 2015, \mnras, 449, 1691

\bibitem[{{Carrier} \& {Bourban}(2003)}]{CarrierBourban03}
{Carrier}, F. \& {Bourban}, G. 2003, \aap, 406, L23

\bibitem[{{Carter-Bond} {et~al.}(2012){Carter-Bond}, {O'Brien}, \&
  {Raymond}}]{CarterBond12}
{Carter-Bond}, J.~C., {O'Brien}, D.~P., \& {Raymond}, S.~N. 2012, \apj, 760, 44

\bibitem[{{Chaplin} {et~al.}(2014){Chaplin}, {Basu}, {Huber}, {Serenelli},
  {Casagrande}, {Silva Aguirre}, {Ball}, {Creevey}, {Gizon}, {Handberg},
  {Karoff}, {Lutz}, {Marques}, {Miglio}, {Stello}, {Suran}, {Pricopi},
  {Metcalfe}, {Monteiro}, {Molenda-{\.Z}akowicz}, {Appourchaux},
  {Christensen-Dalsgaard}, {Elsworth}, {Garc{\'{\i}}a}, {Houdek}, {Kjeldsen},
  {Bonanno}, {Campante}, {Corsaro}, {Gaulme}, {Hekker}, {Mathur}, {Mosser},
  {R{\'e}gulo}, \& {Salabert}}]{Chaplin2014}
{Chaplin}, W.~J., {Basu}, S., {Huber}, D., {Serenelli}, A., {Casagrande}, L.,
  {Silva Aguirre}, V., {Ball}, W.~H., {Creevey}, O.~L., {Gizon}, L.,
  {Handberg}, R., {Karoff}, C., {Lutz}, R., {Marques}, J.~P., {Miglio}, A.,
  {Stello}, D., {Suran}, M.~D., {Pricopi}, D., {Metcalfe}, T.~S., {Monteiro},
  M.~J.~P.~F.~G., {Molenda-{\.Z}akowicz}, J., {Appourchaux}, T.,
  {Christensen-Dalsgaard}, J., {Elsworth}, Y., {Garc{\'{\i}}a}, R.~A.,
  {Houdek}, G., {Kjeldsen}, H., {Bonanno}, A., {Campante}, T.~L., {Corsaro},
  E., {Gaulme}, P., {Hekker}, S., {Mathur}, S., {Mosser}, B., {R{\'e}gulo}, C.,
  \& {Salabert}, D. 2014, \apjs, 210, 1

\bibitem[{{Chatterjee} {et~al.}(2008){Chatterjee}, {Ford}, {Matsumura}, \&
  {Rasio}}]{Chatterjee08}
{Chatterjee}, S., {Ford}, E.~B., {Matsumura}, S., \& {Rasio}, F.~A. 2008, \apj,
  686, 580

\bibitem[{{Checlair} {et~al.}(2017){Checlair}, {Menou}, \&
  {Abbot}}]{Checlair17}
{Checlair}, J., {Menou}, K., \& {Abbot}, D.~S. 2017, \apj, 845, 132

\bibitem[{{Ciesla} {et~al.}(2015){Ciesla}, {Mulders}, {Pascucci}, \&
  {Apai}}]{Ciesla15}
{Ciesla}, F.~J., {Mulders}, G.~D., {Pascucci}, I., \& {Apai}, D. 2015, \apj,
  804, 9

\bibitem[{{Cincunegui} {et~al.}(2007){Cincunegui}, {D{\'{\i}}az}, \&
  {Mauas}}]{Cincunegui07}
{Cincunegui}, C., {D{\'{\i}}az}, R.~F., \& {Mauas}, P.~J.~D. 2007, \aap, 461,
  1107

\bibitem[{{Cohen} {et~al.}(2014){Cohen}, {Drake}, {Glocer}, {Garraffo},
  {Poppenhaeger}, {Bell}, {Ridley}, \& {Gombosi}}]{Cohen14}
{Cohen}, O., {Drake}, J.~J., {Glocer}, A., {Garraffo}, C., {Poppenhaeger}, K.,
  {Bell}, J.~M., {Ridley}, A.~J., \& {Gombosi}, T.~I. 2014, \apj, 790, 57

\bibitem[{{Cohen} {et~al.}(2015){Cohen}, {Ma}, {Drake}, {Glocer}, {Garraffo},
  {Bell}, \& {Gombosi}}]{Cohen15}
{Cohen}, O., {Ma}, Y., {Drake}, J.~J., {Glocer}, A., {Garraffo}, C., {Bell},
  J.~M., \& {Gombosi}, T.~I. 2015, \apj, 806, 41

\bibitem[{{Coleman} {et~al.}(2017){Coleman}, {Nelson}, {Paardekooper},
  {Dreizler}, {Giesers}, \& {Anglada-Escud{\'e}}}]{Coleman17}
{Coleman}, G.~A.~L., {Nelson}, R.~P., {Paardekooper}, S.~J., {Dreizler}, S.,
  {Giesers}, B., \& {Anglada-Escud{\'e}}, G. 2017, \mnras

\bibitem[{{Colombo}(1966)}]{Colombo1966}
{Colombo}, G. 1966, \aj, 71, 891

\bibitem[{{Correia} {et~al.}(2008){Correia}, {Levrard}, \&
  {Laskar}}]{Correia08}
{Correia}, A.~C.~M., {Levrard}, B., \& {Laskar}, J. 2008, \aap, 488, L63

\bibitem[{{Damasso} \& {Del Sordo}(2017)}]{DamassoDelSordo17}
{Damasso}, M. \& {Del Sordo}, F. 2017, \aap, 599, A126

\bibitem[{{Darwin}(1880)}]{Darwin1880}
{Darwin}, G.~H. 1880, Royal Society of London Philosophical Transactions Series
  I, 171, 713

\bibitem[{{Davenport} {et~al.}(2016){Davenport}, {Kipping}, {Sasselov},
  {Matthews}, \& {Cameron}}]{Davenport16}
{Davenport}, J.~R.~A., {Kipping}, D.~M., {Sasselov}, D., {Matthews}, J.~M., \&
  {Cameron}, C. 2016, \apjl, 829, L31

\bibitem[{{Deitrick} {et~al.}(2017){Deitrick}, {Barnes}, {Quinn}, {Armstrong},
  {Charnay}, \& {Wilhelm}}]{Deitrick2018}
{Deitrick}, R., {Barnes}, R., {Quinn}, T.~R., {Armstrong}, J., {Charnay}, B.,
  \& {Wilhelm}, C. 2017, ArXiv e-prints

\bibitem[{{Delfosse} {et~al.}(2000){Delfosse}, {Forveille}, {S{\'e}gransan},
  {Beuzit}, {Udry}, {Perrier}, \& {Mayor}}]{Delfosse00}
{Delfosse}, X., {Forveille}, T., {S{\'e}gransan}, D., {Beuzit}, J.-L., {Udry},
  S., {Perrier}, C., \& {Mayor}, M. 2000, \aap, 364, 217

\bibitem[{{Demory} {et~al.}(2009){Demory}, {S{\'e}gransan}, {Forveille},
  {Queloz}, {Beuzit}, {Delfosse}, {di Folco}, {Kervella}, {Le Bouquin},
  {Perrier}, {Benisty}, {Duvert}, {Hofmann}, {Lopez}, \& {Petrov}}]{Demory09}
{Demory}, B.-O., {S{\'e}gransan}, D., {Forveille}, T., {Queloz}, D., {Beuzit},
  J.-L., {Delfosse}, X., {di Folco}, E., {Kervella}, P., {Le Bouquin}, J.-B.,
  {Perrier}, C., {Benisty}, M., {Duvert}, G., {Hofmann}, K.-H., {Lopez}, B., \&
  {Petrov}, R. 2009, \aap, 505, 205

\bibitem[{{Des Marais} {et~al.}(2002){Des Marais}, {Harwit}, {Jucks},
  {Kasting}, {Lin}, {Lunine}, {Schneider}, {Seager}, {Traub}, \&
  {Woolf}}]{DesMarais2002}
{Des Marais}, D.~J., {Harwit}, M.~O., {Jucks}, K.~W., {Kasting}, J.~F., {Lin},
  D.~N.~C., {Lunine}, J.~I., {Schneider}, J., {Seager}, S., {Traub}, W.~A., \&
  {Woolf}, N.~J. 2002, Astrobiology, 2, 153

\bibitem[{{Desidera} \& {Barbieri}(2007)}]{DesideraBarbieri07}
{Desidera}, S. \& {Barbieri}, M. 2007, \aap, 462, 345

\bibitem[{{Dickey} {et~al.}(1994){Dickey}, {Bender}, {Faller}, {Newhall},
  {Ricklefs}, {Ries}, {Shelus}, {Veillet}, {Whipple}, {Wiant}, {Williams}, \&
  {Yoder}}]{Dickey94}
{Dickey}, J.~O., {Bender}, P.~L., {Faller}, J.~E., {Newhall}, X.~X.,
  {Ricklefs}, R.~L., {Ries}, J.~G., {Shelus}, P.~J., {Veillet}, C., {Whipple},
  A.~L., {Wiant}, J.~R., {Williams}, J.~G., \& {Yoder}, C.~F. 1994, Science,
  265, 482

\bibitem[{{Dittman} {et~al.}(2017){Dittman}, {Irwin}, {Charbonneau}, \&
  {Bonfils}}]{Dittman17}
{Dittman}, J., {Irwin}, J., {Charbonneau}, D., \& {Bonfils}, X. 2017, {\it
  Nature}, 333

\bibitem[{{Dole}(1964)}]{Dole64}
{Dole}, S.~H. 1964, {Habitable planets for man}

\bibitem[{{Dotter} {et~al.}(2008){Dotter}, {Chaboyer}, {Jevremovi{\'c}},
  {Kostov}, {Baron}, \& {Ferguson}}]{Dartmouth08}
{Dotter}, A., {Chaboyer}, B., {Jevremovi{\'c}}, D., {Kostov}, V., {Baron}, E.,
  \& {Ferguson}, J.~W. 2008, \apjs, 178, 89

\bibitem[{{Downes} {et~al.}(2015){Downes}, {Rom{\'a}n-Z{\'u}{\~n}iga},
  {Ballesteros-Paredes}, {Mateu}, {Brice{\~n}o}, {Hern{\'a}ndez},
  {Petr-Gotzens}, {Calvet}, {Hartmann}, \& {Mauco}}]{Downes15}
{Downes}, J.~J., {Rom{\'a}n-Z{\'u}{\~n}iga}, C., {Ballesteros-Paredes}, J.,
  {Mateu}, C., {Brice{\~n}o}, C., {Hern{\'a}ndez}, J., {Petr-Gotzens}, M.~G.,
  {Calvet}, N., {Hartmann}, L., \& {Mauco}, K. 2015, \mnras, 450, 3490

\bibitem[{{Dressing} \& {Charbonneau}(2013)}]{DressingCharbonneau13}
{Dressing}, C.~D. \& {Charbonneau}, D. 2013, \apj, 767, 95

\bibitem[{{Driscoll} \& {Bercovici}(2014)}]{DriscollBercovici14}
{Driscoll}, P. \& {Bercovici}, D. 2014, Physics of the Earth and Planetary
  Interiors, 236, 36

\bibitem[{{Driscoll} \& {Barnes}(2015)}]{DriscollBarnes15}
{Driscoll}, P.~E. \& {Barnes}, R. 2015, Astrobiology, 15, 739

\bibitem[{{Dye}(2010)}]{Dye10}
{Dye}, S.~T. 2010, Earth and Planetary Science Letters, 297, 1

\bibitem[{{Efroimsky} \& {Makarov}(2013)}]{EfroimskyMakarov13}
{Efroimsky}, M. \& {Makarov}, V.~V. 2013, \apj, 764, 26

\bibitem[{{Endl} \& {K{\"u}rster}(2008)}]{EndlKurster08}
{Endl}, M. \& {K{\"u}rster}, M. 2008, \aap, 488, 1149

\bibitem[{{Erkaev} {et~al.}(2007){Erkaev}, {Kulikov}, {Lammer}, {Selsis},
  {Langmayr}, {Jaritz}, \& {Biernat}}]{Erkaev07}
{Erkaev}, N.~V., {Kulikov}, Y.~N., {Lammer}, H., {Selsis}, F., {Langmayr}, D.,
  {Jaritz}, G.~F., \& {Biernat}, H.~K. 2007, A\&A, 472, 329

\bibitem[{{Ferraz-Mello} {et~al.}(2008){Ferraz-Mello}, {Rodr{\'{\i}}guez}, \&
  {Hussmann}}]{FerrazMello08}
{Ferraz-Mello}, S., {Rodr{\'{\i}}guez}, A., \& {Hussmann}, H. 2008, Celestial
  Mechanics and Dynamical Astronomy, 101, 171

\bibitem[{{Ford} {et~al.}(2000){Ford}, {Kozinsky}, \& {Rasio}}]{Ford2000}
{Ford}, E.~B., {Kozinsky}, B., \& {Rasio}, F.~A. 2000, \apj, 535, 385

\bibitem[{{Foreman-Mackey} {et~al.}(2013){Foreman-Mackey}, {Hogg}, {Lang}, \&
  {Goodman}}]{ForemanMackey13}
{Foreman-Mackey}, D., {Hogg}, D.~W., {Lang}, D., \& {Goodman}, J. 2013, PASP,
  125, 306

\bibitem[{{Garc{\'{\i}}a-S{\'a}nchez}
  {et~al.}(2001){Garc{\'{\i}}a-S{\'a}nchez}, {Weissman}, {Preston}, {Jones},
  {Lestrade}, {Latham}, {Stefanik}, \& {Paredes}}]{Garciasanchez2001}
{Garc{\'{\i}}a-S{\'a}nchez}, J., {Weissman}, P.~R., {Preston}, R.~A., {Jones},
  D.~L., {Lestrade}, J.-F., {Latham}, D.~W., {Stefanik}, R.~P., \& {Paredes},
  J.~M. 2001, \aap, 379, 634

\bibitem[{{Garraffo} {et~al.}(2016){Garraffo}, {Drake}, \&
  {Cohen}}]{Garraffo16}
{Garraffo}, C., {Drake}, J.~J., \& {Cohen}, O. 2016, \apjl, 833, L4

\bibitem[{{Gillon} {et~al.}(2017){Gillon}, {Triaud}, {Demory}, {Jehin}, {Agol},
  {Deck}, {Lederer}, {de Wit}, {Burdanov}, {Ingalls}, {Bolmont}, {Leconte},
  {Raymond}, {Selsis}, {Turbet}, {Barkaoui}, {Burgasser}, {Burleigh}, {Carey},
  {Chaushev}, {Copperwheat}, {Delrez}, {Fernandes}, {Holdsworth}, {Kotze}, {Van
  Grootel}, {Almleaky}, {Benkhaldoun}, {Magain}, \& {Queloz}}]{Gillon17}
{Gillon}, M., {Triaud}, A.~H.~M.~J., {Demory}, B.-O., {Jehin}, E., {Agol}, E.,
  {Deck}, K.~M., {Lederer}, S.~M., {de Wit}, J., {Burdanov}, A., {Ingalls},
  J.~G., {Bolmont}, E., {Leconte}, J., {Raymond}, S.~N., {Selsis}, F.,
  {Turbet}, M., {Barkaoui}, K., {Burgasser}, A., {Burleigh}, M.~R., {Carey},
  S.~J., {Chaushev}, A., {Copperwheat}, C.~M., {Delrez}, L., {Fernandes},
  C.~S., {Holdsworth}, D.~L., {Kotze}, E.~J., {Van Grootel}, V., {Almleaky},
  Y., {Benkhaldoun}, Z., {Magain}, P., \& {Queloz}, D. 2017, \nat, 542, 456

\bibitem[{{Goldblatt}(2015)}]{Goldblatt15}
{Goldblatt}, C. 2015, Astrobiology, 15, 362

\bibitem[{{Goldreich}(1966)}]{Goldreich66}
{Goldreich}, P. 1966, \aj, 71, 1

\bibitem[{{Gomes} {et~al.}(2005){Gomes}, {Levison}, {Tsiganis}, \&
  {Morbidelli}}]{Gomes05}
{Gomes}, R., {Levison}, H.~F., {Tsiganis}, K., \& {Morbidelli}, A. 2005, \nat,
  435, 466

\bibitem[{{Greenberg}(2009)}]{Greenberg09}
{Greenberg}, R. 2009, Astrophys.~J., 698, L42

\bibitem[{{Haldane}(1929)}]{Haldane29}
{Haldane}, J.~B.~S. 1929, The Rationalist Annal., 3

\bibitem[{{Hamilton} \& {Ward}(2004)}]{Hamilton2004}
{Hamilton}, D.~P. \& {Ward}, W.~R. 2004, \aj, 128, 2510

\bibitem[{{Harrington}(1968)}]{Harrington1968}
{Harrington}, R.~S. 1968, \aj, 73, 190

\bibitem[{{Hayden} {et~al.}(2015){Hayden}, {Bovy}, {Holtzman}, {Nidever},
  {Bird}, {Weinberg}, {Andrews}, {Majewski}, {Allende Prieto}, {Anders},
  {Beers}, {Bizyaev}, {Chiappini}, {Cunha}, {Frinchaboy},
  {Garc{\'{\i}}a-Her{\'n}andez}, {Garc{\'{\i}}a P{\'e}rez}, {Girardi},
  {Harding}, {Hearty}, {Johnson}, {M{\'e}sz{\'a}ros}, {Minchev}, {O'Connell},
  {Pan}, {Robin}, {Schiavon}, {Schneider}, {Schultheis}, {Shetrone},
  {Skrutskie}, {Steinmetz}, {Smith}, {Wilson}, {Zamora}, \&
  {Zasowski}}]{Hayden15}
{Hayden}, M.~R., {Bovy}, J., {Holtzman}, J.~A., {Nidever}, D.~L., {Bird},
  J.~C., {Weinberg}, D.~H., {Andrews}, B.~H., {Majewski}, S.~R., {Allende
  Prieto}, C., {Anders}, F., {Beers}, T.~C., {Bizyaev}, D., {Chiappini}, C.,
  {Cunha}, K., {Frinchaboy}, P., {Garc{\'{\i}}a-Her{\'n}andez}, D.~A.,
  {Garc{\'{\i}}a P{\'e}rez}, A.~E., {Girardi}, L., {Harding}, P., {Hearty},
  F.~R., {Johnson}, J.~A., {M{\'e}sz{\'a}ros}, S., {Minchev}, I., {O'Connell},
  R., {Pan}, K., {Robin}, A.~C., {Schiavon}, R.~P., {Schneider}, D.~P.,
  {Schultheis}, M., {Shetrone}, M., {Skrutskie}, M., {Steinmetz}, M., {Smith},
  V., {Wilson}, J.~C., {Zamora}, O., \& {Zasowski}, G. 2015, \apj, 808, 132

\bibitem[{{Heisler} \& {Tremaine}(1986)}]{Heisler1986}
{Heisler}, J. \& {Tremaine}, S. 1986, \icarus, 65, 13

\bibitem[{{Heisler} {et~al.}(1987){Heisler}, {Tremaine}, \&
  {Alcock}}]{Heisler1987}
{Heisler}, J., {Tremaine}, S., \& {Alcock}, C. 1987, \icarus, 70, 269

\bibitem[{{Heller} {et~al.}(2010){Heller}, {Jackson}, {Barnes}, {Greenberg}, \&
  {Homeier}}]{Heller10}
{Heller}, R., {Jackson}, B., {Barnes}, R., {Greenberg}, R., \& {Homeier}, D.
  2010, \aap, 514, A22

\bibitem[{{Heller} {et~al.}(2011){Heller}, {Leconte}, \& {Barnes}}]{Heller11}
{Heller}, R., {Leconte}, J., \& {Barnes}, R. 2011, Astro.~\& Astrophys., 528,
  A27+

\bibitem[{{Henning} {et~al.}(2009){Henning}, {O'Connell}, \&
  {Sasselov}}]{Henning09}
{Henning}, W.~G., {O'Connell}, R.~J., \& {Sasselov}, D.~D. 2009, Astrophys.~J.,
  707, 1000

\bibitem[{{Hern{\'a}ndez} {et~al.}(2007){Hern{\'a}ndez}, {Hartmann}, {Megeath},
  {Gutermuth}, {Muzerolle}, {Calvet}, {Vivas}, {Brice{\~n}o}, {Allen},
  {Stauffer}, {Young}, \& {Fazio}}]{Hernandez07}
{Hern{\'a}ndez}, J., {Hartmann}, L., {Megeath}, T., {Gutermuth}, R.,
  {Muzerolle}, J., {Calvet}, N., {Vivas}, A.~K., {Brice{\~n}o}, C., {Allen},
  L., {Stauffer}, J., {Young}, E., \& {Fazio}, G. 2007, \apj, 662, 1067

\bibitem[{Hevey \& Sanders(2006)}]{HeveySanders06}
Hevey, P.~J. \& Sanders, I.~S. 2006, Meteoritics \& Planetary Science, 41, 95

\bibitem[{{Hinkel} \& {Kane}(2013)}]{HinkelKane13}
{Hinkel}, N.~R. \& {Kane}, S.~R. 2013, \mnras, 432, 36

\bibitem[{{Holmberg} \& {Flynn}(2000)}]{Holmberg2000}
{Holmberg}, J. \& {Flynn}, C. 2000, \mnras, 313, 209

\bibitem[{{Huang} {et~al.}(2013){Huang}, {Chubakov}, {Mantovani}, {Rudnick}, \&
  {McDonough}}]{Huang13}
{Huang}, Y., {Chubakov}, V., {Mantovani}, F., {Rudnick}, R.~L., \& {McDonough},
  W.~F. 2013, Geochemistry, Geophysics, Geosystems, 14, 2003

\bibitem[{{Hunten}(1973)}]{Hunten73}
{Hunten}, D.~M. 1973, Journal of Atmospheric Sciences, 30, 1481

\bibitem[{{Hunten} {et~al.}(1987){Hunten}, {Pepin}, \& {Walker}}]{Hunten87}
{Hunten}, D.~M., {Pepin}, R.~O., \& {Walker}, J.~C.~G. 1987, Icarus, 69, 532

\bibitem[{{Ida} \& {Lin}(2005)}]{Ida2005}
{Ida}, S. \& {Lin}, D.~N.~C. 2005, \apj, 626, 1045

\bibitem[{{Innes}(1915)}]{Innes1915}
{Innes}, R.~T.~A. 1915, Circular of the Union Observatory Johannesburg, 30, 235

\bibitem[{{Jackson} {et~al.}(2012){Jackson}, {Davis}, \&
  {Wheatley}}]{Jackson2012}
{Jackson}, A.~P., {Davis}, T.~A., \& {Wheatley}, P.~J. 2012, \mnras, 422, 2024

\bibitem[{{Jackson} {et~al.}(2008{\natexlab{a}}){Jackson}, {Barnes}, \&
  {Greenberg}}]{Jackson08c}
{Jackson}, B., {Barnes}, R., \& {Greenberg}, R. 2008{\natexlab{a}}, \mnras,
  391, 237

\bibitem[{{Jackson} {et~al.}(2009){Jackson}, {Barnes}, \&
  {Greenberg}}]{Jackson09}
---. 2009, Astrophys.~J., 698, 1357

\bibitem[{{Jackson} {et~al.}(2008{\natexlab{b}}){Jackson}, {Greenberg}, \&
  {Barnes}}]{Jackson08a}
{Jackson}, B., {Greenberg}, R., \& {Barnes}, R. 2008{\natexlab{b}},
  Astrophys.~J., 678, 1396

\bibitem[{Jaupart {et~al.}(2015)Jaupart, Labrosse, Lucazaeu, \&
  Mareschal}]{Jaupart15}
Jaupart, C., Labrosse, S., Lucazaeu, F., \& Mareschal, J.~C. {Temperatures,
  Heat and Energy in the Mantle of the Earth}, 2nd edn., 253--305

\bibitem[{{Johnson} \& {Apps}(2009)}]{Johnson2009}
{Johnson}, J.~A. \& {Apps}, K. 2009, \apj, 699, 933

\bibitem[{{Joshi} \& {Haberle}(2012)}]{JoshiHaberle12}
{Joshi}, M.~M. \& {Haberle}, R.~M. 2012, Astrobiology, 12, 3

\bibitem[{{Joshi} {et~al.}(1997){Joshi}, {Haberle}, \& {Reynolds}}]{Joshi97}
{Joshi}, M.~M., {Haberle}, R.~M., \& {Reynolds}, R.~T. 1997, \icarus, 129, 450

\bibitem[{{Kaib} {et~al.}(2013){Kaib}, {Raymond}, \& {Duncan}}]{Kaib13}
{Kaib}, N.~A., {Raymond}, S.~N., \& {Duncan}, M. 2013, \nat, 493, 381

\bibitem[{{Kasting}(1988)}]{Kasting88}
{Kasting}, J.~F. 1988, Icarus, 74, 472

\bibitem[{{Kasting} {et~al.}(1993){Kasting}, {Whitmire}, \&
  {Reynolds}}]{Kasting93}
{Kasting}, J.~F., {Whitmire}, D.~P., \& {Reynolds}, R.~T. 1993, \icarus, 101,
  108

\bibitem[{{Kervella} {et~al.}(2017){Kervella}, {Th{\'e}venin}, \&
  {Lovis}}]{Kervella17}
{Kervella}, P., {Th{\'e}venin}, F., \& {Lovis}, C. 2017, \aap, 598, L7

\bibitem[{{Kinoshita}(1975)}]{Kinoshita1975}
{Kinoshita}, H. 1975, SAO Special Report, 364

\bibitem[{{Kinoshita}(1977)}]{Kinoshita1977}
---. 1977, Celestial Mechanics, 15, 277

\bibitem[{{Kipping} {et~al.}(2017){Kipping}, {Cameron}, {Hartman}, {Davenport},
  {Matthews}, {Sasselov}, {Rowe}, {Siverd}, {Chen}, {Sandford}, {Bakos},
  {Jord{\'a}n}, {Bayliss}, {Henning}, {Mancini}, {Penev}, {Csubry}, {Bhatti},
  {Da Silva Bento}, {Guenther}, {Kuschnig}, {Moffat}, {Rucinski}, \&
  {Weiss}}]{Kipping17}
{Kipping}, D.~M., {Cameron}, C., {Hartman}, J.~D., {Davenport}, J.~R.~A.,
  {Matthews}, J.~M., {Sasselov}, D., {Rowe}, J., {Siverd}, R.~J., {Chen}, J.,
  {Sandford}, E., {Bakos}, G.~{\'A}., {Jord{\'a}n}, A., {Bayliss}, D.,
  {Henning}, T., {Mancini}, L., {Penev}, K., {Csubry}, Z., {Bhatti}, W., {Da
  Silva Bento}, J., {Guenther}, D.~B., {Kuschnig}, R., {Moffat}, A.~F.~J.,
  {Rucinski}, S.~M., \& {Weiss}, W.~W. 2017, \aj, 153, 93

\bibitem[{{Kopparapu} {et~al.}(2013){Kopparapu}, {Ramirez}, {Kasting}, {Eymet},
  {Robinson}, {Mahadevan}, {Terrien}, {Domagal-Goldman}, {Meadows}, \&
  {Deshpande}}]{Kopparapu13}
{Kopparapu}, R.~K., {Ramirez}, R., {Kasting}, J.~F., {Eymet}, V., {Robinson},
  T.~D., {Mahadevan}, S., {Terrien}, R.~C., {Domagal-Goldman}, S., {Meadows},
  V., \& {Deshpande}, R. 2013, \apj, 765, 131

\bibitem[{{Kopparapu} {et~al.}(2014){Kopparapu}, {Ramirez}, {SchottelKotte},
  {Kasting}, {Domagal-Goldman}, \& {Eymet}}]{Kopparapu14}
{Kopparapu}, R.~K., {Ramirez}, R.~M., {SchottelKotte}, J., {Kasting}, J.~F.,
  {Domagal-Goldman}, S., \& {Eymet}, V. 2014, \apjl, 787, L29

\bibitem[{{Kopparapu} {et~al.}(2016){Kopparapu}, {Wolf}, {Haqq-Misra}, {Yang},
  {Kasting}, {Meadows}, {Terrien}, \& {Mahadevan}}]{Kopparapu16}
{Kopparapu}, R.~k., {Wolf}, E.~T., {Haqq-Misra}, J., {Yang}, J., {Kasting},
  J.~F., {Meadows}, V., {Terrien}, R., \& {Mahadevan}, S. 2016, \apj, 819, 84

\bibitem[{{Kordopatis} {et~al.}(2015){Kordopatis}, {Binney}, {Gilmore}, {Wyse},
  {Belokurov}, {McMillan}, {Hatfield}, {Grebel}, {Steinmetz}, {Navarro},
  {Seabroke}, {Minchev}, {Chiappini}, {Bienaym{\'e}}, {Bland-Hawthorn},
  {Freeman}, {Gibson}, {Helmi}, {Munari}, {Parker}, {Reid}, {Siebert},
  {Siviero}, \& {Zwitter}}]{Kordopatis15}
{Kordopatis}, G., {Binney}, J., {Gilmore}, G., {Wyse}, R.~F.~G., {Belokurov},
  V., {McMillan}, P.~J., {Hatfield}, P., {Grebel}, E.~K., {Steinmetz}, M.,
  {Navarro}, J.~F., {Seabroke}, G., {Minchev}, I., {Chiappini}, C.,
  {Bienaym{\'e}}, O., {Bland-Hawthorn}, J., {Freeman}, K.~C., {Gibson}, B.~K.,
  {Helmi}, A., {Munari}, U., {Parker}, Q., {Reid}, W.~A., {Siebert}, A.,
  {Siviero}, A., \& {Zwitter}, T. 2015, \mnras, 447, 3526

\bibitem[{{Korenaga}(2003)}]{Korenaga03}
{Korenaga}, J. 2003, \grl, 30, 20

\bibitem[{{Kreidberg} \& {Loeb}(2016)}]{KreidbergLoeb16}
{Kreidberg}, L. \& {Loeb}, A. 2016, \apjl, 832, L12

\bibitem[{{Laird}(1985)}]{Laird85}
{Laird}, J.~B. 1985, \apj, 289, 556

\bibitem[{{Lammer} {et~al.}(2013){Lammer}, {Erkaev}, {Odert}, {Kislyakova},
  {Leitzinger}, \& {Khodachenko}}]{Lammer2013}
{Lammer}, H., {Erkaev}, N.~V., {Odert}, P., {Kislyakova}, K.~G., {Leitzinger},
  M., \& {Khodachenko}, M.~L. 2013, \mnras, 430, 1247

\bibitem[{{Laskar}(1986)}]{Laskar1986}
{Laskar}, J. 1986, \aap, 157, 59

\bibitem[{{Laskar} {et~al.}(1993{\natexlab{a}}){Laskar}, {Joutel}, \&
  {Boudin}}]{Laskar1993a}
{Laskar}, J., {Joutel}, F., \& {Boudin}, F. 1993{\natexlab{a}}, \aap, 270, 522

\bibitem[{{Laskar} {et~al.}(1993{\natexlab{b}}){Laskar}, {Joutel}, \&
  {Robutel}}]{Laskar1993b}
{Laskar}, J., {Joutel}, F., \& {Robutel}, P. 1993{\natexlab{b}}, \nat, 361, 615

\bibitem[{{Lehmer} \& {Catling}(2017)}]{LehmerCatling17}
{Lehmer}, O.~R. \& {Catling}, D.~C. 2017, \apj, 845, 130

\bibitem[{Li {et~al.}(2017)Li, Stefansson, Robertson, Monson, Cañas, \&
  Mahadevan}]{Li17}
Li, Y., Stefansson, G., Robertson, P., Monson, A., Cañas, C., \& Mahadevan, S.
  2017, Research Notes of the AAS, 1, 49

\bibitem[{{Lissauer}(2007)}]{Lissauer07}
{Lissauer}, J.~J. 2007, \apjl, 660, L149

\bibitem[{Liu {et~al.}(2018)Liu, Jiang, Huang, Yu, Yang, Jia, Awiphan, Pan,
  Liu, Zhang, Wang, Li, Du, Li, Lu, Zhang, Tian, Li, Ji, Zhang, Shi, Wang,
  Zhou, \& Zhou}]{Liu18}
Liu, H.-G., Jiang, P., Huang, X., Yu, Z.-Y., Yang, M., Jia, M., Awiphan, S.,
  Pan, X., Liu, B., Zhang, H., Wang, J., Li, Z., Du, F., Li, X., Lu, H., Zhang,
  Z., Tian, Q.-G., Li, B., Ji, T., Zhang, S., Shi, X., Wang, J., Zhou, J.-L.,
  \& Zhou, H. 2018, The Astronomical Journal, 155, 12

\bibitem[{{Loebman} {et~al.}(2016){Loebman}, {Debattista}, {Nidever}, {Hayden},
  {Holtzman}, {Clarke}, {Ro{\v s}kar}, \& {Valluri}}]{Loebman16}
{Loebman}, S.~R., {Debattista}, V.~P., {Nidever}, D.~L., {Hayden}, M.~R.,
  {Holtzman}, J.~A., {Clarke}, A.~J., {Ro{\v s}kar}, R., \& {Valluri}, M. 2016,
  \apjl, 818, L6

\bibitem[{{Lopez} \& {Fortney}(2014)}]{LopezFortney14}
{Lopez}, E.~D. \& {Fortney}, J.~J. 2014, \apj, 792, 1

\bibitem[{{Lopez} {et~al.}(2012){Lopez}, {Fortney}, \& {Miller}}]{Lopez12}
{Lopez}, E.~D., {Fortney}, J.~J., \& {Miller}, N. 2012, \apj, 761, 59

\bibitem[{{Lovelock}(1965)}]{Lovelock65}
{Lovelock}, J.~E. 1965, \nat, 207, 568

\bibitem[{{Lovis} {et~al.}(2017){Lovis}, {Snellen}, {Mouillet}, {Pepe},
  {Wildi}, {Astudillo-Defru}, {Beuzit}, {Bonfils}, {Cheetham}, {Conod},
  {Delfosse}, {Ehrenreich}, {Figueira}, {Forveille}, {Martins}, {Quanz},
  {Santos}, {Schmid}, {S{\'e}gransan}, \& {Udry}}]{Lovis17}
{Lovis}, C., {Snellen}, I., {Mouillet}, D., {Pepe}, F., {Wildi}, F.,
  {Astudillo-Defru}, N., {Beuzit}, J.-L., {Bonfils}, X., {Cheetham}, A.,
  {Conod}, U., {Delfosse}, X., {Ehrenreich}, D., {Figueira}, P., {Forveille},
  T., {Martins}, J.~H.~C., {Quanz}, S.~P., {Santos}, N.~C., {Schmid}, H.-M.,
  {S{\'e}gransan}, D., \& {Udry}, S. 2017, \aap, 599, A16

\bibitem[{{Luger} \& {Barnes}(2015)}]{LugerBarnes15}
{Luger}, R. \& {Barnes}, R. 2015, Astrobiology, 15, 119

\bibitem[{{Luger} {et~al.}(2015){Luger}, {Barnes}, {Lopez}, {Fortney},
  {Jackson}, \& {Meadows}}]{Luger15}
{Luger}, R., {Barnes}, R., {Lopez}, E., {Fortney}, J., {Jackson}, B., \&
  {Meadows}, V. 2015, Astrobiology, 15, 57

\bibitem[{{Luger} {et~al.}(2017){Luger}, {Lustig-Yaeger}, {Fleming}, {Tilley},
  {Agol}, {Meadows}, {Deitrick}, \& {Barnes}}]{Luger17}
{Luger}, R., {Lustig-Yaeger}, J., {Fleming}, D.~P., {Tilley}, M.~A., {Agol},
  E., {Meadows}, V.~S., {Deitrick}, R., \& {Barnes}, R. 2017, \apj, 837, 63

\bibitem[{{Luhman}(2012)}]{Luhman12}
{Luhman}, K.~L. 2012, \araa, 50, 65

\bibitem[{{Lundkvist} {et~al.}(2014){Lundkvist}, {Kjeldsen}, \& {Silva
  Aguirre}}]{Lundkvist14}
{Lundkvist}, M., {Kjeldsen}, H., \& {Silva Aguirre}, V. 2014, \aap, 566, A82

\bibitem[{{Lurie} {et~al.}(2014){Lurie}, {Henry}, {Jao}, {Quinn}, {Winters},
  {Ianna}, {Koerner}, {Riedel}, \& {Subasavage}}]{Lurie14}
{Lurie}, J.~C., {Henry}, T.~J., {Jao}, W.-C., {Quinn}, S.~N., {Winters}, J.~G.,
  {Ianna}, P.~A., {Koerner}, D.~W., {Riedel}, A.~R., \& {Subasavage}, J.~P.
  2014, \aj, 148, 91

\bibitem[{{MacDonald}(1964)}]{MacDonald64}
{MacDonald}, G.~J.~F. 1964, Reviews of Geophysics and Space Physics, 2, 467

\bibitem[{{Makarov} {et~al.}(2016){Makarov}, {Frouard}, \&
  {Dorland}}]{Makarov16}
{Makarov}, V.~V., {Frouard}, J., \& {Dorland}, B. 2016, \mnras, 456, 665

\bibitem[{{Malmberg} {et~al.}(2007){Malmberg}, {de Angeli}, {Davies}, {Church},
  {Mackey}, \& {Wilkinson}}]{Malmberg07}
{Malmberg}, D., {de Angeli}, F., {Davies}, M.~B., {Church}, R.~P., {Mackey},
  D., \& {Wilkinson}, M.~I. 2007, \mnras, 378, 1207

\bibitem[{{Matthews} \& {Gilmore}(1993)}]{MatthewsGilmore93}
{Matthews}, R. \& {Gilmore}, G. 1993, \mnras, 261, L5

\bibitem[{{Matvienko} \& {Orlov}(2014)}]{MatvienkoOrlov14}
{Matvienko}, A.~S. \& {Orlov}, V.~V. 2014, Astrophysical Bulletin, 69, 205

\bibitem[{{Meadows}(2017)}]{Meadows17}
{Meadows}, V.~S. 2017, Astrobiology, 17, 1022

\bibitem[{{Meadows} {et~al.}(2018){Meadows}, {Arney}, {Schwieterman},
  {Lustig-Yaeger}, {Lincowski}, {Robinson}, {Domagal-Goldman}, {Deitrick},
  {Barnes}, {Fleming}, {Luger}, {Driscoll}, {Quinn}, \& {Crisp}}]{Meadows18}
{Meadows}, V.~S., {Arney}, G.~N., {Schwieterman}, E.~W., {Lustig-Yaeger}, J.,
  {Lincowski}, A.~P., {Robinson}, T., {Domagal-Goldman}, S.~D., {Deitrick}, R.,
  {Barnes}, R.~K., {Fleming}, D.~P., {Luger}, R., {Driscoll}, P., {Quinn}, T.,
  \& {Crisp}, D. 2018, Astrobiology, 18, 133

\bibitem[{{Mesa} {et~al.}(2017){Mesa}, {Zurlo}, {Milli}, {Gratton}, {Desidera},
  {Langlois}, {Vigan}, {Bonavita}, {Antichi}, {Avenhaus}, {Baruffolo},
  {Biller}, {Boccaletti}, {Bruno}, {Cascone}, {Chauvin}, {Claudi}, {De Caprio},
  {Fantinel}, {Farisato}, {Girard}, {Giro}, {Hagelberg}, {Incorvaia}, {Janson},
  {Kral}, {Lagadec}, {Lagrange}, {Lessio}, {Meyer}, {Peretti}, {Perrot},
  {Salasnich}, {Schlieder}, {Schmid}, {Scuderi}, {Sissa}, {Thalmann}, \&
  {Turatto}}]{Mesa17}
{Mesa}, D., {Zurlo}, A., {Milli}, J., {Gratton}, R., {Desidera}, S.,
  {Langlois}, M., {Vigan}, A., {Bonavita}, M., {Antichi}, J., {Avenhaus}, H.,
  {Baruffolo}, A., {Biller}, B., {Boccaletti}, A., {Bruno}, P., {Cascone}, E.,
  {Chauvin}, G., {Claudi}, R.~U., {De Caprio}, V., {Fantinel}, D., {Farisato},
  G., {Girard}, J., {Giro}, E., {Hagelberg}, J., {Incorvaia}, S., {Janson}, M.,
  {Kral}, Q., {Lagadec}, E., {Lagrange}, A.-M., {Lessio}, L., {Meyer}, M.,
  {Peretti}, S., {Perrot}, C., {Salasnich}, B., {Schlieder}, J., {Schmid},
  H.-M., {Scuderi}, S., {Sissa}, E., {Thalmann}, C., \& {Turatto}, M. 2017,
  \mnras, 466, L118

\bibitem[{{Minchev} {et~al.}(2012){Minchev}, {Famaey}, {Quillen}, {Dehnen},
  {Martig}, \& {Siebert}}]{Minchev2012}
{Minchev}, I., {Famaey}, B., {Quillen}, A.~C., {Dehnen}, W., {Martig}, M., \&
  {Siebert}, A. 2012, \aap, 548, A127

\bibitem[{{Misra} {et~al.}(2015){Misra}, {Krissansen-Totton}, {Koehler}, \&
  {Sholes}}]{Misra15}
{Misra}, A., {Krissansen-Totton}, J., {Koehler}, M.~C., \& {Sholes}, S. 2015,
  Astrobiology, 15, 462

\bibitem[{{Mulders} {et~al.}(2015){Mulders}, {Ciesla}, {Min}, \&
  {Pascucci}}]{Mulders15}
{Mulders}, G.~D., {Ciesla}, F.~J., {Min}, M., \& {Pascucci}, I. 2015, \apj,
  807, 9

\bibitem[{{Mumford}(1909)}]{Mumford09}
{Mumford}, N.~W. 1909, Popular Astronomy, 17, 497

\bibitem[{{Murray} \& {Dermott}(1999)}]{MurrayDermott99}
{Murray}, C.~D. \& {Dermott}, S.~F. 1999, {Solar system dynamics}

\bibitem[{{Murray-Clay} {et~al.}(2009){Murray-Clay}, {Chiang}, \&
  {Murray}}]{MurrayClay09}
{Murray-Clay}, R.~A., {Chiang}, E.~I., \& {Murray}, N. 2009, \apj, 693, 23

\bibitem[{{Neuforge-Verheecke} \& {Magain}(1997)}]{NeuforgeMagain97}
{Neuforge-Verheecke}, C. \& {Magain}, P. 1997, \aap, 328, 261

\bibitem[{{Noll} {et~al.}(1995){Noll}, {Weaver}, \& {Gonnella}}]{Noll95}
{Noll}, K.~S., {Weaver}, H.~A., \& {Gonnella}, A.~M. 1995, \jgr, 100, 19057

\bibitem[{{Oparin}(1924)}]{Oparin24}
{Oparin}, A. 1924, {The Origin of Life} (Moscow), tr. in J. D. Bernal, The
  Origin of Life, Cleveland: World, 1967

\bibitem[{{Owen} \& {Mohanty}(2016)}]{OwenMohanty16}
{Owen}, J.~E. \& {Mohanty}, S. 2016, \mnras, 459, 4088

\bibitem[{{Owen} \& {Wu}(2013)}]{OwenWu13}
{Owen}, J.~E. \& {Wu}, Y. 2013, ArXiv e-prints

\bibitem[{{Owen} \& {Wu}(2017)}]{OwenWu17}
---. 2017, \apj, 847, 29

\bibitem[{{Peale} {et~al.}(1979){Peale}, {Cassen}, \& {Reynolds}}]{Peale79}
{Peale}, S.~J., {Cassen}, P., \& {Reynolds}, R.~T. 1979, Science, 203, 892

\bibitem[{{Pierrehumbert} \& {Gaidos}(2011)}]{PierrehumbertGaidos11}
{Pierrehumbert}, R. \& {Gaidos}, E. 2011, \apjl, 734, L13

\bibitem[{{Pierrehumbert}(2011)}]{Pierrehumbert11}
{Pierrehumbert}, R.~T. 2011, \apjl, 726, L8

\bibitem[{{Pourbaix} \& {Boffin}(2016)}]{PourbaixBoffin16}
{Pourbaix}, D. \& {Boffin}, H.~M.~J. 2016, \aap, 586, A90

\bibitem[{{Pourbaix} {et~al.}(2002){Pourbaix}, {Nidever}, {McCarthy}, {Butler},
  {Tinney}, {Marcy}, {Jones}, {Penny}, {Carter}, {Bouchy}, {Pepe}, {Hearnshaw},
  {Skuljan}, {Ramm}, \& {Kent}}]{Pourbaix02}
{Pourbaix}, D., {Nidever}, D., {McCarthy}, C., {Butler}, R.~P., {Tinney},
  C.~G., {Marcy}, G.~W., {Jones}, H.~R.~A., {Penny}, A.~J., {Carter}, B.~D.,
  {Bouchy}, F., {Pepe}, F., {Hearnshaw}, J.~B., {Skuljan}, J., {Ramm}, D., \&
  {Kent}, D. 2002, \aap, 386, 280

\bibitem[{{Poveda} {et~al.}(1996){Poveda}, {Allen}, {Herrera}, {Cordero}, \&
  {Lavalley}}]{Poveda96}
{Poveda}, A., {Allen}, C., {Herrera}, M.~A., {Cordero}, G., \& {Lavalley}, C.
  1996, \aap, 308, 55

\bibitem[{{Raghavan} {et~al.}(1998){Raghavan}, {Schoenert}, {Enomoto},
  {Shirai}, {Suekane}, \& {Suzuki}}]{Raghavan98}
{Raghavan}, R.~S., {Schoenert}, S., {Enomoto}, S., {Shirai}, J., {Suekane}, F.,
  \& {Suzuki}, A. 1998, Physical Review Letters, 80, 635

\bibitem[{{Rauch} \& {Hamilton}(2002)}]{RauchHamilton02}
{Rauch}, K.~P. \& {Hamilton}, D.~P. 2002, in Bulletin of the American
  Astronomical Society, Vol.~34, AAS/Division of Dynamical Astronomy Meeting
  \#33, 938

\bibitem[{{Raymond} {et~al.}(2008){Raymond}, {Barnes}, \&
  {Mandell}}]{Raymond06}
{Raymond}, S.~N., {Barnes}, R., \& {Mandell}, A.~M. 2008, \mnras, 384, 663

\bibitem[{{Raymond} {et~al.}(2004){Raymond}, {Quinn}, \& {Lunine}}]{Raymond04}
{Raymond}, S.~N., {Quinn}, T., \& {Lunine}, J.~I. 2004, Icarus, 168, 1

\bibitem[{{Raymond} {et~al.}(2007){Raymond}, {Scalo}, \& {Meadows}}]{Raymond07}
{Raymond}, S.~N., {Scalo}, J., \& {Meadows}, V.~S. 2007, \apj, 669, 606

\bibitem[{{Reid} {et~al.}(2002){Reid}, {Gizis}, \& {Hawley}}]{Reid2002}
{Reid}, I.~N., {Gizis}, J.~E., \& {Hawley}, S.~L. 2002, \aj, 124, 2721

\bibitem[{{Reiners} \& {Basri}(2008)}]{ReinersBasri08}
{Reiners}, A. \& {Basri}, G. 2008, \aap, 489, L45

\bibitem[{{Remy} \& {Mignard}(1985)}]{Remy1985}
{Remy}, F. \& {Mignard}, F. 1985, \icarus, 63, 1

\bibitem[{{Ribas} {et~al.}(2016){Ribas}, {Bolmont}, {Selsis}, {Reiners},
  {Leconte}, {Raymond}, {Engle}, {Guinan}, {Morin}, {Turbet}, {Forget}, \&
  {Anglada-Escud{\'e}}}]{Ribas16}
{Ribas}, I., {Bolmont}, E., {Selsis}, F., {Reiners}, A., {Leconte}, J.,
  {Raymond}, S.~N., {Engle}, S.~G., {Guinan}, E.~F., {Morin}, J., {Turbet}, M.,
  {Forget}, F., \& {Anglada-Escud{\'e}}, G. 2016, \aap, 596, A111

\bibitem[{{Ribas} {et~al.}(2005){Ribas}, {Guinan}, {G{\"u}del}, \&
  {Audard}}]{Ribas05}
{Ribas}, I., {Guinan}, E.~F., {G{\"u}del}, M., \& {Audard}, M. 2005,
  Astrophys.~J., 622, 680

\bibitem[{{Rickman} {et~al.}(2008){Rickman}, {Fouchard}, {Froeschl{\'e}}, \&
  {Valsecchi}}]{Rickman2008}
{Rickman}, H., {Fouchard}, M., {Froeschl{\'e}}, C., \& {Valsecchi}, G.~B. 2008,
  Celestial Mechanics and Dynamical Astronomy, 102, 111

\bibitem[{{Rickman} {et~al.}(2005){Rickman}, {Fouchard}, {Valsecchi}, \&
  {Froeschl{\'e}}}]{Rickman2005}
{Rickman}, H., {Fouchard}, M., {Valsecchi}, G.~B., \& {Froeschl{\'e}}, C. 2005,
  Earth Moon and Planets, 97, 411

\bibitem[{{Robinson} {et~al.}(2010){Robinson}, {Meadows}, \&
  {Crisp}}]{Robinson10}
{Robinson}, T.~D., {Meadows}, V.~S., \& {Crisp}, D. 2010, \apjl, 721, L67

\bibitem[{{Rodr{\'{\i}}guez} {et~al.}(2012){Rodr{\'{\i}}guez}, {Callegari},
  {Michtchenko}, \& {Hussmann}}]{Rodriguez12}
{Rodr{\'{\i}}guez}, A., {Callegari}, N., {Michtchenko}, T.~A., \& {Hussmann},
  H. 2012, \mnras, 427, 2239

\bibitem[{{Rogers}(2015)}]{Rogers15}
{Rogers}, L.~A. 2015, \apj, 801, 41

\bibitem[{{Roskar}(2010)}]{Roskar2010}
{Roskar}, R. 2010, PhD thesis, University of Washington

\bibitem[{{Ro{\v s}kar} {et~al.}(2012){Ro{\v s}kar}, {Debattista}, {Quinn}, \&
  {Wadsley}}]{Roskar2012}
{Ro{\v s}kar}, R., {Debattista}, V.~P., {Quinn}, T.~R., \& {Wadsley}, J. 2012,
  \mnras, 426, 2089

\bibitem[{{Sahu} {et~al.}(2014){Sahu}, {Bond}, {Anderson}, \&
  {Dominik}}]{Sahu14}
{Sahu}, K.~C., {Bond}, H.~E., {Anderson}, J., \& {Dominik}, M. 2014, \apj, 782,
  89

\bibitem[{{Schaefer} {et~al.}(2016){Schaefer}, {Wordsworth}, {Berta-Thompson},
  \& {Sasselov}}]{Schaefer16}
{Schaefer}, L., {Wordsworth}, R.~D., {Berta-Thompson}, Z., \& {Sasselov}, D.
  2016, \apj, 829, 63

\bibitem[{{Schwieterman} {et~al.}(2016){Schwieterman}, {Meadows},
  {Domagal-Goldman}, {Deming}, {Arney}, {Luger}, {Harman}, {Misra}, \&
  {Barnes}}]{Schwieterman16}
{Schwieterman}, E.~W., {Meadows}, V.~S., {Domagal-Goldman}, S.~D., {Deming},
  D., {Arney}, G.~N., {Luger}, R., {Harman}, C.~E., {Misra}, A., \& {Barnes},
  R. 2016, \apjl, 819, L13

\bibitem[{{S{\'e}gransan} {et~al.}(2003){S{\'e}gransan}, {Kervella},
  {Forveille}, \& {Queloz}}]{Segransan2003}
{S{\'e}gransan}, D., {Kervella}, P., {Forveille}, T., \& {Queloz}, D. 2003,
  \aap, 397, L5

\bibitem[{{Sellwood} \& {Binney}(2002)}]{SellwoodBinney02}
{Sellwood}, J.~A. \& {Binney}, J.~J. 2002, \mnras, 336, 785

\bibitem[{{Selsis} {et~al.}(2007){Selsis}, {Kasting}, {Levrard}, {Paillet},
  {Ribas}, \& {Delfosse}}]{Selsis07}
{Selsis}, F., {Kasting}, J.~F., {Levrard}, B., {Paillet}, J., {Ribas}, I., \&
  {Delfosse}, X. 2007, Astro.~\& Astrophys., 476, 1373

\bibitem[{{Shapley}(1951)}]{Shapley51}
{Shapley}, H. 1951, Proceedings of the National Academy of Science, 37, 15

\bibitem[{{Shields} {et~al.}(2016){Shields}, {Barnes}, {Agol}, {Charnay},
  {Bitz}, \& {Meadows}}]{Shields16}
{Shields}, A.~L., {Barnes}, R., {Agol}, E., {Charnay}, B., {Bitz}, C., \&
  {Meadows}, V.~S. 2016, Astrobiology, 16, 443

\bibitem[{{Shields} {et~al.}(2013){Shields}, {Meadows}, {Bitz},
  {Pierrehumbert}, {Joshi}, \& {Robinson}}]{Shields13}
{Shields}, A.~L., {Meadows}, V.~S., {Bitz}, C.~M., {Pierrehumbert}, R.~T.,
  {Joshi}, M.~M., \& {Robinson}, T.~D. 2013, Astrobiology, 13, 715

\bibitem[{{Shkolnik} \& {Barman}(2014)}]{Shkolnik2014}
{Shkolnik}, E.~L. \& {Barman}, T.~S. 2014, \aj, 148, 64

\bibitem[{{Sotin} {et~al.}(2007){Sotin}, {Grasset}, \& {Mocquet}}]{Sotin07}
{Sotin}, C., {Grasset}, O., \& {Mocquet}, A. 2007, Icarus, 191, 337

\bibitem[{{Spada} {et~al.}(2013){Spada}, {Demarque}, {Kim}, \&
  {Sills}}]{YonseiYale13}
{Spada}, F., {Demarque}, P., {Kim}, Y.-C., \& {Sills}, A. 2013, \apj, 776, 87

\bibitem[{{Storch} \& {Lai}(2014)}]{StorchLai14}
{Storch}, N.~I. \& {Lai}, D. 2014, \mnras, 438, 1526

\bibitem[{{Strangeway} {et~al.}(2010){Strangeway}, {Russell}, \&
  {Luhmann}}]{Strangeway10}
{Strangeway}, R.~J., {Russell}, C.~T., \& {Luhmann}, J.~G. 2010, in European
  Planetary Science Congress 2010, 334

\bibitem[{{Takeda} \& {Rasio}(2005)}]{TakedaRasio05}
{Takeda}, G. \& {Rasio}, F.~A. 2005, \apj, 627, 1001

\bibitem[{{Th{\'e}venin} {et~al.}(2002){Th{\'e}venin}, {Provost}, {Morel},
  {Berthomieu}, {Bouchy}, \& {Carrier}}]{Thevenin02}
{Th{\'e}venin}, F., {Provost}, J., {Morel}, P., {Berthomieu}, G., {Bouchy}, F.,
  \& {Carrier}, F. 2002, \aap, 392, L9

\bibitem[{{Thoul} {et~al.}(2003){Thoul}, {Scuflaire}, {Noels}, {Vatovez},
  {Briquet}, {Dupret}, \& {Montalban}}]{Thoul03}
{Thoul}, A., {Scuflaire}, R., {Noels}, A., {Vatovez}, B., {Briquet}, M.,
  {Dupret}, M.-A., \& {Montalban}, J. 2003, \aap, 402, 293

\bibitem[{{Tian}(2015)}]{Tian15}
{Tian}, F. 2015, Earth and Planetary Science Letters, 432, 126

\bibitem[{{Touma} \& {Wisdom}(1994)}]{ToumaWisdom94}
{Touma}, J. \& {Wisdom}, J. 1994, \aj, 108, 1943

\bibitem[{{Turbet} {et~al.}(2016){Turbet}, {Leconte}, {Selsis}, {Bolmont},
  {Forget}, {Ribas}, {Raymond}, \& {Anglada-Escud{\'e}}}]{Turbet16}
{Turbet}, M., {Leconte}, J., {Selsis}, F., {Bolmont}, E., {Forget}, F.,
  {Ribas}, I., {Raymond}, S.~N., \& {Anglada-Escud{\'e}}, G. 2016, \aap, 596,
  A112

\bibitem[{{Veeder} {et~al.}(1994){Veeder}, {Matson}, {Johnson}, {Blaney}, \&
  {Goguen}}]{Veeder94}
{Veeder}, G.~J., {Matson}, D.~L., {Johnson}, T.~V., {Blaney}, D.~L., \&
  {Goguen}, J.~D. 1994, \jgr, 99, 17095

\bibitem[{{Vidotto} {et~al.}(2013){Vidotto}, {Jardine}, {Morin}, {Donati},
  {Lang}, \& {Russell}}]{Vidotto13}
{Vidotto}, A.~A., {Jardine}, M., {Morin}, J., {Donati}, J.-F., {Lang}, P., \&
  {Russell}, A.~J.~B. 2013, \aap, 557, A67

\bibitem[{{Volk} \& {Gladman}(2015)}]{VolkGladman15}
{Volk}, K. \& {Gladman}, B. 2015, \apjl, 806, L26

\bibitem[{{Vo{\^u}te}(1917)}]{Voute1917}
{Vo{\^u}te}, J. 1917, \mnras, 77, 650

\bibitem[{{Walker}(1981)}]{Walker81}
{Walker}, A.~R. 1981, \mnras, 195, 1029

\bibitem[{{Wargelin} {et~al.}(2017){Wargelin}, {Saar}, {Pojma{\'n}ski},
  {Drake}, \& {Kashyap}}]{Wargelin17}
{Wargelin}, B.~J., {Saar}, S.~H., {Pojma{\'n}ski}, G., {Drake}, J.~J., \&
  {Kashyap}, V.~L. 2017, \mnras, 464, 3281

\bibitem[{{Watson} {et~al.}(1981){Watson}, {Donahue}, \& {Walker}}]{Watson81}
{Watson}, A.~J., {Donahue}, T.~M., \& {Walker}, J.~C.~G. 1981, Icarus, 48, 150

\bibitem[{{Weiss} \& {Marcy}(2014)}]{WeissMarcy14}
{Weiss}, L.~M. \& {Marcy}, G.~W. 2014, \apjl, 783, L6

\bibitem[{{Wertheimer} \& {Laughlin}(2006)}]{WertheimerLaughlin06}
{Wertheimer}, J.~G. \& {Laughlin}, G. 2006, \aj, 132, 1995

\bibitem[{{West} {et~al.}(2008){West}, {Hawley}, {Bochanski}, {Covey}, {Reid},
  {Dhital}, {Hilton}, \& {Masuda}}]{West08}
{West}, A.~A., {Hawley}, S.~L., {Bochanski}, J.~J., {Covey}, K.~R., {Reid},
  I.~N., {Dhital}, S., {Hilton}, E.~J., \& {Masuda}, M. 2008, \aj, 135, 785

\bibitem[{{Williams} {et~al.}(1978){Williams}, {Sinclair}, \&
  {Yoder}}]{Williams78}
{Williams}, J.~G., {Sinclair}, W.~S., \& {Yoder}, C.~F. 1978, \grl, 5, 943

\bibitem[{{Williams} \& {Cieza}(2011)}]{WilliamsCieza11}
{Williams}, J.~P. \& {Cieza}, L.~A. 2011, \araa, 49, 67

\bibitem[{{Willson} {et~al.}(1981){Willson}, {Gulkis}, {Janssen}, {Hudson}, \&
  {Chapman}}]{Willson81}
{Willson}, R.~C., {Gulkis}, S., {Janssen}, M., {Hudson}, H.~S., \& {Chapman},
  G.~A. 1981, Science, 211, 700

\bibitem[{{Wood} {et~al.}(2001){Wood}, {Linsky}, {M{\"u}ller}, \&
  {Zank}}]{Wood01}
{Wood}, B.~E., {Linsky}, J.~L., {M{\"u}ller}, H.-R., \& {Zank}, G.~P. 2001,
  \apjl, 547, L49

\bibitem[{{Wordsworth} {et~al.}(2011){Wordsworth}, {Forget}, {Selsis},
  {Millour}, {Charnay}, \& {Madeleine}}]{Wordsworth11}
{Wordsworth}, R.~D., {Forget}, F., {Selsis}, F., {Millour}, E., {Charnay}, B.,
  \& {Madeleine}, J.-B. 2011, \apjl, 733, L48

\bibitem[{{Yadav} {et~al.}(2016){Yadav}, {Christensen}, {Wolk}, \&
  {Poppenhaeger}}]{Yadav16}
{Yadav}, R.~K., {Christensen}, U.~R., {Wolk}, S.~J., \& {Poppenhaeger}, K.
  2016, \apjl, 833, L28

\bibitem[{{Yang} {et~al.}(2013){Yang}, {Cowan}, \& {Abbot}}]{Yang13}
{Yang}, J., {Cowan}, N.~B., \& {Abbot}, D.~S. 2013, \apjl, 771, L45

\bibitem[{{Yang} {et~al.}(2014){Yang}, {Liu}, {Hu}, \& {Abbot}}]{Yang14}
{Yang}, J., {Liu}, Y., {Hu}, Y., \& {Abbot}, D.~S. 2014, \apjl, 796, L22

\bibitem[{{Yoder}(1995)}]{Yoder95}
{Yoder}, C.~F. 1995, in Global Earth Physics: A Handbook of Physical Constants,
  ed. {T.~J.~Ahrens}, 1--31

\bibitem[{{Young} {et~al.}(2014){Young}, {Desch}, {Anbar}, {Barnes}, {Hinkel},
  {Kopparapu}, {Madhusudhan}, {Monga}, {Pagano}, {Riner}, {Scannapieco},
  {Shim}, \& {Truitt}}]{Young14}
{Young}, P.~A., {Desch}, S.~J., {Anbar}, A.~D., {Barnes}, R., {Hinkel}, N.~R.,
  {Kopparapu}, R., {Madhusudhan}, N., {Monga}, N., {Pagano}, M.~D., {Riner},
  M.~A., {Scannapieco}, E., {Shim}, S.-H., \& {Truitt}, A. 2014, Astrobiology,
  14, 603

\bibitem[{{Zanazzi} \& {Lai}(2017)}]{ZanazziLai17}
{Zanazzi}, J.~J. \& {Lai}, D. 2017, ArXiv e-prints

\bibitem[{{Zhang} \& {Hamilton}(2008)}]{ZhangHamilton08}
{Zhang}, K. \& {Hamilton}, D.~P. 2008, \icarus, 193, 267

\end{thebibliography}

\end{document}